\documentclass[reprint,amsmath,amssymb,aps,prx,longbibliography,floatfix]{revtex4-2} 

\usepackage[dvips]{graphicx}
\usepackage{dcolumn}
\usepackage{bm}
\usepackage{hyperref}

\makeatletter
\let\MYcaption\@makecaption
\makeatother
\usepackage[subrefformat=parens]{subcaption}
\makeatletter
\let\@makecaption\MYcaption
\makeatother
\usepackage{ulem}
\usepackage {pifont}
\usepackage{color}
\newcommand{\red}[1]{\textcolor{red}{#1}}
\newcommand{\blue}[1]{\textcolor{blue}{#1}}
\newcommand{\green}[1]{\textcolor{green}{#1}}

\begin{document}

\title{Light and  Slow-Neutron Diffraction by Nanodiamond-Dispersed \\Nanocomposite Holographic Gratings}

\author{Yasuo Tomita,$^1$} 
\email{ytomita@uec.ac.jp}
\author{Akihisa Kageyama,$^1$ Yuko Iso,$^1$ Koichi Umemoto,$^2$ Atsushi Kume,$^2$ Ming Liu$^2$\\  Christian Pruner,$^3$
 Tobias Jenke,$^4$ Stephanie Roccia,$^4$ Peter Geltenbort,$^4$ Martin Fally,$^5$\\ and J\"urgen Klepp$^5$}
\email{juergen.klepp@univie.ac.at}
\affiliation{%
$^1$ Department of Engineering Science, University of Electro-Communications,
1-5-1 Chofugaoka, Chofu, Tokyo 182-8585, Japan\\
$^2$ Central Research Center, Daicel Corp., 1239 Shinzaike, Aboshi, Himeji, Hyogo 671-1283, Japan\\
$^3$ Department of Material Science and Physics, University of Salzburg, A-5020 Salzburg, Austria\\
$^4$ Institut Laue-Langevin, 71 avenue des Martyrs, CS 20156, 38042 Grenoble Cedex 9, France\\ 
$^5$ Faculty of Physics, University of Vienna, Boltzmanngasse 5, A-1090 Wien, Austria
}%

\date{\today}

\begin{abstract}
We demonstrate the use of nanodiamond  in constructing holographic nanoparticle-polymer composite  transmission gratings with large saturated refractive index modulation amplitudes at both optical and slow-neutron wavelengths, resulting in efficient control of light and slow-neutron beams. Nanodiamond possesses a  high refractive index at optical wavelengths and  large coherent and  small incoherent scattering cross sections with low absorption at slow-neutron wavelengths. We describe the synthesis of nanodiamond, the preparation of photopolymerizable nanodiamond-polymer composite  films, the construction of  transmission gratings in nanodiamond-polymer composite films and light optical diffraction experiments. Results of slow-neutron diffraction from such gratings are also presented.
\end{abstract}

\maketitle

\section{\label{sec:level1}INTRODUCTION}
Nanodiamond (ND), an intriguing allotrope of carbon, is a carbon nanoparticle with truncated octahedral architecture~\cite{krueger2008,mochalin2012,dolmatov2007}. 
ND has been of considerable interest because they are considered to possess  superior mechanical, thermal, electrical and optical properties of diamonds such as high hardness, high Young's modulus, low coefficient of friction, high thermal conductivity, low thermal expansion, low  heat capacity,  high insulation, large bandgap,  and   high refractive index. These advantages could potentially  lead to various applications  including  novel composite materials, additive,  abrasive, lubricant, plating, heat dissipation materials, biomedical/medicine materials for drug delivery systems and biosensing/imaging, electronics, magnetic recording, coating, luminescent materials, nonlinear optical materials, and single photon emitters/spin detectors for quantum optics and photonics~\cite{zhang2018,mochalin2017,aharonovich2011,radulaski2019}. 
Here we report on the realization of ND-dispersed nanocomposite holographic gratings for efficient control of light and slow-neutron beams,  formed in the so-called 
photopolymerizable nanoparticle-polymer composite (NPC)~\cite{tomita2016}, where nanoparticles (ND particles in this case) are uniformly dispersed in monomer capable of radical-mediated chain-growth or step-growth photopolymerization.

Since our first report on volume holographic recording in an NPC dispersed with TiO$_2$ nanoparticles~\cite{suzuki2002}, NPC gratings using various types of inorganic  and organic nanoparticles such as SiO$_2$~\cite{suzuki2004,hata2011}, ZrO$_2$~\cite{suzuki2006,sakhno2007,garnweitner2007,fujii2014}, ZnS~\cite{peng2015}, luminescent nanoparticles/nanorods~\cite{sakhno2008,peng2019}, nanotubes~\cite{guo2020}, zeolites~\cite{leite2009}, semiconductor quantum dots~\cite{liu2009,sakhno2009} and 
hyperbranched polymer~\cite{tomita_hbp2006,tomita2019} were reported.
The dispersion of nanoparticles  in recorded holographic gratings results in large saturated refractive index modulation amplitudes ($\Delta n_{\rm sat}$) and simultaneously makes their mechanical and thermal stability higher in comparison to those recorded in a conventional all-organic photopolymer~\cite{tomita2008,hata2010}. It was also shown that optical nonlinearities were induced in NPCs dispersed with  nanoparticles possessing large optical nonlinearities including nonlinear Bragg diffraction~\cite{smirnova2005}, optical multi-stability~\cite{tomita2005a}, light-induced transparency~\cite{liu2010} and  nonlinear multi-wave mixing~\cite{liu2012} due to cascaded high-order optical nonlinearities associated with the local-field enhancement~\cite{dolgaleva2009}. NPCs using various nanoparticles can be used for photonic applications such as holographic data storage~\cite{momose2012,takayama2014}, volume holographic optical elements for wearable headsets~\cite{tomita_hbp2016,fernandez2019}, security holograms~\cite{peng2019} and distributed feedback plastic lasers~\cite{smirnova2014}.

Moreover, in quest for the realization of efficient slow-neutron  beam-control and ultrahigh precision measurements using neutron interferometers~\cite{rauch2015}, we have also demonstrated various manipulations of slow-neutron beams at a few nm wavelengths with NPC transmission gratings dispersed with SiO$_2$ or ZrO$_2$ nanoparticles. These include two and multi-beam splitting and beam deflection with up to 90\,\% efficiency~\cite{fally2010,klepp2012a, klepp2012b}, i.e., deflective mirror-like functionality. In order to realize such high diffraction efficiency, it was necessary to effectively increase the interaction length of an NPC transmission grating of one hundred  microns  thickness by tilting it at a large angle (up to 70$^\circ$) about an axis parallel to the grating vector. Such a large increase in the effective grating thickness is, however, detrimental  to some device applications
since it introduces a substantive increase in incoherent scattering and absorption loss of a slow-neutron beam propagating in supporting glass substrates and a host polymer material~\cite{blaickner2019}. It is also better to avoid a large grating tilt in order not to cause a substantial decrease in the diffraction efficiency  due to the increased angular selectivity of the tilted grating for a slow-neutron beam with the finite beam divergence of the order of 1\,mrad. 
 In order to circumvent such disadvantages, one needs to increase $\Delta n_{\rm sat}$ for slow neutrons, $\Delta n_{\rm sat,\text{\tiny{N}}}$, instead of increasing the effective grating thickness. 
This calls for a new type of nanoparticles possessing stronger interactions with slow neutrons than SiO$_2$ and ZrO$_2$ nanoparticles.

In this paper we describe new NPC gratings dispersed with ND for light and slow-neutron diffraction.  
High hardness and low thermal expansion as well as high refractive index of ND in ultraviolet and visible spectral regions provide the realization of highly efficient NPC phase gratings with large $\Delta n_{\rm sat}$ and high environmental stability.  
ND is also considered useful in neutron optics for possessing very large coherent and very small incoherent scattering cross sections with low absorption for neutrons~\cite{klepp2012c}. For example, quasi-specular reflection of cold neutrons from ND powders was  demonstrated~\cite{nesvizhevskyMat2010,nesvizhevsky2018}. ND also provides a possibility for efficient slow-neutron beam manipulation by NPC phase gratings of only a few tens of microns thick as we demonstrated  recently~\cite{tomita2019a}.  
Here we describe the synthesis of ND, the preparation of the NPC, recording of ND-dispersed transmission NPC gratings and the light optical diffraction experiments.  We conclude with results on slow-neutron diffraction from ND-dispersed NPC gratings. 

\section{\label{sec:level2}HOLOGRAPHIC GRATING FORMATION IN NPC MEDIA}
\begin{figure}[htbp]
\centerline{\includegraphics[width=8.6cm]{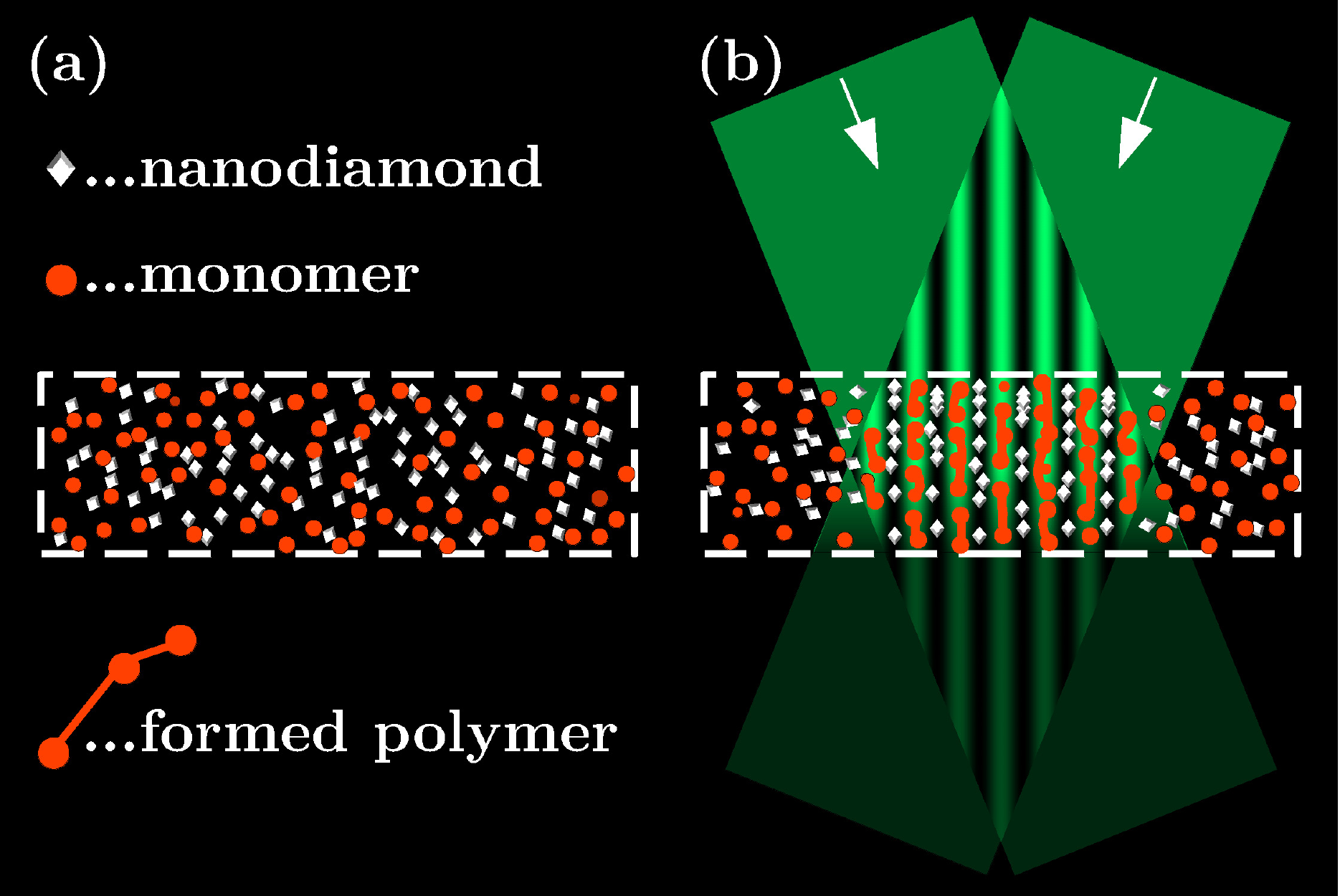}}
\caption{Schematic of distributions of constituents (monomer, the formed polymer and nanoparticles)  (a) before and (b)  during holographic exposure to describe the process of holographic assembly of nanoparticles in the formed polymer.}
\label{fig:f1}
\end{figure}

The basic idea of using nanoparticles as secondary species  stems from the fact that  an increase in the refractive index change per one diffusing monomer molecule requires a high refractive index
of the monomer in organic multi-component photopolymer systems~\cite{tomlinson1976}.  
The incorporation of inactive nanoparticles in host photopolymer  gives a much higher refractive index modulation amplitude ($\Delta n$) in holographic recording as compared with conventional binder-based photopolymers when a difference in refractive index between   nanoparticles and the formed polymer is sufficiently large. The inclusion of nanoparticles also contributes to the suppression of polymerization shrinkage and the improvement of thermal stability~\cite{tomita2008,hata2011}.
The basic mechanism of holographic grating formation can be described as follows: Suppose that nanoparticles (ND particles in our case) are uniformly dispersed in a host monomer capable of radical photopolymerization  as shown in Fig.\,\ref{fig:f1}(a).  Spatially non-uniform light illumination  produces free radicals by dissociation of initiators and the subsequent reaction of free radicals with monomer molecules leads to the polymerization reaction between monomer radicals and individual monomer molecules in the bright 
regions. This polymerization process lowers the chemical potential of the monomer in the bright regions, leading to the migration (diffusion) of monomer molecules from the dark to the bright regions. On the other hand, photo-insensitive  nanoparticles experience counterdiffusion from the bright to the dark regions  since they are not consumed by light exposure and their chemical potential increases in the bright regions due to the consumption of the monomer. Such a polymerization-driven mutual diffusion process essentially continues until the photopolymerization completes. 
In this way  the mutual  diffusion of monomer molecules and nanoparticles results in the  periodic assembly of nanoparticles during holographic exposure [see Fig.\,\ref{fig:f1}(b)] and a refractive index grating (a phase hologram) is created owing to a compositional and density difference between the bright and the dark regions. 
Indeed, it was  confirmed experimentally that the distribution of dispersed inorganic nanoparticles followed the light-intensity interference pattern in the formed polymer: A phase shift of $\pi$ between the periodic density distributions  of formed polymer and inorganic nanoparticles was induced as a result of the mutual diffusion~\cite{tomita2005,tomita2006,suzukitomita2006}.

Generally, NPC volume gratings with large $\Delta n_{\rm sat}$ at
short grating spacings are required for light and slow-neutron
optics applications for high diffraction efficiency~\cite{tomita2016}. 
In particular, thin volume holographic gratings with very large $\Delta n_{\rm sat}$ ($>2\times$10$^{-2}$) are required for wearable headsets for augmented and mixed reality applications in visible spectral regions in order to obtain high diffraction efficiency and wide fields of view~\cite{piao2014,yeom2015}. 
In slow-neutron optics  a  diffractive mirror with $\eta\sim$\,100\,\% requires large $\Delta n_{\rm sat,\text{\tiny{N}}}$ at slow-neutron wavelengths, 
moderate film thickness, and short grating spacing~\cite{klepp2012a}.
Regardless of wavelengths considered, $\Delta n_{\rm sat}$ of a holographic phase grating with the 1st-order (Bragg-matched) periodic spatial modulation formed in NPC media is approximately given by (see Appendix A for the detailed derivation) 
\begin{equation}
\Delta n_{\rm sat}=a_1\Delta f |n_{np}-n_p|,
\label{eq:eq1}
\end{equation}
where $n_{np}$\,($n_p$) is the refractive index of a nanoparticle\,(the formed host polymer) 
at either a light or a slow-neutron wavelength, and $\Delta f$ is the amplitude of  a sinusoidally modulated spatial density of dispersed nanoparticles at their spatially dependent volume fraction $f_{np}$  (see Appendix A). Here $\Delta f$ is  either 
smaller than or equals to  spatially averaged volume fraction $\bar{f}_{np}$.
The parameter $a_1$ is twice the magnitude of the 1st-order Fourier component of 
the refractive index modulation of a phase grating, {\it i.e.}, it is  
unity for a pure sinusoidal waveform and is $4\sin (r\pi)/\pi$ for a rectangular 
waveform, with the duty ratio $r$ of the concentration distribution in volume between 
nanoparticle-rich and -poor regions.

Equation~(\ref{eq:eq1}) implies the following chief ingredients for the enhancement of $\Delta n_{\rm sat}$:
(i) A near-sinusoidal form of $f_{np}$ to minimize the high order diffraction signals, (ii) Large $\Delta f$ by efficient mutual diffusion of nanoparticles and monomer under holographic exposure, and  
(iii) Large $\vert n_{np}-n_p\vert$.
The item (i) is usually difficult to achieve since the photopolymerization-driven mutual diffusion process is nonlinear, so that the spatial waveform of $f_{np}$ would contain high order components~\cite{tomita2005spie}. The item (ii) can be realized by large $f_{np}$ and the facilitation of mutual diffusion under appropriate viscosity of the NPC system. The item (iii) is possible in the visible light spectral region since $n_p$ is generally of the order of 1.5 but $n_{np}$ can be much larger than $n_p$.  ND is a good choice of  nanoparticle species for the enhancement of  $\Delta n_{\rm sat}$ at visible light wavelengths since it possesses a large value of $n_{np}$  ($\approx2.42$ for bulk diamond). 
 In neutron optics the neutron refractive index ($ n_\text{\tiny{N}}$) at a neutron wavelength ($\lambda_\text{\tiny{N}}$) in matter is approximately given by the following formula~\cite{sears1989}:
\begin{equation}
n_\text{\tiny{N}}=1-\frac{\lambda_\text{\tiny{N}}^2}{2\pi}b_c\rho,
\label{eq:eq2}
\end{equation}
where $b_c$ and  $\rho$ are the mean bound coherent scattering length 
and  the atomic number density of matter, respectively.  
The primary requirement for the enhancement of  
$\Delta n_{\rm sat,\text{\tiny{N}}}$ at $\lambda_\text{\tiny{N}}$ can be translated into a large absolute difference in the scattering length density (SLD) $b_c\rho$ between a nanoparticle and the formed polymer.  In other words, 
nanoparticles having a large value for SLD  are  desired. Indeed, bulk diamond has a very large value of SLD to be about $11.8\times10^{-6}$\,\AA$^{-2}$~\cite{avdeev2013}, as compared with $2.08\times10^{-6}$\,\AA$^{-2}$ for bulk Si, $3.64\times10^{-6}$\,\AA$^{-2}$ for bulk SiO$_2$ and $7.3\times10^{-6}$\,\AA$^{-2}$ for bulk graphite \cite{ILLNeutronDataBooklet2003}, all of them used as materials for
neutron-beam diffraction at slow-neutron wavelengths and neutron optics in general. 
Even if the value for ND may be reduced to some extent as compared to that of bulk diamond due to several contaminations of graphite and other organic/inorganic materials on the surface of an ND core~\cite{mochalin2012}, the use of ND is highly favorable for constructing NPC gratings at slow-neutron wavelengths since 
such an NPC grating would possess $\Delta n_{sat}$ up to three times as large as that dispersed with SiO$_2$ nanoparticles. Thus,  
 high $\eta$ at slow-neutron wavelengths at relaxed angular selectivity and small tilt angles is achievable. 
The latter features are very desirable to make slow-neutron beam experiments at limited intensities and with a finite beam divergence of several mrad.
The use of ND for NPC gratings also provides other advantages such as very low incoherent scattering loss and low absorption for neutrons. Other approaches for fabricating (optically) thin diffraction gratings that are also used in neutron optics have been put forward in, for instance, Refs.\,\cite{gruenzweigRScIntr2008,miaoNL2014}.  
 
Note that there is no reason why Eq.\,(\ref{eq:eq1}) should not also hold for neutrons. In light and slow-neutron diffraction experiments $\eta$ depends on the product of $\Delta n_{\rm sat}$ being proportional to $a_1\Delta f$ [see  Eq.\,(\ref{eq:eq1})] and the grating thickness ($d$). 
Because it is possible to estimate  $a_1\Delta f$  and $d$ from light diffraction measurements only,   one can assess the performance of  a recorded NPC grating at slow-neutron wavelengths from measured values for $a_1\Delta f d$ and a known value for 
$|(b_c\rho)_{np}-(b_c\rho)_p|$, or alternatively one can estimate $|(b_c\rho)_{np}-(b_c\rho)_p|$ from optically estimated values for $a_1\Delta f d$  and the measurement of $\eta$ in slow-neutron diffraction experiments. Such discussions will be given in Sec.\,\ref{subSec:SlowNeutronDiffr}.  

\section{EXPERIMENTAL RESULTS AND DISCUSSIONS} 
\subsection{Synthesis and Analysis of Nanodiamonds}\label{subsec:SampleSyn}
\subsubsection{Synthesis of surface-modified nanodiamonds}
\begin{figure}[t]
\begin{center}
    \begin{tabular}{c}
    \hspace{-10mm} 
 \begin{minipage}[htbp]{0.5\hsize}
  \begin{center}
 \includegraphics[clip, width=25mm]{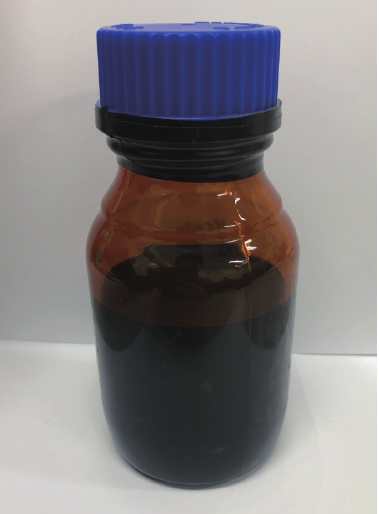}
\subcaption{}
  \end{center}
 \end{minipage}
 \begin{minipage}[htbp]{0.5\hsize}
  \begin{center}
 \includegraphics[clip, width=44mm]{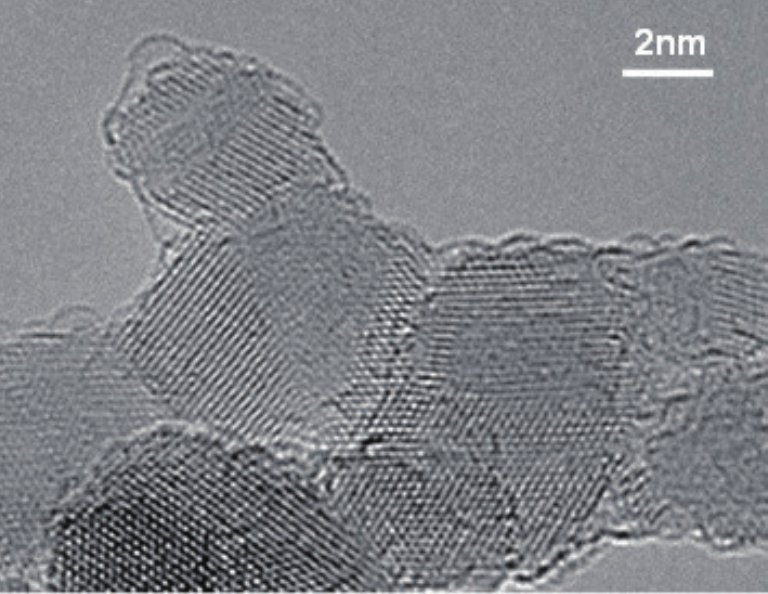}
\subcaption{}
  \end{center}
 \end{minipage}
    \vspace{3mm}
 \end{tabular}
 \caption{\label{fig:f2}(a) Photograph of   an  ND sol in a vial. 
 (b) TEM image of surface-modified ND.}
 \end{center} 
\end{figure}

We synthesized a cluster of ND by the detonation of trinitrotoluene/hexogen charges between 40/60 and 60/40 in wt.\%. The purification process used to recover the detonated ND particles from the detonation soot involves the acid treatment in water solvent for the removal of metal oxides originated from a container, followed by the decantation treatment to remove residual graphite having sp$^2$ orbitals. Organic modification on the surface of each ND core was made with silane coupling agent in an organic solvent  after the ultrasonication of ND aggregates with ceramic (ZrO$_2$) microbeads used as crushing agents. ND suspension sol  was finally prepared in a solution of methyl isobutyl ketone (MIBK). Figure~\ref{fig:f2}(a) illustrates an ND sol in a vial. The opacity indicates non-negligible broadband absorption in the visible as a result of remaining sp$^2$ graphite layers on the surface of an ND core. Figure~\ref{fig:f2}(b) shows a transmission electron microscopy (TEM) image of surface-modified ND deposited on carbon-coated grids after evaporating a tetrahydrofuran suspension of the ND sol. It can be observed that each ND has the  size of the core possessing the sp$^3$ structure  to be approximately 4\,nm on average and that they aggregate due to their lateral diffusion during the evaporation process. Note that remaining sp$^2$ layers were not detected in the TEM image. 
\subsubsection{Analysis of surface-modified nanodiamonds}
\begin{figure}[b]
\centerline{\includegraphics[clip, width=65mm]{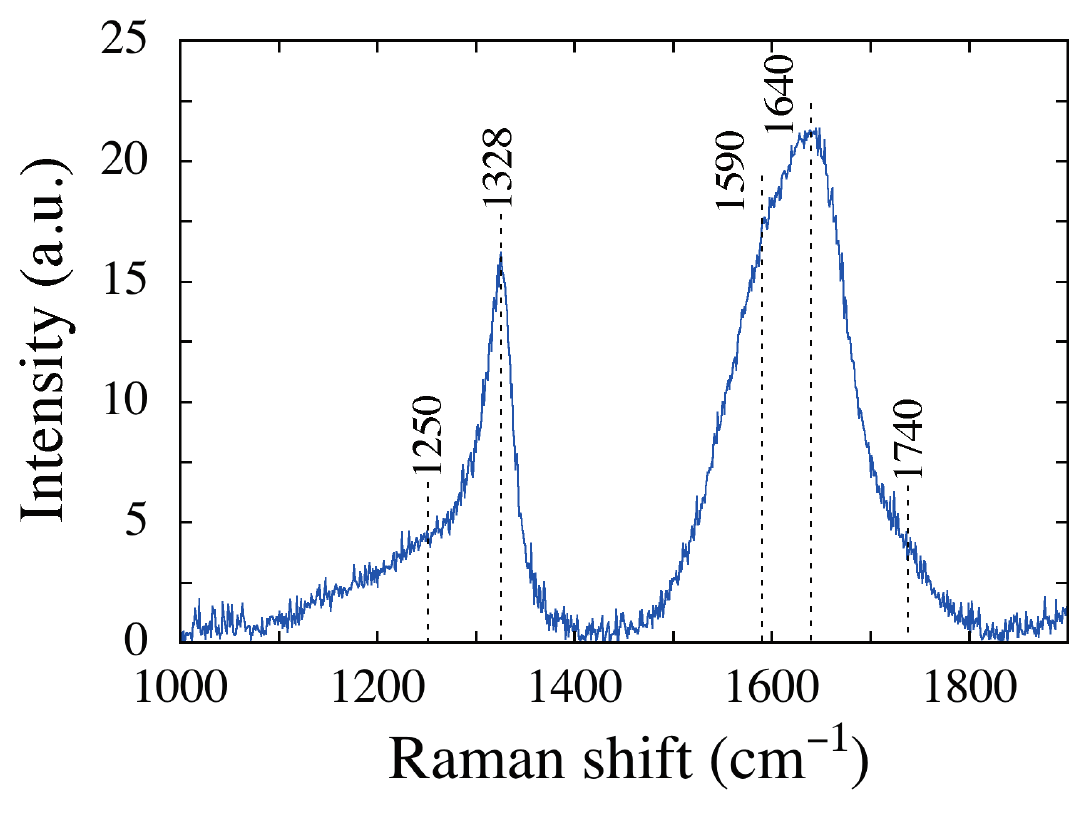}}
\caption{Raman spectrum of surface-modified ND after the purification process.}
\label{fig:f3}
\end{figure}

In order to quantitatively examine remaining graphite components in surface-modified ND, we performed Raman spectroscopy by using a commercial Raman microscope (LabRAM HR Evolution LabSpec 6, HORIBA Scientific). 
Figure~\ref{fig:f3} shows the UV-Raman spectrum of synthesized ND after the purification process at an excitation laser wavelength of 325\,nm and at excitation laser power of 0.8\,mW. Clear Raman peaks can be attributed to a down-shifted peak at 1328\,cm$^{-1}$ from ND  with respect to that from bulk diamond (1332\,cm$^{-1})$ and to an O-H bending peak at 1640\,cm$^{-1}$ from one of the H$_2$O vibrational modes, respectively~\cite{mochalin2012,mochalin2008}. Other possible peaks from 
contributions from smaller ND particles or smaller coherent scattering domains separated by defects in larger ND particles at 1250\,cm$^{-1}$, 
the G-band peak of graphitic sp$^2$ carbon at 1590\,cm$^{-1}$  and a C=O stretching peak at 1740\,cm$^{-1}$ that is coming from surface functional groups~\cite{mochalin2008} were not noticeable. 
However, it is known that the formation of non-diamond carbons occurs after the growth of the diamond sp$^3$ structure as a result of a decrease in pressure below the diamond-graphite equilibrium during detonation~\cite{mochalin2008}. Therefore, such a non-carbon layer must remain on the ND surface even after the purification process, as is also confirmed from the opacity of the ND sol shown in Fig.\,\ref{fig:f2}(a). The content of graphitic sp$^2$ carbon is considered to be more or less a few wt.\% in our surface-modified ND as the relative Raman peak at 1590\,cm$^{-1}$ partially hidden in the broad peak centered at 1640\,cm$^{-1}$ is very similar to the result in Ref.~\cite{mochalin2008}.

\begin{figure}[htbp]
\centerline{\includegraphics[clip, width=65mm]{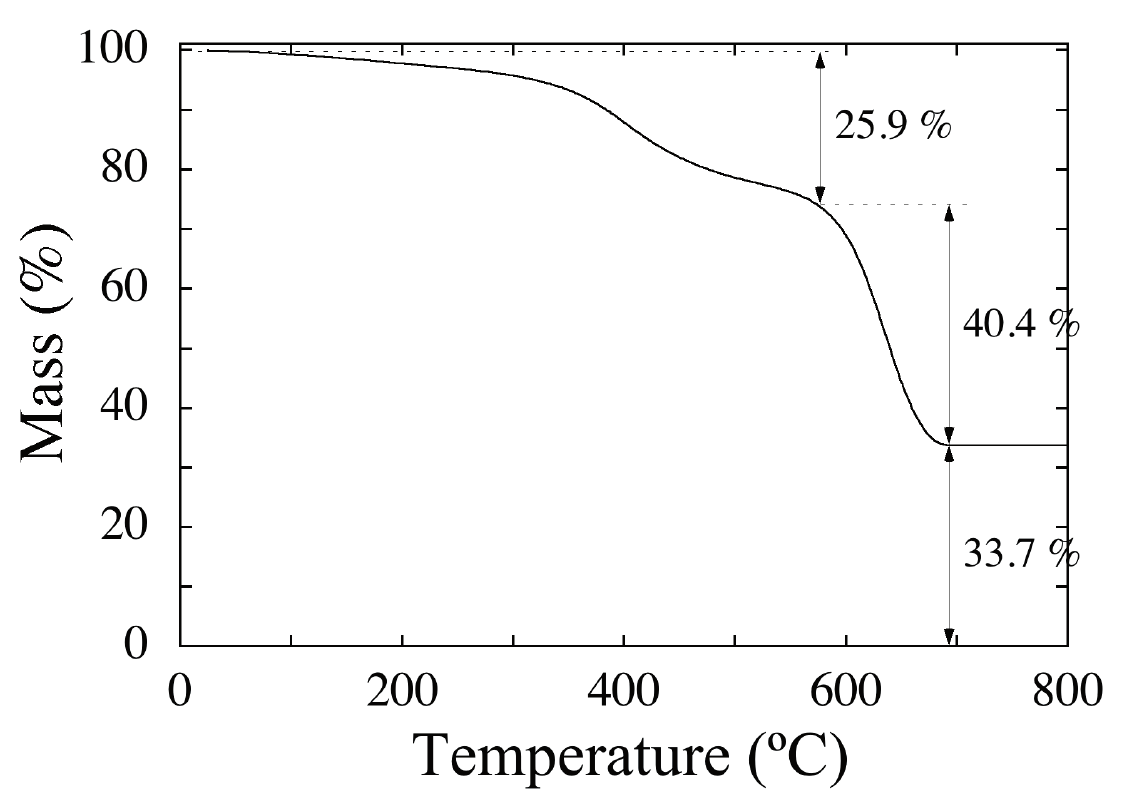}}
\caption{TGA curve of surface-modified ND under an air flow.}
\label{fig:f4}
\end{figure}

We also carried out the thermogravimetric analysis (TGA)  by using a commercial TGA/DTA system (EXSTAR TG/DTA6200, Seiko Instruments) 
to quantify the ratio of the components in the surface-modified ND. It was done at a heating rate of 20$^\circ$\,C/min from 25$^\circ$\,C to 815$^\circ$\,C under air flow. The mass ratio was calculated using the weight loss \% at the onset temperatures of the decomposition of grafted organics, diamond and non-diamond carbon contents and inorganic species. The result is shown in Fig.\,\ref{fig:f4}.  It was found that grafted organics at 25.9\,wt.\% attached by the silane coupling agent was decomposed first (see the second shoulder near 600$^\circ$\,C) followed by the decomposition of the diamond and non-diamond carbon 
contents (ND and graphite) at 40.4\,wt.\% (see the abrupt change near 680$^\circ$\,C), leaving behind the remaining metals and oxides such as ZrO$_2$ at 33.7\,wt.\% introduced during the crushing and surface modification processes.

\begin{figure}[t]
\centerline{\includegraphics[clip, width=65mm]{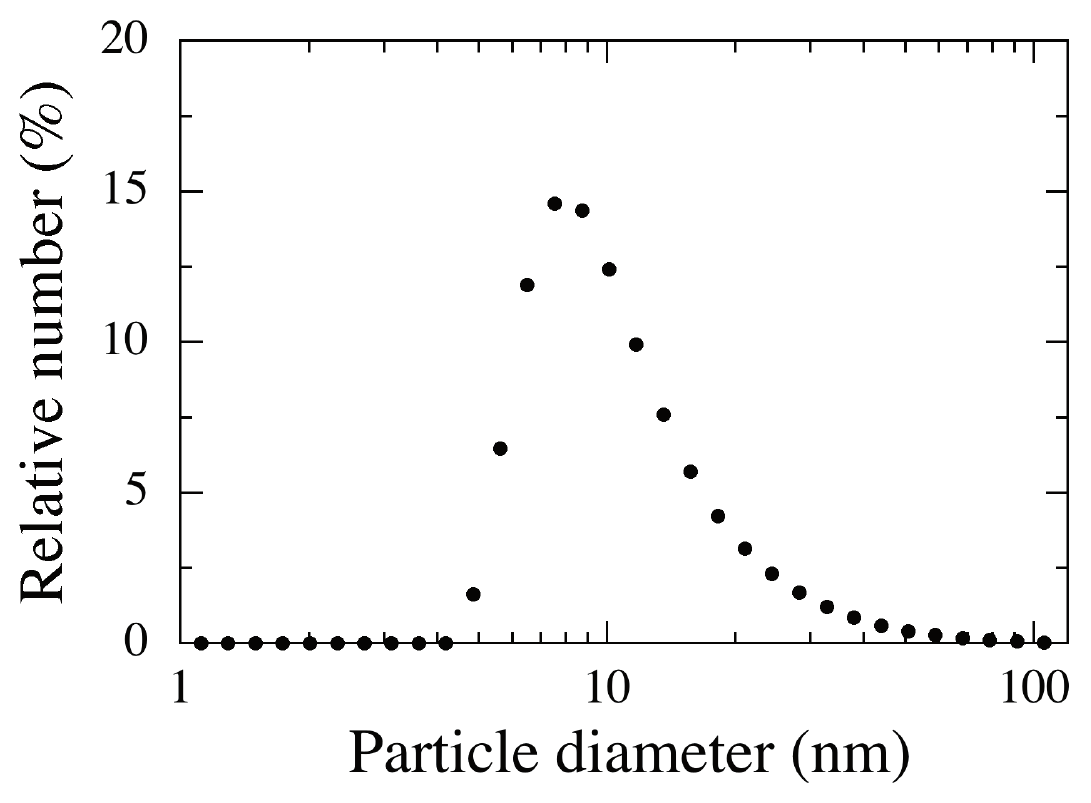}}
\caption{Size distribution of surface-modified ND in tetrahydrofuran solution.}
\label{fig:f5}
\end{figure}

We further measured the hydrodynamic size dispersion of the surface-modified ND. Figure~\ref{fig:f5} shows the size dispersion of surface-modified ND in tetrahydrofuran solution  obtained  by means of a dynamic light scattering instrument (Zetasizer, Malvern Panalytical). It is found that the size distribution exhibits a typical broad peak stemming from aggregates as reported for colloidal solutions of SiO$_2$ nanoparticle and ND~\cite{ahn2004,terada2019}. The hydrodynamic median-average size (d50)  is approximately 9.6\,nm that must include the outer layers of the surface-modified ND. 
This result indicates that the main volume of the surface-modified ND is occupied by soot (mainly the contents of graphite and ZrO$_2$) and organic components.
From the results shown in Figs.~\ref{fig:f2}(b)--\ref{fig:f5} we may deduce the geometric structure of a multilayered concentric sphere of our surface-modified ND by use of the core radius of the sp$^3$ ND (2\,nm) and the radius of the surface-modified ND (4.8\,nm) provided that a given wt.\% value of the graphite component in the surface-modified ND is used. Let us assume that the weight fraction of the ND core and graphite components are 35.4 and 5\,wt.\%, respectively, resulting in the weight fraction of the diamond and non-diamond contents of 40.4\,wt.\%, which would be 
an appropriate assumption for well purified ND~\cite{duan2019}. It is then straightforward to approximately estimate the relative volume fractions of the ND core and the shell portion (soot and organic layers). Using known numerical values for densities of bulk diamond (3.51\,g/cm$^3$),  graphite (2.\,g/cm$^3$), ZrO$_2$ (5.68\,g/cm$^3$) and acrylate (1.18\,g/cm$^3$), we find 7.2, 6.8, 18.3 and 67.7\,vol.\% for 
ND, graphite, ZrO$_2$ and grafted organics, respectively.  We can then estimate the layer thickness of each component measured from the surface of the ND core to be 
0.5, 0.8 and 1.5\,nm for graphite, ZrO$_2$ and grafted organics, respectively.
\subsection{Preparation of Samples}\label{subsec:SamplePrep}
We dispersed the ND sol to a blend of two acrylate monomers, tri/tetrapentaerythritol acrylate (PETIA, Daicel-ALLNEX LTD.) of refractive indices of $n_D$=1.487 and 1.515 in liquid and solid phases, respectively,
\begin{figure}[htbp]
\begin{center}
 \includegraphics[clip, width=60mm]{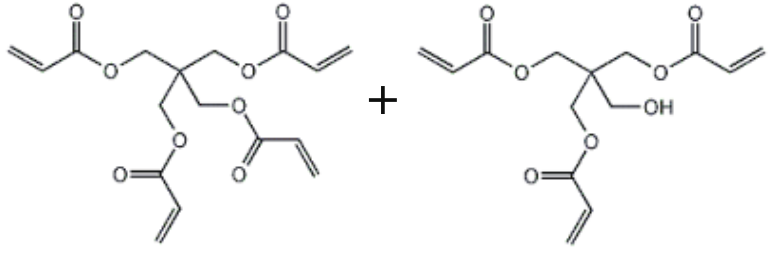}
 \subcaption{}
 \includegraphics[clip, width=50mm]{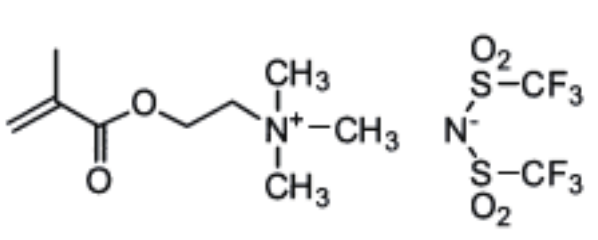}
  \subcaption{}
 \includegraphics[clip, width=30mm]{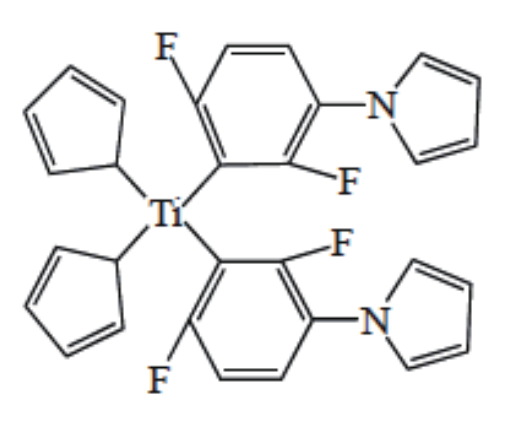}
  \subcaption{}
   \end{center} 
 \caption{Chemical structures of (a) PETIA, (b) MOE-200T and (c) Irgacure784.}
 \label{fig:f6}
\end{figure}
and a single functional ionic liquid monomer (MOE-200T, Piotrek) of refractive indices of  $n_D$=1.43 and 1.45 in liquid and solid phases, respectively 
 at various ratios  in wt.\%. The concentration of ND after evaporation of MIBK was as high as 22\,vol.\%. As shown later, we found that the unity wt.\% ratio of PETIA to MOE-200T gave the highest value for $\Delta n_{\rm sat}$. Titanocene (Irgacure784, Ciba) with the refractive index of $n_D$=1.76 was used as a radical photoinitiator  to provide the photosensitivity in the green. The concentration of the photoinitiator was 4.3 wt.\% with respect to the monomer blend. These chemical structures are shown in Fig.\,\ref{fig:f6}.  MIBK was removed from the mixed syrup  containing the ND sol. The syrup was kept in a vial at 60$^\circ$\,C for 6 hours and then 
cast on a spacer-loaded glass substrate. Finally, it was covered with another glass plate  to be used as NPC film samples. 

\begin{figure}[b]
\centerline{\includegraphics[clip, width=70mm]{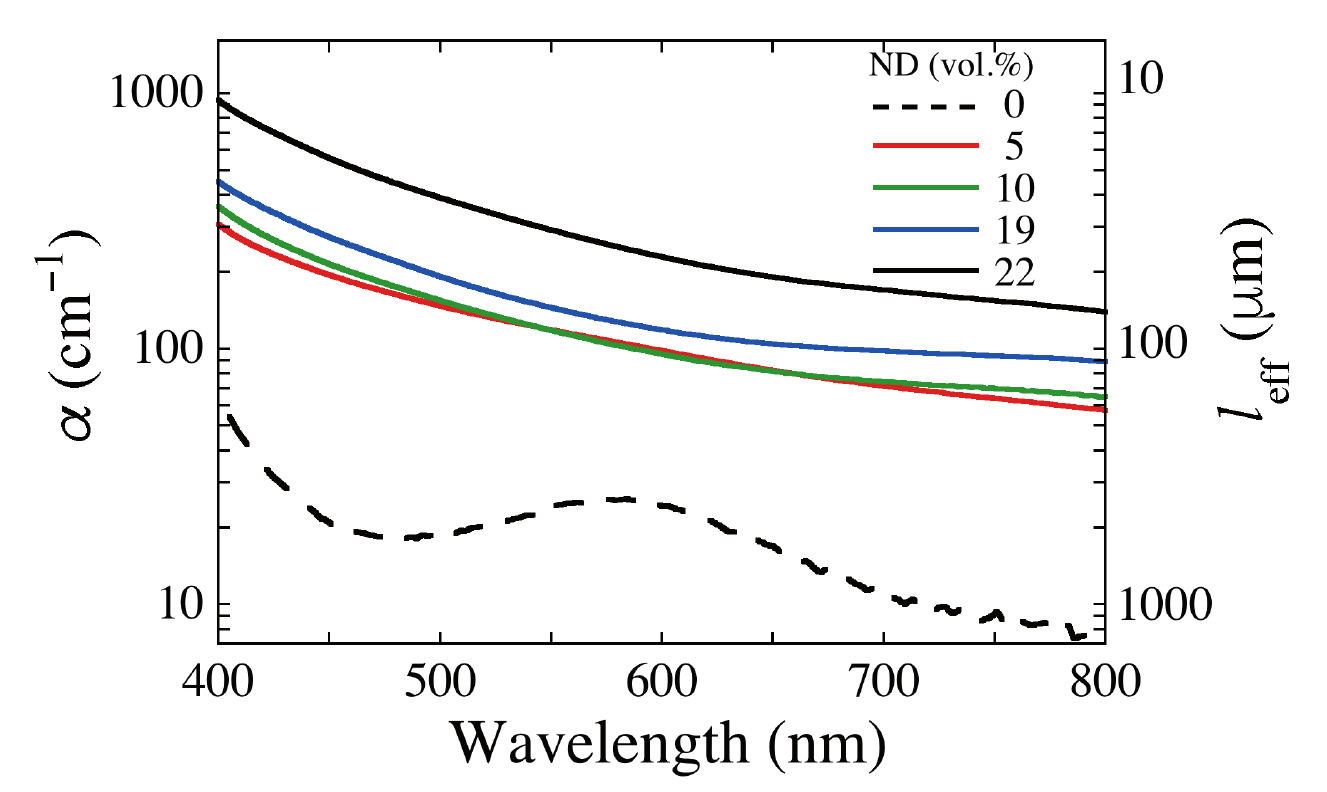}}
\caption{Spectral absorption coefficients and the corresponding effective thickness for uniformly cured NPC film samples at various doping concentrations of ND.}
\label{fig:f7}
\vspace*{5.mm}
\end{figure}

Figure\,\ref{fig:f7} shows spectral absorption coefficients $\alpha$ and the effective thickness [$\ell_{\rm eff}(\equiv 1/\alpha$)] of uniformly cured NPC film samples without and with ND at different doping concentrations. It can be seen  that the optical absorption due to the inclusion of ND extends over the whole visible spectral range as seen in Fig.\,\ref{fig:f2}(a). However, 
it can also be seen that values for 
$\ell_{\rm eff}$ at a wavelength of 532\,nm are of the order of 100\,$\mu$m at ND concentrations lower than 20 vol.\%, substantially thick enough for our use as holographic gratings.  
\subsection{Holographic Recording and Light Diffraction Experiments}\label{subsec:HologrRecLightDiffrExp}
\subsubsection{Experimental setup for holographic recording and a method for data analysis}
In a holographic recording experiment we used a two-beam interference setup to record unslanted and plane-wave transmission gratings in  ND-dispersed NPC films at grating spacing\,($\Lambda$) of 0.5\,$\mu$m and 1\,$\mu$m by two mutually coherent beams of equal intensities  from a diode-pumped frequency-doubled Nd:YVO$_4$ laser operating at a wavelength of 532\,nm. A low-intensity He-Ne laser beam operating at a wavelength of 633\,nm was employed as a readout beam to monitor the buildup dynamics of a recorded plane-wave transmission NPC grating since  the photoinitiator  employed was insensitive in the red. All the beams were s-polarized. We characterized NPC gratings by measuring the buildup dynamics of the diffraction efficiency 
($\eta$)  at a probe wavelength of 633\,nm. Here, $\eta$ was defined as $I_1/(I_0+I_1)$, where $I_1$ is the 1st order diffracted probe intensity and $I_0$ is the transmitted probe intensity. Then, $d$  was estimated by measuring the saturated $\eta$ ($\eta_{\rm sat}$) as a function of Bragg-angle detuning ($\Delta\theta_B$) from a phase-matched Bragg angle probed at 633\,nm after the completion of the recording process 
and then by fitting the data to Kogelnik's coupled-wave formula that describes light diffraction in the Bragg (two-wave coupling) regime~\cite{kogelnik1969,uchida1973,yeh1993} as given by
\begin{equation}
\eta_{\rm sat} (\Delta\theta_B) = \frac{\sin^2\biggl
\{\frac{\pi\Delta n_{\rm sat}d}{\lambda\cos\theta_B}\biggl[1+\biggl(\frac{\lambda\cos\theta_B\sin\Delta\theta_B}{\Lambda\Delta n_{\rm sat}}\biggr)^2\biggr]^{1/2}\biggr\}}{1+\biggl(\frac{\lambda\cos\theta_B\sin\Delta\theta_B}{\Lambda\Delta n_{\rm sat}}\biggr)^2},
\label{eq:kogelnik}
\end{equation}
where $\lambda$ is a readout wavelength in vacuum, $\Lambda$ is grating spacing and $\theta_B$ is a phase-matched Bragg angle inside a sample. 
 
\subsubsection{Light diffraction properties}
The buildup dynamics of $\Delta n$  at 532\,nm  were evaluated by using the Kogelnik's formula with the already measured buildup dynamics of $\eta$ and $d$ at 633\,nm 
by multiplying the time series of $\Delta n$  evaluated at 633\,nm with a
factor being the ratio of $\Delta n_{\rm sat}$ measured at 532\,nm to that measured at 633\,nm. 
The rationale of using the Kogelnik's formula for evaluating  $d$ and $\Delta n_{\rm sat}$ will be discussed later in this subsection. 

\begin{figure}[t]
\begin{tabular}{cc}
 \hspace{-4mm} 
 \begin{minipage}[htbp]{0.5\hsize}
  \begin{center}
 \includegraphics[clip, width=30mm]{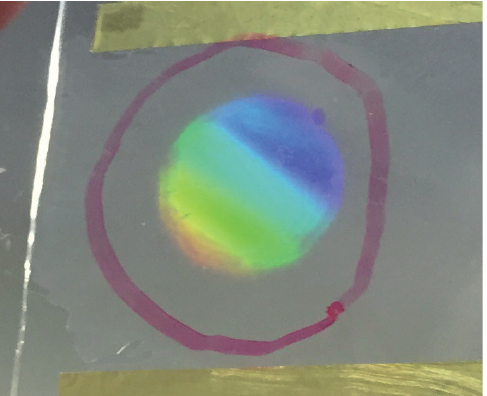}
 \subcaption{}
  \end{center}
 \end{minipage}
 \begin{minipage}[htbp]{0.5\hsize}
  \begin{center}
 \includegraphics[clip, width=38mm]{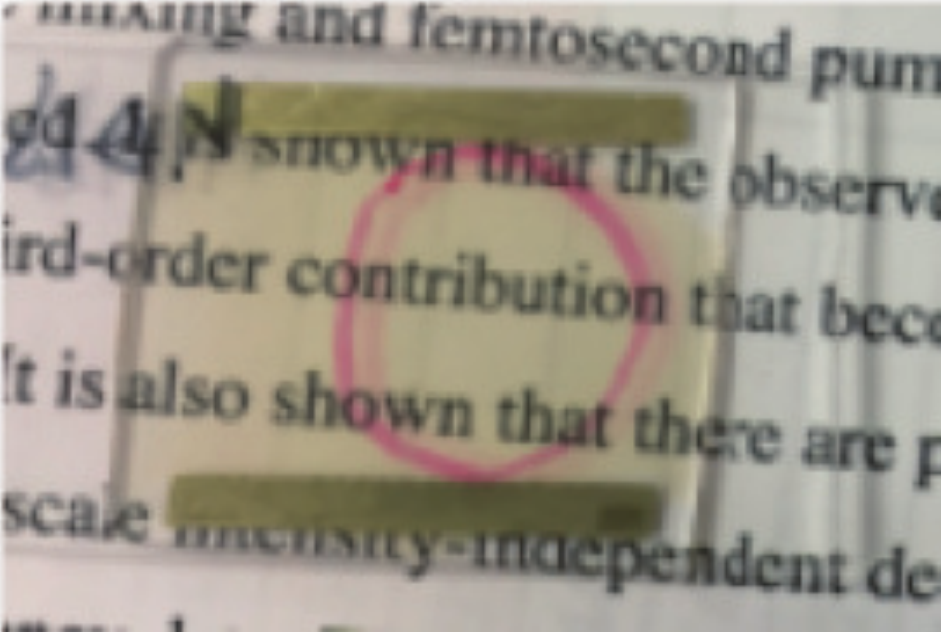}
\subcaption{}
  \end{center}
 \end{minipage}
  \vspace{4mm} \\
 \begin{minipage}[htbp]{0.5\hsize}
  \begin{center}
 \includegraphics[clip, width=32mm]{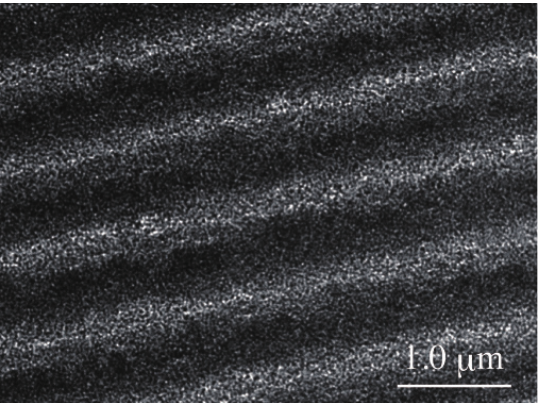}
\subcaption{}
  \end{center}
 \end{minipage}
 \end{tabular}
 \caption{(a) A photograph of a recorded NPC transmission plane-wave grating (circularly marked in red) at grating spacing of 1\,$\mu$m under white light illumination from a fluorescent lamp. (b) A photograph of the same grating viewed from the top.  (c) A TEM  image of the cross section of the NPC grating.}
 \label{fig:f8}
\end{figure}

Figure~\ref{fig:f8}(a) illustrates a photograph of an ND composite grating (approximately 10\,mm in diameter) recorded in an NPC film sample dispersed with 19\,vol.\% ND. Figure~\ref{fig:f8}(b) shows the same grating viewed from the top. Good uniformity and high transparency of the grating are seen. 
The shallow brownish color is a result of absorption by the residual graphite component attached to the peripheral of ND particles. Figure~\ref{fig:f8}(c) shows a TEM image of the cross section of the recorded grating. The dark (bright) banded areas correspond to high concentration portions of ND (the formed polymer), confirming holographic assembly~\cite{tomita2005}  of ND in the formed polymer.  

\begin{figure}[t]
\begin{center}
 \includegraphics[clip, width=65mm]{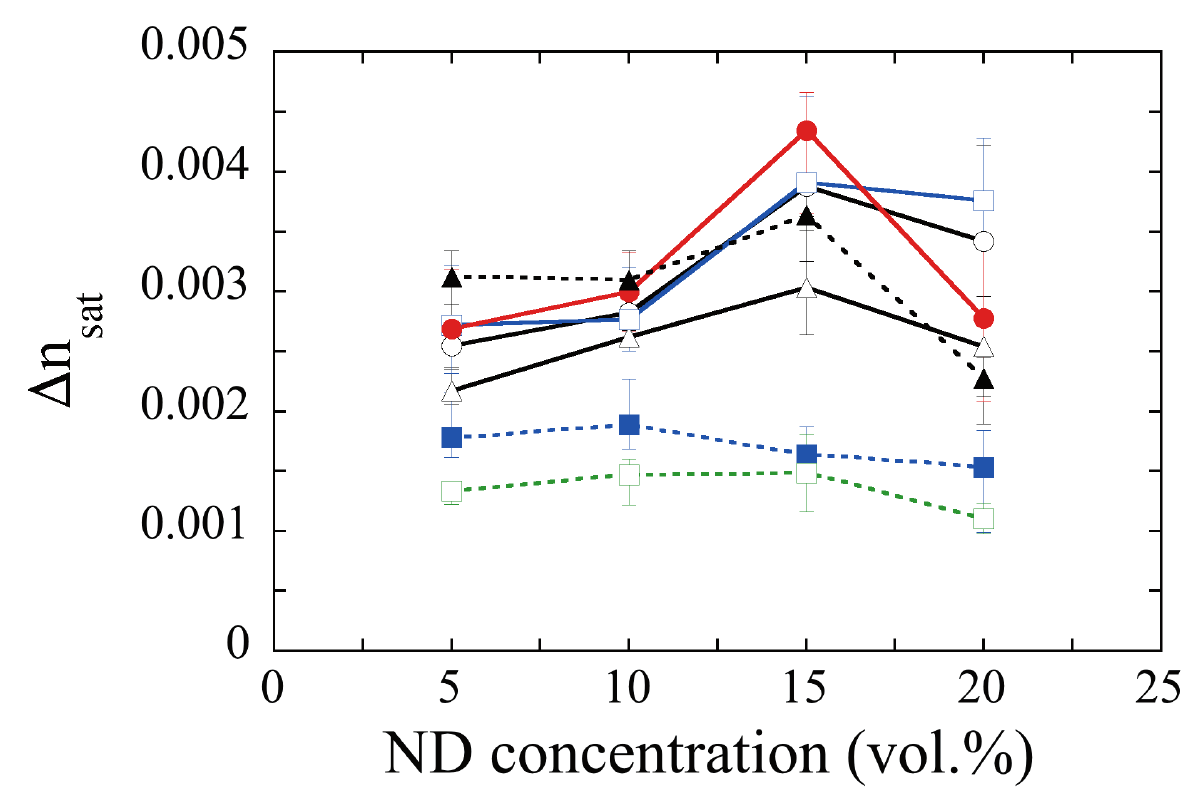}
\subcaption{}
 \includegraphics[clip, width=65mm]{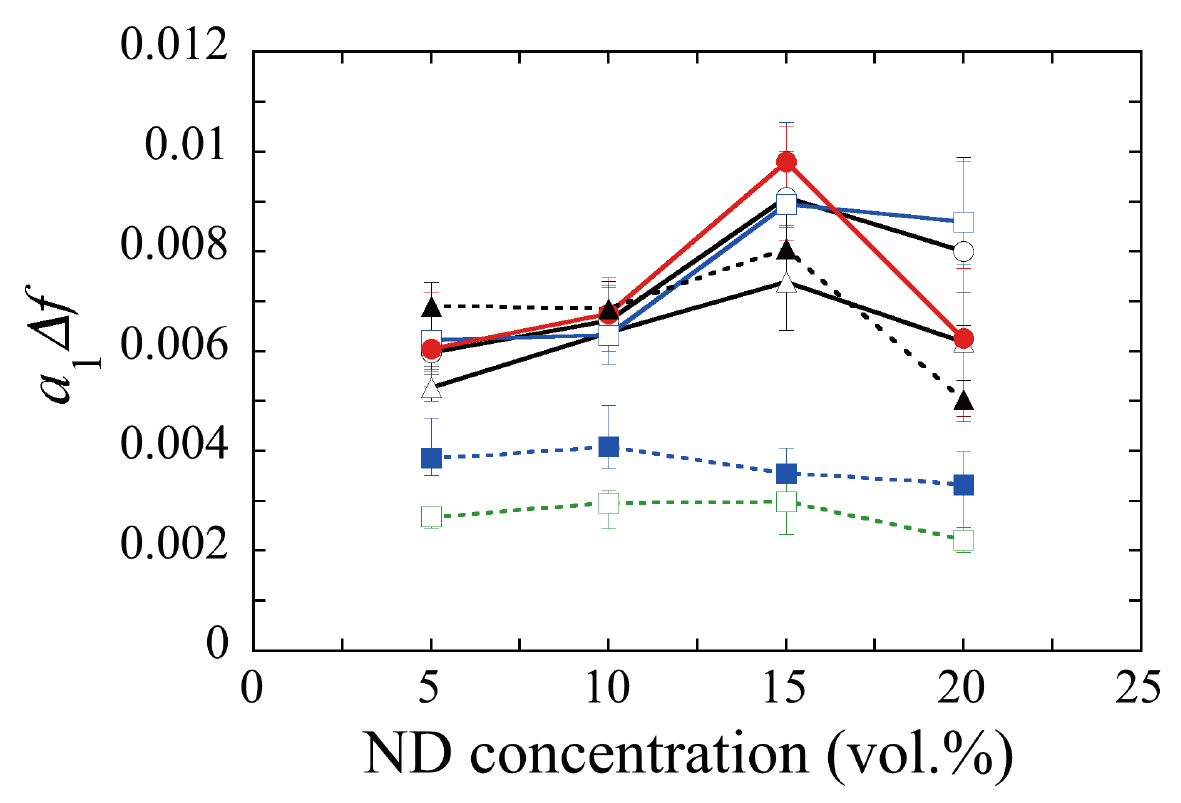}
\subcaption{}
 \end{center} 
\vspace*{-4.mm}
 \caption{Dependences of (a) $\Delta n_{\rm sat}$ and (b)   $a_1\Delta f$  on concentration of dispersed ND in an NPC grating of 1\,$\mu$m spacing at different ratios of [MOE-200T]   with respect to the sum of  [PETIA] and [MOE-200T]. All symbols shown in the figures correspond to 0 ($\triangle$), 0.25 ($\circ$), 0.40 (\blue{$\square$}), 
 0.50 (\red{$\bullet$}),  0.60 ($\blacktriangle$), 0.75 (\blue{$\blacksquare$}), 1 (\green{$\square$}), respectively.}
 \label{fig:f9}
 \end{figure}

Figure~\ref{fig:f9}(a) shows  the  dependence on ND concentration  of $\Delta n_{\rm sat}$ evaluated at a wavelength of 532\,nm for an NPC grating of 1\,$\mu$m spacing at different ratios of [MOE-200T] with respect to the sum of [PETIA] and [MOE-200T] all measured in wt.\%. The recording intensity at 75\,mW/cm$^2$ was chosen so as to maximize
$\Delta n_{\rm sat}$. It can be seen that $\Delta n_{\rm sat}$ peaks at the ND concentration of 15\,vol.\% and at [MOE]/([PETIA]+[MOE])=0.50, {\it i.e.}, [MOE]=[PETIA] in wt.\%.   
The observed decrease in $\Delta n_{\rm sat}$ at ND concentrations higher than 15\,vol.\% is caused by increased light scattering that results in 
 a decrease in intensity-interference fringe contrast. 
The optimum condition of [MOE]=[PETIA] maximizing $\Delta n_{\rm sat}$ is 
found by the following consideration: an increase in [MOE-200T] lowers the viscosity of the blend monomer, facilitating the mutual diffusion of unreacted monomer and ND so that 
 $a_1\Delta f$ in Eq.\,(\ref{eq:eq1}) increases, as seen in Fig.\,\ref{fig:f9}(b).
Too much addition of MOE-200T, however, inhibits the development of cross-linking network structures during holographic exposure, resulting in a decrease in 
 $a_1\Delta f$ and thus in $\Delta n_{\rm sat}$. It can be seen in Fig.\,\ref{fig:f9}(b) that $a_1\Delta f$ peaks at the ND concentration of 15\,vol.\%  ($\bar{f}_{np}$=0.15) and at [MOE-200T]/([PETIA]+[MOE-200T])=0.50. But the peak value of 0.01 is much smaller than the ND concentration of 15\,vol.\% ({\it i.e.}, 0.15). We speculate that because the spatial density distribution of ND after holographic exposure is more or less rectangular with $r\approx 0.55$  [see Fig.\,\ref{fig:f8}(c)],  such a small value ($\approx$ 0.07) for $a_1\Delta f/\bar{f}_{np}$ is caused by insufficient mutual diffusion of ND and monomer molecules during holographic exposure. A further increase in $\Delta f$ by facilitating the mutual diffusion is necessary to increase $\Delta n_{\rm sat}$.
 An increase in $|n_{np}-n_p|$ due to a decrease in $n_p$ at higher concentration of MOE-200T also contributes to an increase in $\Delta n_{\rm sat}$ but only marginally.  
  In order to estimate  $a_1\Delta f$  from extracted values for $\Delta n_{\rm sat}$ [see Eq.\,(\ref{eq:eq1})], we measured values for  $n_p$ of uniformly cured  polymer films without ND dispersion at various ratios of [MOE-200T] to [PETIA]  by means of an Abbe refractometer (DR-M2, ATAGO). 
The refractive index of a surface-modified ND, $n_{np}$, was evaluated from a refractive index measurement of an ND-dispersed polymer film after uniform curing by the so-called m-line spectrometer by a prism coupler~\cite{sharda2003}, from which $n_{np}$ was determined by 1.9420 via a volume fractional estimation using 
Eq.\,(\ref{eq:A1}) with help of the known $n_p$.
A simple volume fractional estimation of $n_{np}$ by using our estimated volume fractions of the constituents shown in the previous subsection was also carried out. It was found that $n_{np}$ was 1.7713 with refractive indices of 2.4173~\cite{phillip1964}, 2.7733~\cite{djurisic1999}, 2.1745~\cite{bodurov2016} and 1.492~\cite{kasarova2007} for diamond, graphite, ZrO$_2$ and acrylate, respectively. The calculated value is lower than the measured one. We speculate that the main discrepancy results from the use of the simple volume fractional calculation: The geometric structure ({\it i.e.}, a multilayered concentric sphere) must be taken  into account in the calculation~\cite{poularikas1985}, which would be left for our future investigation.
 
\begin{figure}[b]
\begin{center}
 \includegraphics[clip, width=65mm]{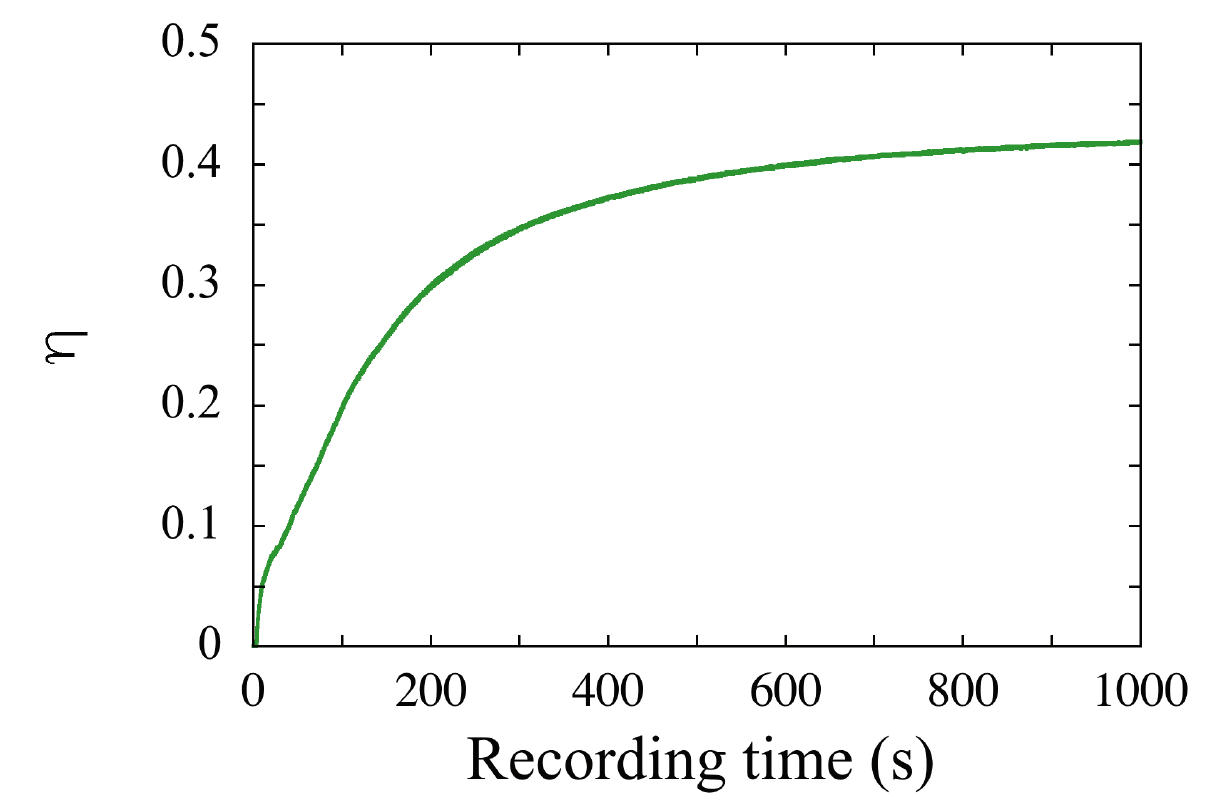}
\subcaption{}
 \includegraphics[clip, width=62mm]{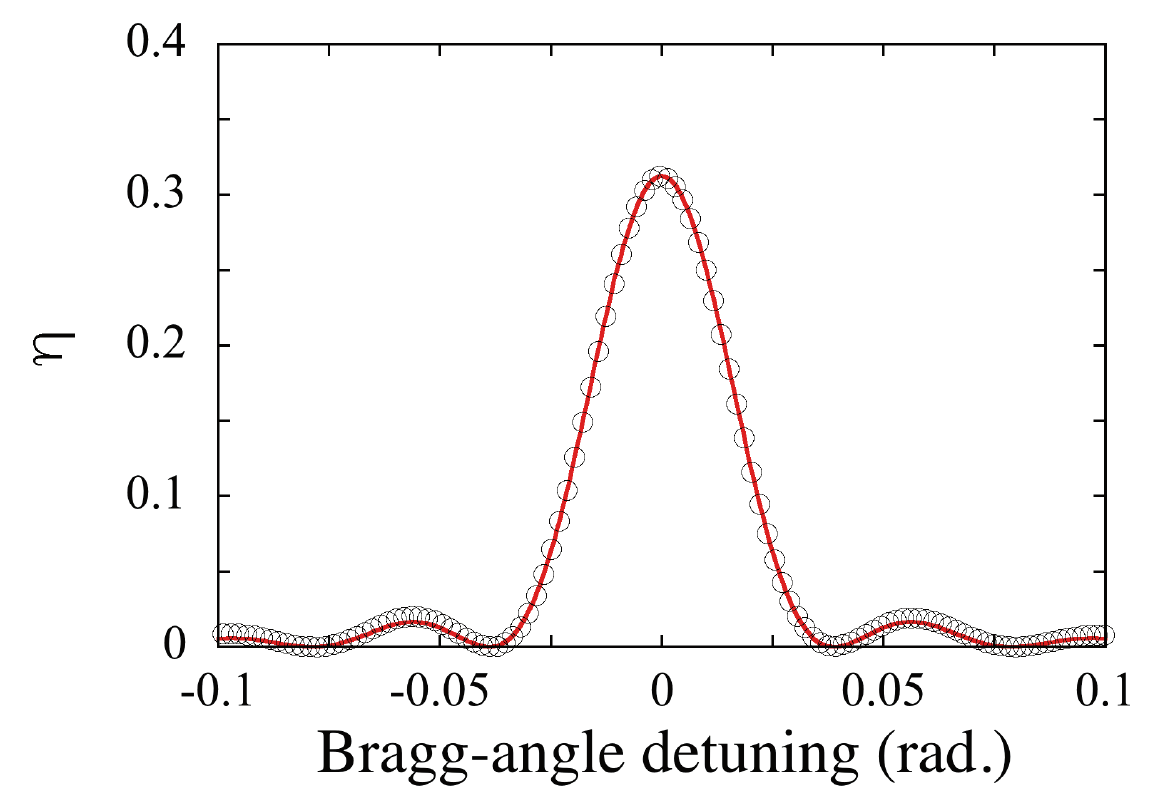}
\subcaption{}
 \includegraphics[clip, width=62mm]{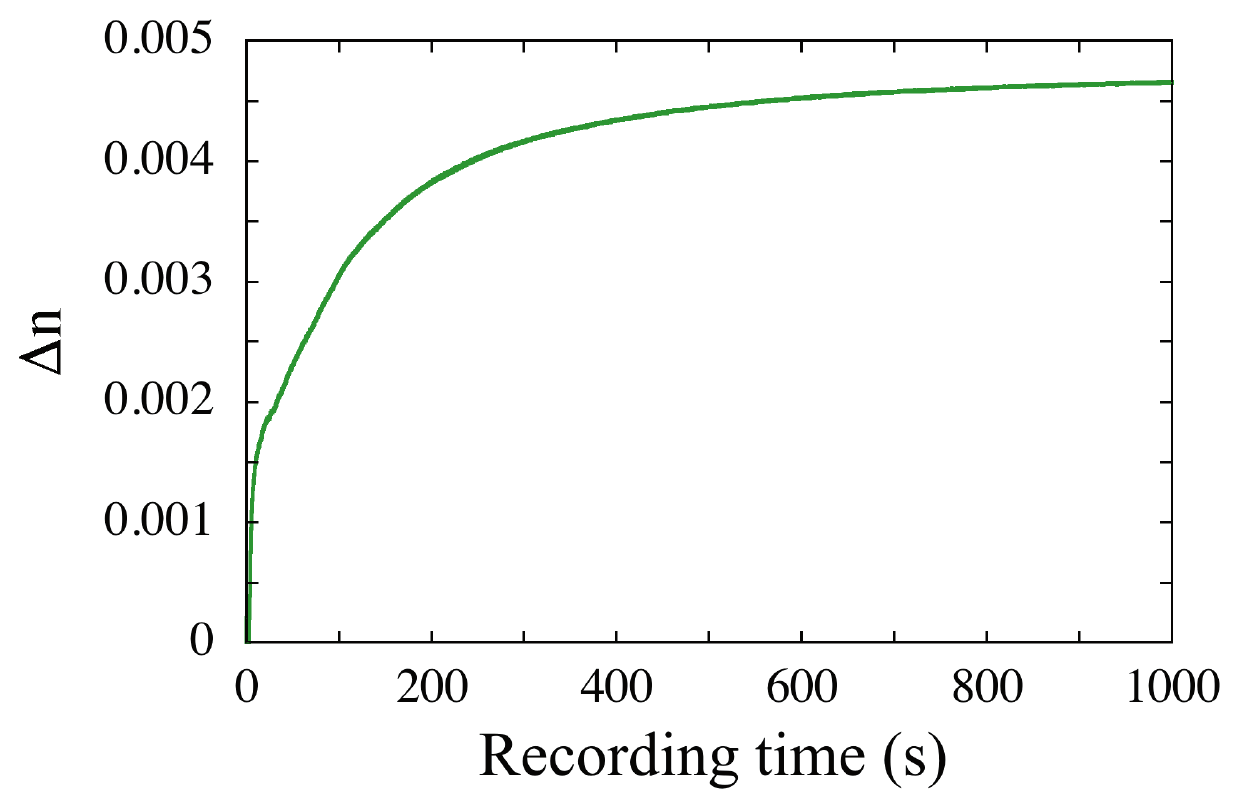}
\subcaption{}
  \end{center}
    \vspace{3mm}
 \caption{(a)  Extracted buildup dynamics of $\eta$   at 532\,nm from  an ND composite grating of 1\,$\mu$m spacing. (b) Bragg-angle detuning curve at 633\,nm. The solid line is the least-squares fit to the measured data.  (c) Extracted buildup dynamics of $\Delta n$ at 532\,nm.}
 \label{fig:f10}
\end{figure}

Figure~\ref{fig:f10}(a) shows the buildup dynamics of $\eta$ at 532\,nm, that is extracted from that at 633\,nm, for an unslanted and plane-wave transmission NPC   
grating of 1\,$\mu$m spacing at the  optimum ND concentration (15\,vol.\%)  and the optimum recording intensity (75\,mW/cm$^2)$. Furthermore, we used the optimum ratio [MOE]/([PETIA]+[MOE]) of 0.5. We refer to this sample as the NPC grating \#1. 
We note that the measured curve of $\eta$ shown in Fig.\,\ref{fig:f10}(a) is presented at 532\,nm in order to evaluate the performance of light diffraction at the same wavelength as a recording one (532\,nm) in order for our present results to be applicable to a more generic case of non-plane wave recording where   readout has to be done at the same Bragg-matched wavelength.
 Figure~\ref{fig:f10}(b) shows the Bragg-angle detuning curve probed at 633\,nm after the saturation of $\eta$. The solid curve in red denotes the least-squares curve fit of the data to the Kogelnik's formula from which we find 
$d$ to be 25.29$\pm0.02$\,$\mu$m. It can be seen that the curve fit follows well with the measured data, indicating that the recorded NPC grating is uniform along the film's thickness direction~\cite{uchida1973}. 
Figure~\ref{fig:f10}(c) shows the corresponding buildup dynamics of $\Delta n$ at 532\,nm. 
 It can be seen that the NPC grating \#1 has  $\Delta n_{\rm sat}$ of approximately 4.7$\times$10$^{-3}$. This value is similar to that reached by SiO$_2$ nanoparticle-dispersed NPC gratings at  the same  grating spacing   used  in previous slow-neutron diffraction measurements~\cite{fally2010,klepp2012b}. We also find values for  $a_1\Delta f$ and $a_1\Delta f d$ to be approximately 0.011 and 
0.268\,$\mu$m, respectively, with $n_p=1.4980$. 

\begin{figure}[t]
\begin{center}
 \includegraphics[clip, width=65mm]{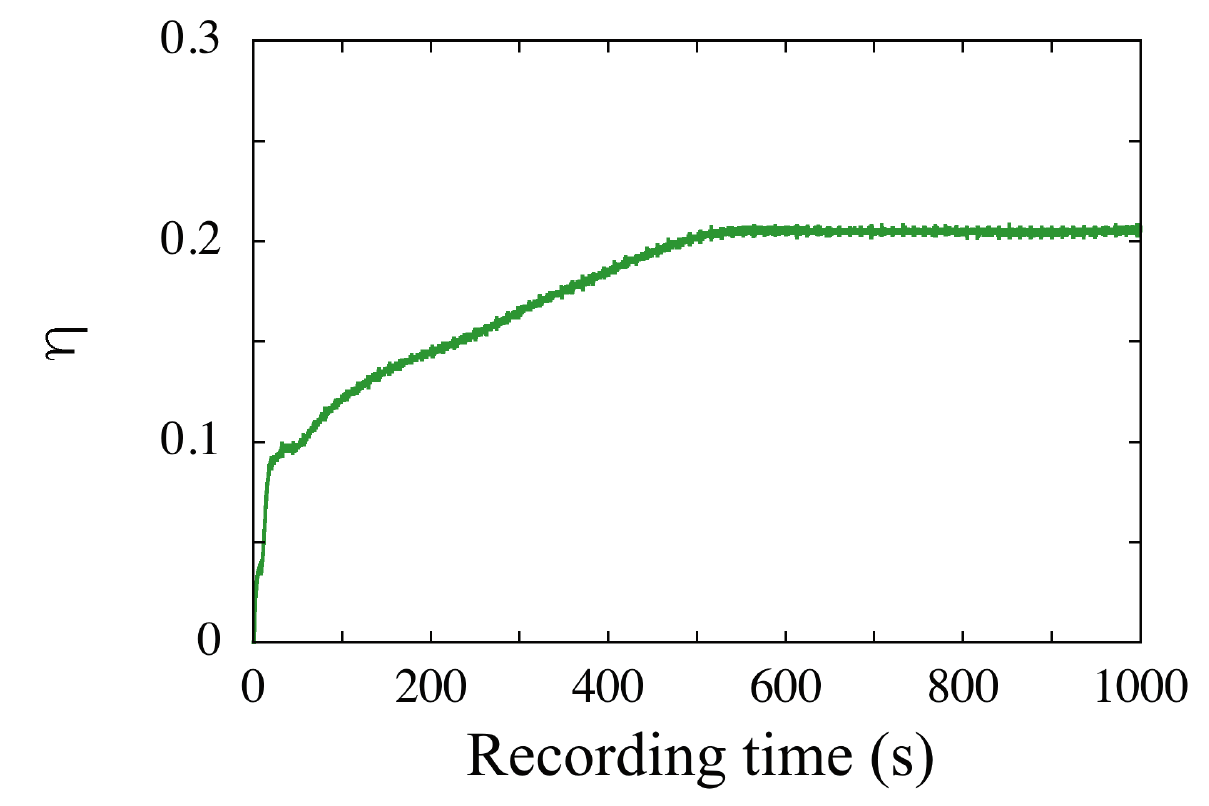}
\subcaption{}
 \includegraphics[clip, width=62mm]{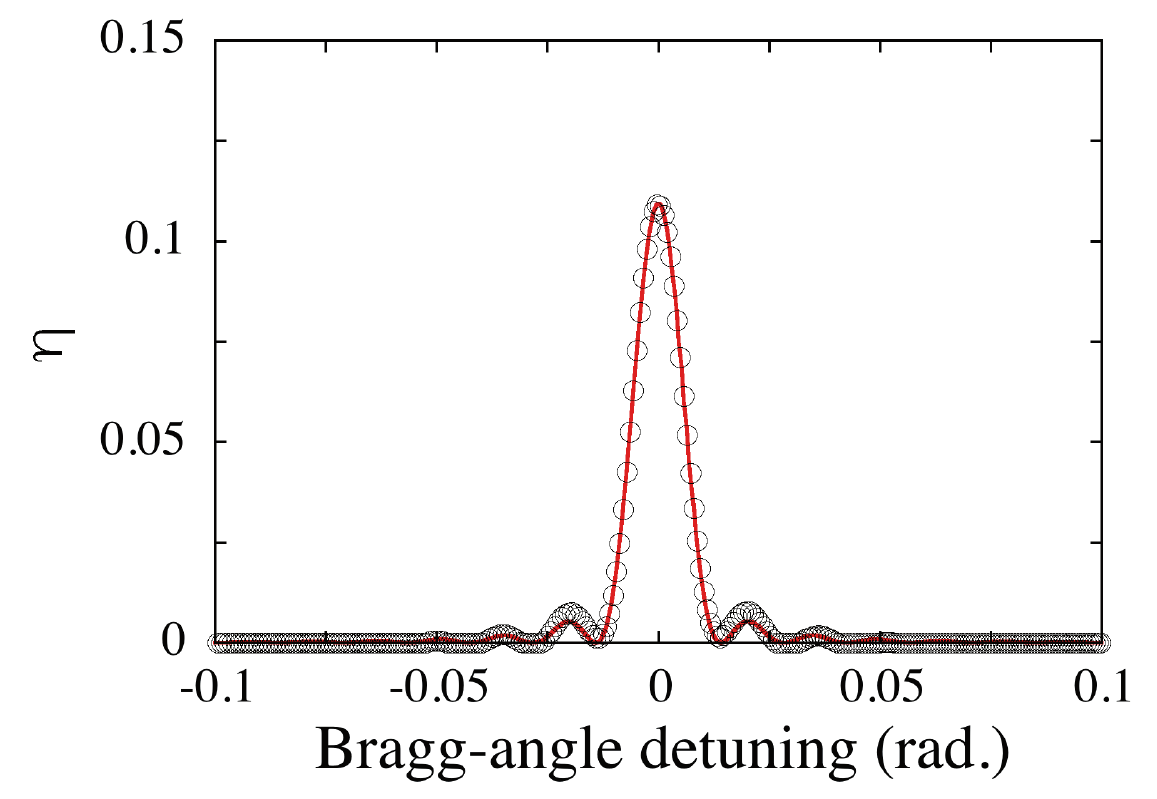}
\subcaption{}
 \includegraphics[clip, width=62mm]{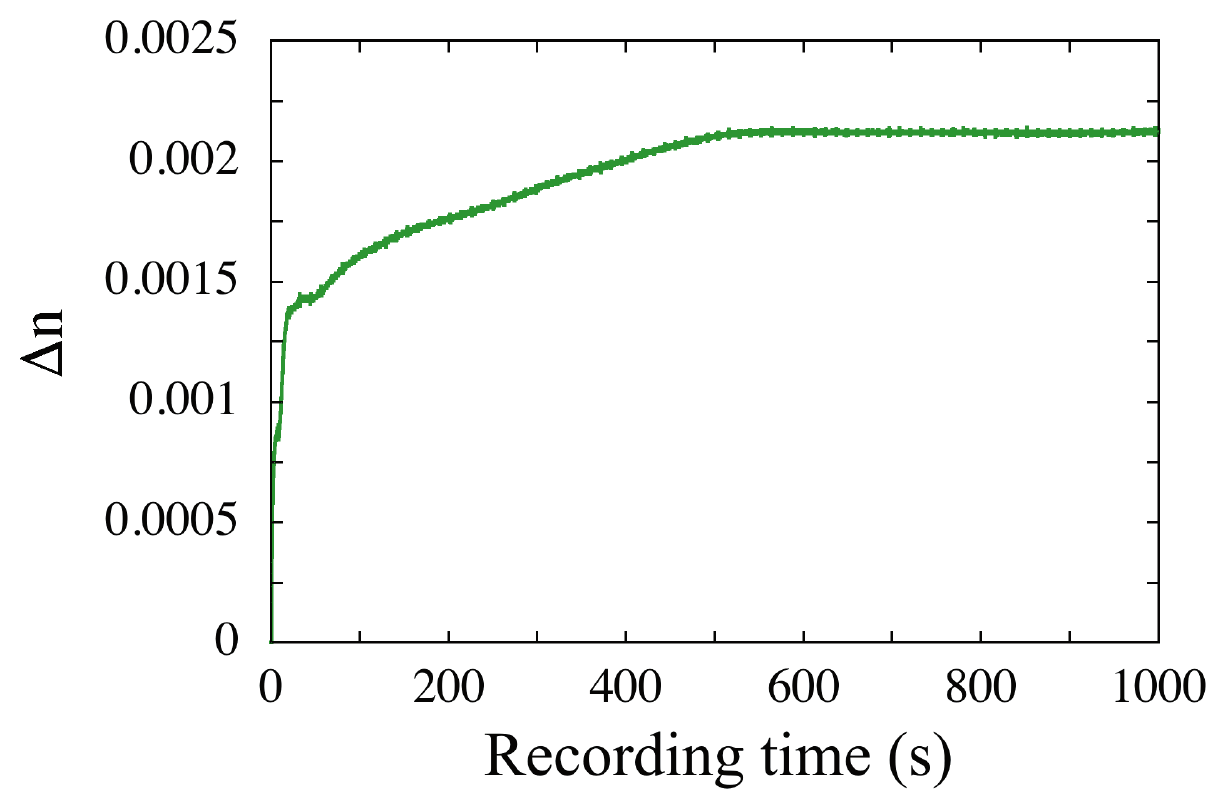}
\subcaption{}
  \end{center}
    \vspace{3mm}
\caption{(a)   Extracted buildup dynamics of $\eta$  at 532\,nm from  an ND composite grating of 0.5\,$\mu$m spacing. (b) Bragg-angle detuning curve at 633\,nm. The solid line is the least-squares fit to the measured data.  (c) Extracted buildup dynamics of $\Delta n$ at 532\,nm.}
\label{fig:f11}
\end{figure}

An unslanted and plane-wave transmission volume grating with a spacing of 0.5\,$\mu$m was also recorded in an ND-dispersed NPC film at the  optimum ND concentration (15\,vol.\%) and at the optimum recording intensity (75\,mW/cm$^2)$. Furthermore, we used the optimum ratio [MOE]/([PETIA]+[MOE]) of 0.5. We refer to that sample as the grating \#2. Figure~\ref{fig:f11}(a) shows the buildup dynamics of $\eta$ at 532\,nm, that is extracted from that at 633\,nm. 
 Figure~\ref{fig:f11}(b) shows the Bragg-angle detuning curve probed at 633\,nm after the saturation of $\eta$ from which we find 
 $d$ to be 35.31$\pm0.04$\,$\mu$m.   Figure~\ref{fig:f11}(c) shows the corresponding buildup dynamics of $\Delta n$ at 532\,nm. It can be seen that the grating \#2 has reached $\Delta n_{\rm sat}$ of approximately 2.2$\times$10$^{-3}$, 
which is smaller than that of the grating \#1.  Such a decrease in  $\Delta n_{\rm sat}$ at shorter grating spacing is a typical trend of multicomponent photopolymer including NPC materials~\cite{tomita2016} as a result of facilitated photoinitiator's counter diffusion to, the cluster formation  of photoinactive species in and the spatial extension of the polymer network into the dark regions. 
 We also found $a_1\Delta f$ and $a_1\Delta f d$ to be approximately 0.005 and 0.175\,$\mu$m, respectively. They are smaller than those of NPC grating \#1. However, such drawback in $a_1\Delta f$ and $a_1\Delta f d$ may be compensated for, to some extent, in slow-neutron diffraction for the following  reasons: 
It is well-known that the Klein-Cook parameter ($Q$) and the grating strength parameter ($\nu$) classify wave diffraction in several diffraction regimes~\cite{klein1967,moharam1978,moharam1980a,moharam1980b}. 
They are given by 
$Q=2\pi\lambda d/n\Lambda^2$ and $\nu=\pi d\Delta n_{\rm sat}/\lambda\cos\theta_B$, respectively, where $\lambda$ is a readout wavelength in vacuum, $n$ is the average refractive index of a phase grating of interest, and $\theta_B$ is the Bragg angle given by $\cos^{-1}\sqrt{1-(\lambda/2n\Lambda)^2}$.  When  the conditions of  $Q\nu/\cos\theta_B>1$ and $Q/\nu\cos\theta_B>20$ are met,  diffraction occurs in the Bragg diffraction (two-coupled waves) regime, {\it i.e.}, two waves propagate in a thick grating, resulting in detectable signal power only in the narrow angular range around the Bragg angle. In contrast, when the conditions of $Q\nu/\cos\theta_B<1$ and $Q/\nu\cos\theta_B<20$ are met, diffraction is in the Raman-Nath diffraction (multiple coupled waves) regime, which features many overlapping diffraction orders almost independent of the angle of incidence in a thin grating. 
Otherwise, diffraction is in the intermediate diffraction regime, for which considerable overlapp of diffraction orders occurs in spite of clearly observable dependence of the diffraction efficiency on Bragg-detuning (or angle of incidence), described by the rigorous coupled-wave analysis (RCWA)~\cite{moharam1981}. 
In our case, the gratings \#1 and \#2 have $(Q\nu/\cos\theta_B,\,\,Q/\nu\cos\theta_B)$ of (39,  77) and (157,  658), respectively, at $\lambda = 532$\,nm and $n=1.5646$. This means that light diffraction by both NPC gratings is in the Bragg diffraction regime, confirming the validity of our analysis in evaluating  $d$ and $\Delta n_{\rm sat}$. 
We also find that  the grating \#2 has factors of approximately 4 and 10 in $Q\nu/\cos\theta_B$ and $Q/\nu\cos\theta_B$, respectively, as compared with those of the  grating \#1. 
The result of our slow-neutron diffraction experiment will be described in the next subsection. 

\subsection{Slow-Neutron Diffraction}\label{subSec:SlowNeutronDiffr}
\vspace*{-3mm}
\subsubsection{Experimental setup and a method for data analysis}
\vspace*{-3mm}
Neutron diffraction experiments were carried out at the beamline PF2/VCN of the Institut Laue-Langevin (ILL) in Grenoble, France.
\begin{figure}[b]
\begin{center}
 \includegraphics[clip, width=55mm]{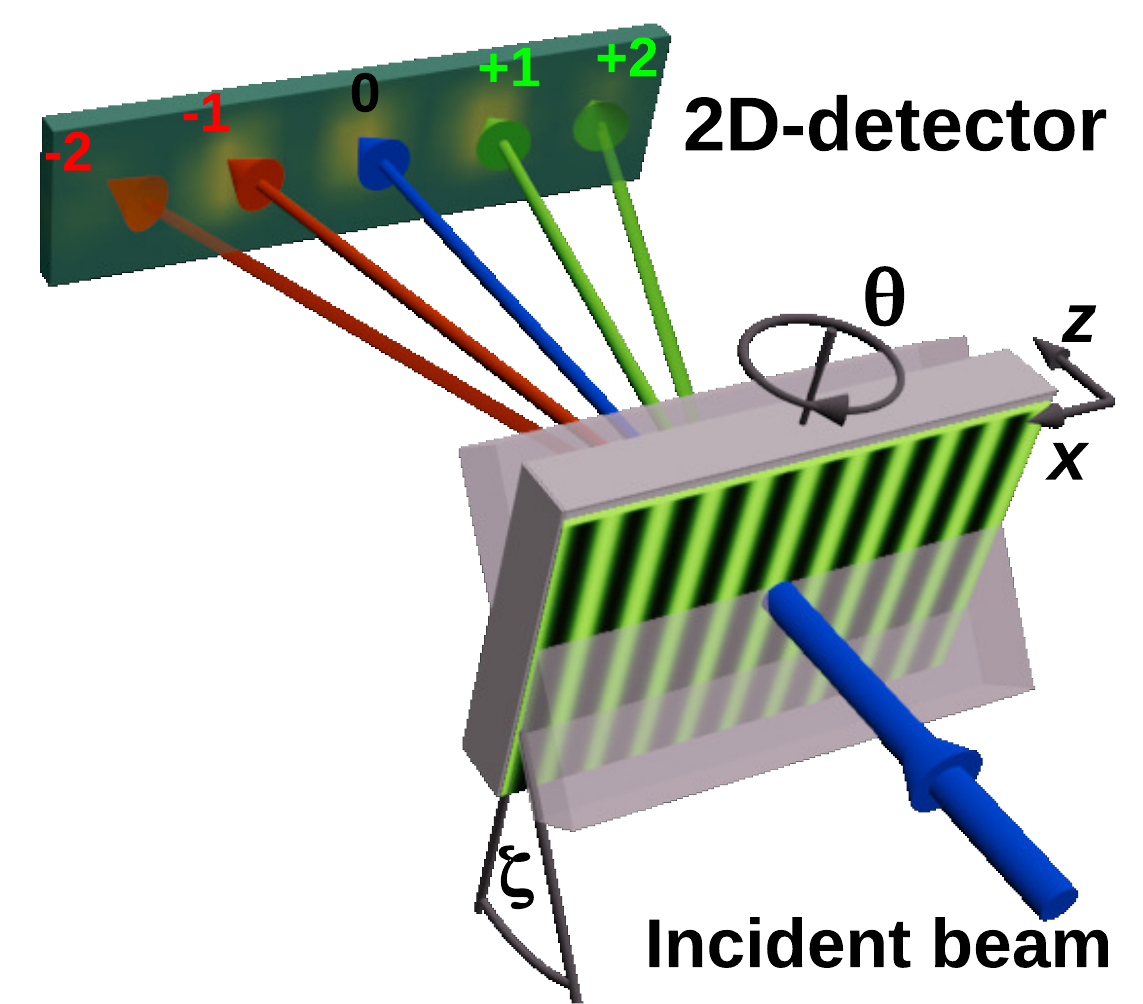}
  \end{center}
    \vspace{-3mm}
\caption{Schematic of the measurement setup for slow-neutron diffraction by an NPC transmission grating.  
Upon step-wise rotation through angles $\theta$ about the $y$ axis, the neutron beam incident from the lower right direction was split into 
several diffracted beams that are measured by a 2D detector. Neutrons arriving at corresponding diffraction-order positions were detected to calculate $\eta_i$ of the $i$-th diffraction order as a function of $\theta$. The NPC transmission grating could be tilted by an angle $\zeta < 90^\circ$ to increase the thickness $d$ by a factor $1/\cos\zeta$.}
\label{fig:f12}
\end{figure}
The measurement principle is illustrated in Fig.\,\ref{fig:f12}. The gratings were -- similarly to the light diffraction measurements -- mounted in  transmission geometry. With the gratings tilted to the angle $\zeta$ around an axis parallel to the grating vector in order to effectively adjust $d$, the incident angle $\theta$ was varied to measure Bragg-angle detuning curves in the vicinity of $\theta_B$ (as given by the Bragg equation $\lambda_\text{\tiny{N}}=2\Lambda\sin\theta_B$) at a slow-neutron wavelength $\lambda_\text{\tiny{N}}$. 
Very cold neutrons with a broad wavelength distribution in the range of about 2.5 to 10\,nm are available at PF2/VCN~\cite{odaNIMA2017}. A Ti/Ni neutron mirror was necessary to redirect the incident neutron beam to make use of the full 
collimation length and to obtain a narrower wavelength distribution like the one shown in Ref.\,\cite{blaickner2019}. The divergence of the beam was about 2\,mrad. 
A 2D detector with $2\times 2$\,mm$^2$ pixel size was used to detect the diffraction spots.
The typical measurement time per incident angular position $\theta$ was about one hour.
Since $\Lambda$ is large compared to $\lambda_\text{\tiny{N}}$, $\theta_B\approx \lambda_\text{\tiny{N}}/(2\Lambda)$ is of the order of 3\,mrad. Given the spatial resolution of the detector system (pixel size $=2\times 2$\,mm$^2$), about one and a half meters distance had to be maintained between the grating and the 2D detector to be able to observe well-separated diffraction spots. 
We define the diffraction efficiency of the $i$th diffraction order as 
$\eta_{i}=I_i/I_{\text{\scriptsize{tot}}}$ ($i=0, \pm1, \pm2, \cdots$), 
where $I_i$ and $I_{\text{\scriptsize{tot}}}$ are the measured intensities of the $i$th and the sum of all measured diffraction orders, respectively. At each incident angle the sum over all pixels in each separated diffraction spot at the 2D detector (see Fig.\,\ref{fig:f12}) -- each associated with one diffraction order -- was calculated and the resulting intensities (corrected for background) plugged into the definition for $\eta_i$.
While the second order diffraction from grating \#1 was clearly detectable, it was marginal for grating \#2 and neglected in the analysis as well as in the data plots (see below). 
\begin{figure}
\begin{center}
 \includegraphics[clip, width=62mm]{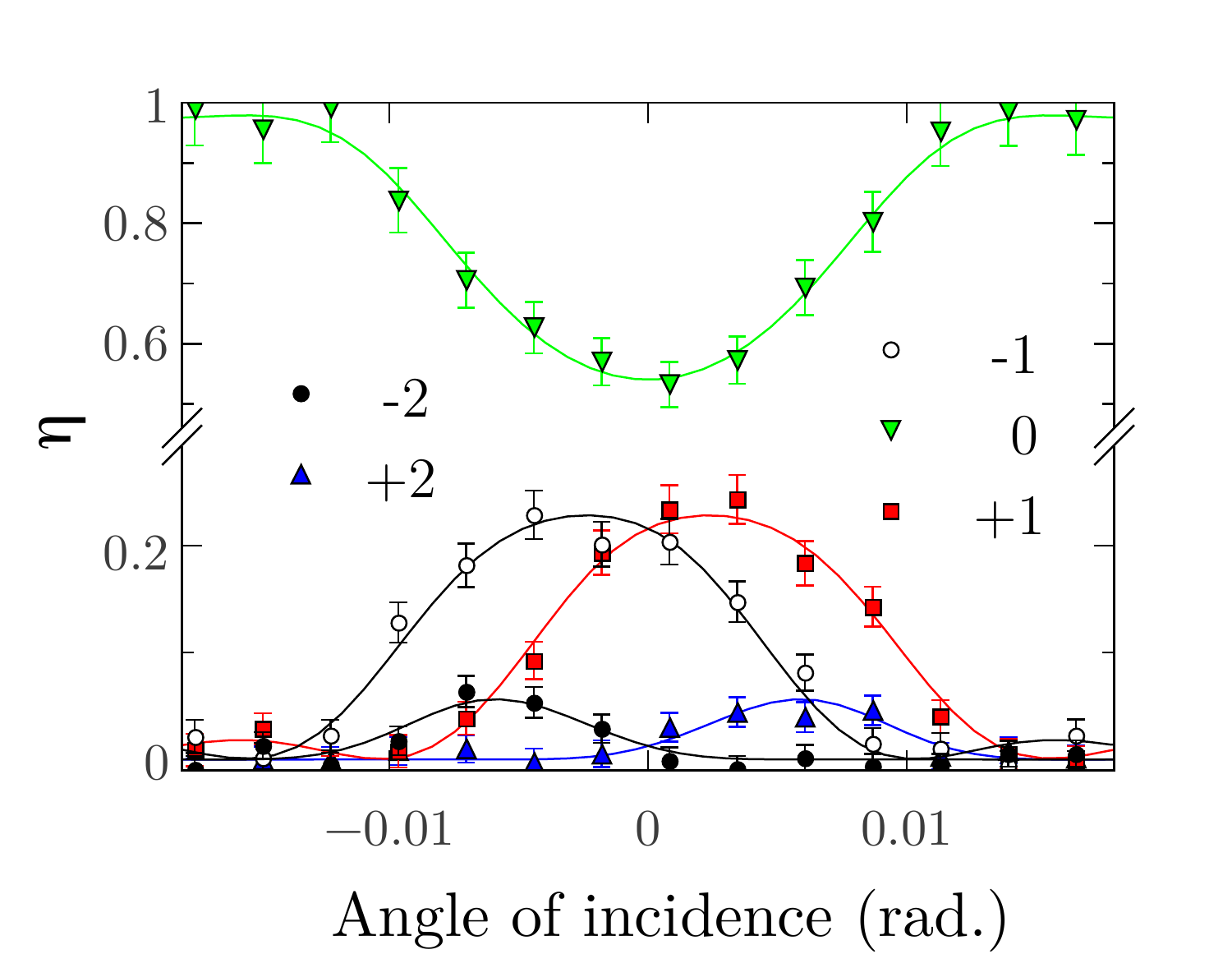}
\subcaption{}
 \includegraphics[clip, width=62mm]{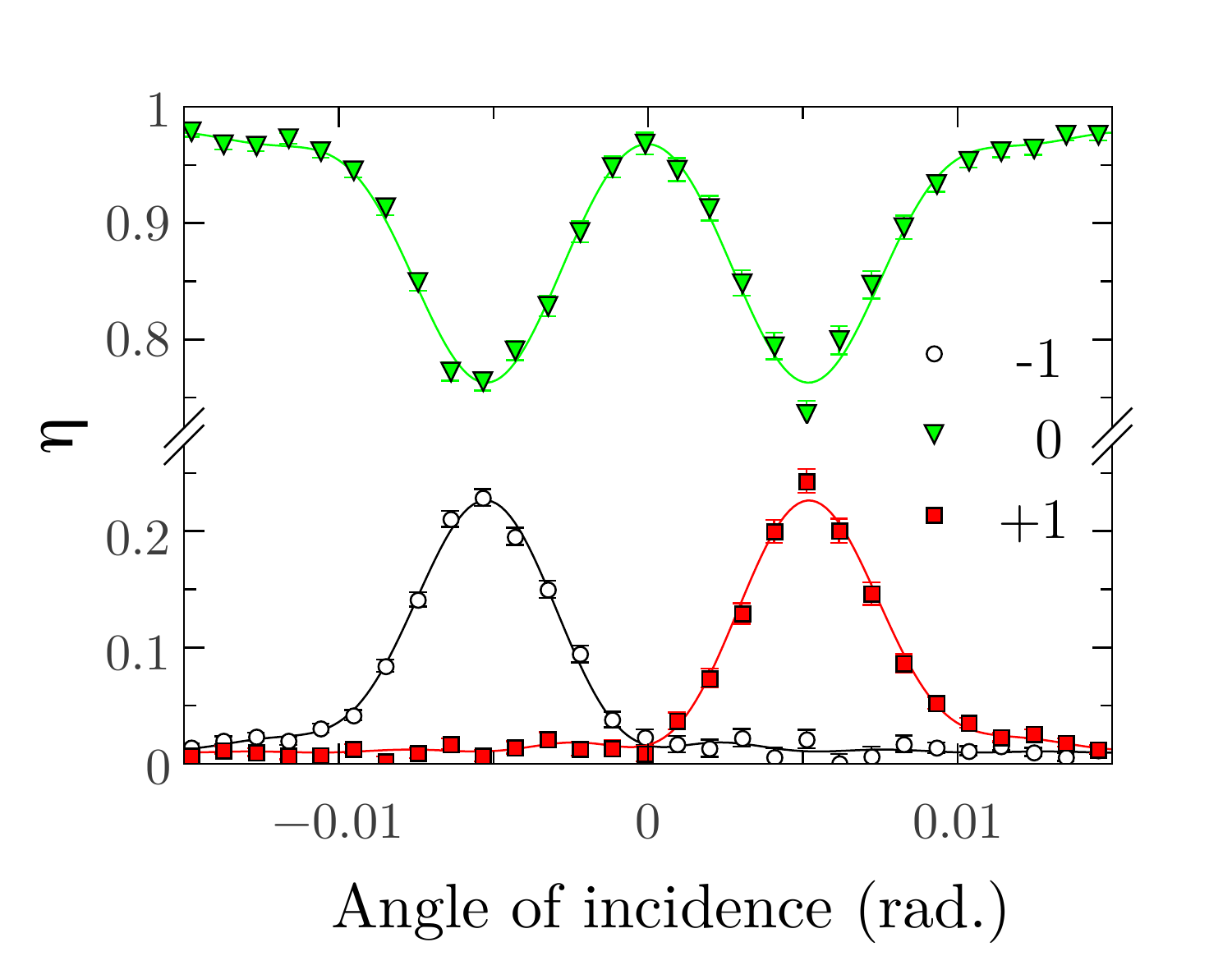}
\subcaption{}
  \end{center}
 \vspace*{-4.mm}
\caption{\label{fig:f13}(a) Bragg-angle detuning curve for grating \#1  at $\bar{\lambda}_\text{\tiny{N}}$ of 4.75\,nm, tilted at $\zeta\approx 70^\circ$. 
Solid curves correspond to least-squares fits of RCWA calculations to the data, additionally taking into account the wavelength distribution by averaging. (b) Bragg-angle detuning curve 
for grating \#2 at $\bar{\lambda}_\text{\tiny{N}}$ of 5\,nm, tilted at $\zeta\approx 70^\circ$. 
Note that a slight difference in $\bar{\lambda}_\text{\tiny{N}}$ from that in Fig.\,\ref{fig:f13} (a) is due to the experiments having taken place during different beamtimes, with slightly different setups. Solid curves correspond to least-squares fits of the Kogelnik's theory to the data, additionally taking into account the wavelength distribution by averaging. 
Data are made available by the ILL (see \cite{dataGrating1,*dataGrating2}).}
\end{figure} 

For neutrons, the Klein-Cook and grating strength parameters can be written as $Q_\text{\tiny{N}}=2\pi d\lambda_\text{\tiny{N}} /(\Lambda^2\cos\zeta)$ and 
$\nu_\text{\tiny{N}} \approx \pi d\Delta n_{\rm sat{,\text{\tiny{N}}}}/(\lambda_n\cos\zeta)$, where we have used that $\cos\theta_B \approx 1$. 
According to the criteria for assigning diffraction regimes at the end of Sec.\,\ref{subsec:HologrRecLightDiffrExp}, neutron diffraction by gratings \#\,1 and \#\,2 should be expected to occur in the RCWA diffraction regime and in the Bragg diffraction regime, respectively, as  ($Q_\text{\tiny{N}}\nu_\text{\tiny{N}}$, $Q_\text{\tiny{N}}/\nu_\text{\tiny{N}}$) are (1.35, 3.53) and (6.50, 23.29) for gratings \#1 and \#2, respectively. 
Consequently, the data for gratings \#\,1 and \#\,2 should be analyzed by the RCWA and the Kogelnik's coupled-wave theory, respectively.
\vspace*{-3mm}
\subsubsection{Results of slow-neutron diffraction}
\vspace*{-3mm} 
In Fig.\,\ref{fig:f13}, Bragg-detuning curves for gratings \#\,1 and \#\,2 are plotted. For the data set of Fig.\,\ref{fig:f13}\,(a), least-squares fits with the RCWA were made (solid curves), whereas for the data set of Fig.\,\ref{fig:f13}\,(b), the Kogelnik's coupled-wave theory was used for the fitting procedure (solid curves). 
Since both gratings are relatively thick (in terms of $Q_\text{\tiny{N}}\nu_\text{\tiny{N}}$ and $Q_\text{\tiny{N}}/\nu_\text{\tiny{N}}$) so that they are expected to exhibit a considerable wavelength selectivity in $\eta$, we performed wavelength averaging by numerical integration of the product of the respective fit model function and a triangular or exponentially-modified-Gaussian shaped wavelength distribution~\cite{blaickner2019}. 
From both fits for gratings \#1 and \#2 we extracted $d$, $n_s$ [for $s=1,2$ in Eq.\,(\ref{eq:A4})], $\bar{\lambda}_\text{\tiny{N}}$ and, $\Delta n_{\rm sat{,\text{\tiny{N}}}}$ [see Eq.\,(\ref{eq:A7})].

A summary of parameter estimations for gratings\,\#\,1 and \#\,2 is given in Table\,\ref{tab:table1}. The values for  $a_1\Delta f$ are obtained from the light diffraction experiments described in the previous subsection.
\begin{table}
  \begin{center}
    \caption{Summary of parameters for gratings\,\#\,1 and \#\,2. (Note that the tilt angle $\zeta$ was nonzero only in the slow-neutron experiment as shown in Fig.\,\ref{fig:f13}.)}  
    \label{tab:table1}
    \begin{tabular}{lcc} 
       \hline\hline\textbf{Grating:} & \textbf{\#\,1}\,($\Lambda\!=\!1\,\mu$m)~ & \textbf{\#\,2}\,($\Lambda\!=\!0.5\,\mu$m)\\
      \hline \textbf{Light diffr.:} && \\  
       $d$\,[$\mu$m] & $25.29\pm 0.02$ & $35.31\pm 0.04$ \\
       $\Delta n_{\rm sat}$ & $\approx 4.7\times 10^{-3}$ & $\approx 2.2\times 10^{-3}$\\ 
       $a_1\Delta f$ & $\approx 0.011$ & $\approx 0.005$ \\
      \hline \textbf{Slow neutron diffr.:} &&  \\
      
      $d$\,[$\mu$m] & $25.0\pm 1.4$ & $33.2\pm 1.4$ \\
      $\overline{\lambda}_\text{\tiny{N}}$\,[nm] & $4.75\pm 0.30$ & $5.030\pm 0.072$ \\      
      $\Delta n_{\rm sat{,\text{\tiny{N}}}}=2|n_1|$\,$\left[\times 10^{-5}\right]$ & $1.278\!\pm\!0.092$ & $0.869\!\pm\! 0.032$\\
      $n_2$\,$\left[\times 10^{-6}\right]$ & $-1.87\!\pm\!0.30$& N/A\,(see text)
      \\ 
      $(b_c\rho)_{np}$\,$\left[\times 10^{-6}\text{\AA}^{-2}\right]$ & $3.4$ & $4.4$\\ 
      \hline\hline  
    \end{tabular}
  \end{center}
\end{table}
As one can see, $d$ as obtained from neutron diffraction agrees well with light diffraction results. 
Considering the magnitudes of $\Delta n_{\rm sat{,\text{\tiny{N}}}}$ and $n_2$ for grating \#\,1 points to the refractive index profile of grating \#\,1 not being completely sinusoidal, as  $|n_2|$ is non-negligible compared to $|n_1|$. The minus sign of $n_2$ indicates that there is a phase shift of $\pi$ between the first and the second Fourier components of the refractive index profile of this grating. This is consistent with our TEM observation of the non-sinusoidal ND density distribution after holographic exposure shown in Fig.\,\ref{fig:f8}. In contrast, the neutron diffraction measurements for grating \#\,2 only showed negligible second order contributions. Thus, the quantity $n_2$ is not available for this sample. 
The extracted value $\Delta n_{\rm sat, \text{\tiny{N}}}$ for slow neutrons is related to the SLD of ND, $(b_c\rho)_{np}$, from Eqs.\,(\ref{eq:eq1}) and (\ref{eq:eq2}) such that
\begin{eqnarray} 
\Delta n_{\rm sat{,\text{\tiny{N}}}}&=&a_1\Delta f \left| n_{np,\text{\tiny{N}}}-n_{p,\text{\tiny{N}}}\right| \nonumber\\ 
&=& a_1\Delta f\left|1-\frac{\overline{\lambda}^2_\text{\tiny{N}} (b_c\rho)_{np}}{2\pi}-1+\frac{\overline{\lambda}^2_\text{\tiny{N}} (b_c\rho)_p}{2\pi}\right|\nonumber\\ 
&=& \frac{a_1\Delta f \overline{\lambda}^2_\text{\tiny{N}}}{2\pi}\left|(b_c\rho)_{np}-(b_c\rho)_p\right|\nonumber\\
&\approx & \frac{a_1\Delta f \overline{\lambda}^2_\text{\tiny{N}}}{2\pi} \vert(b_c\rho)_{np}\vert
\label{eq:eq3},
\end{eqnarray}
where the last step is acceptable for ND since values for SLD of polymer, $\vert(b_c\rho)_p\vert$, are typically around $1\times 10^{-6}$\,\AA$^{-2}$, which is expected to be considerably smaller than $\vert(b_c\rho)_{np}\vert$. 
Rearranging Eq.\,(\ref{eq:eq3}) to obtain
\begin{equation}\label{eq:eq4}
\vert(b_c\rho)_{np}\vert=\frac{2\pi\Delta n_{\rm sat{,\text{\tiny{N}}}}}{a_1\Delta f\overline{\lambda}^2_\text{\tiny{N}}},
\end{equation}
we are now in a position to assess the SLD of ND in structured nanocomposite materials from light and neutron diffraction data. The estimations for SLDs for gratings \#\,1 and \#\,2 resulting form Eq.\,(\ref{eq:eq4}) are given in Table\,\ref{tab:table1}. As expected, the two values for SLDs are not very different from each other and are surely of the correct order of magnitude. The estimation for grating \#\,1 is smaller than that for grating \#\,2. 

For comparison, we also performed an estimation for $\vert(b_c\rho)_{np}\vert$ in an alternative way, i.e., by calculating the volume fractional averaged value 
of the surface-modified ND as done for its effective refractive index described in Sec.\,\ref{subsec:SamplePrep}.
We find $\vert(b_c\rho)_{np}\vert\approx 3.0\times 10^{-6}{\rm\AA}^{-2}$ by using known bulk values for SLDs of diamond, graphite, ZrO$_2$ and grafted organics (acrylate) with their volume fractions in our surface-modified ND. 
As one can see, we obtain the same order of magnitude and the numerical value of the estimation is quite close to the values in Table\,\ref{tab:table1}, suggesting consistency of the two approaches. 
 Obviously, as similar to the case of the effective refractive index calculation, the volume fractional averaging procedure is 
 not theoretically well supported for the structure of a multilayered concentric sphere. Despite this, the calculated value for $\vert(b_c\rho)_{np}\vert$ is in good agreement with values obtained from our slow-neutron diffraction measurement, reflecting the effect of non-diamond components such as graphite and other inorganic/organic materials on the effective SLD of our surface-modified ND.

Since the slow-neutron diffraction efficiency of an NPC grating depends on $\Delta n_{\rm sat{,\text{\tiny{N}}}}$ that is proportional to $a_1\Delta f$ [see Eq.\,(\ref{eq:eq3})], another interesting possibility arising from the analysis in the present paper is the {\it a priori} assessment of slow-neutron diffraction properties of an NPC grating in advance from light diffraction data (assessing $a_1\Delta f$) only. Such quantitative information provides a solid estimation for predicting grating performance in slow-neutron diffraction experiments. In practice, the latter can be a valuable tool to save scarce neutron beamtime that would otherwise be used inefficiently for samples of previously unknown low neutron-optical performance.
 
\section{CONCLUSION}
\vspace*{-3mm}  
We have successfully demonstrated volume holographic recording in NPCs dispersed with ND at a recording wavelength of 532\,nm. ND was synthesized by a detonation method, followed by the removal of residual graphite and the surface treatment. We have found that the surface-modified ND can be uniformly dispersed in an acrylate monomer blend as high as 22\,vol.\%.  
They act as nanoparticles and participate in  the photopolymerization-driven mutual diffusion and
phase separation dynamics of ND and monomer during holographic exposure. 
As a result, such holographic assembly of ND in the formed polymer induces the spatial density modulation of ND, leading to the formation of ND-dispersed holographic phase gratings, which was evidenced by the observation of the cross section of a recorded grating.  
Unslanted and plane-wave transmission volume  gratings with $\Delta n_{\rm sat}$ as high as 4.7$\times$10$^{-3}$ (2.2$\times$10$^{-3}$) at a wavelength of 532\,nm and at grating spacing of 1\,$\mu$m (0.5\,$\mu$m) were recorded in NPC films dispersed with ND of 15\,vol.\%.
Because refractive index differences between the surface-modified ND and the formed polymer are large  at both visible and slow-neutron wavelengths, our proof-of-principle demonstration of recording high contrast gratings paves the way for efficient manipulation of light and slow-neutron beams. 

Our slow-neutron diffraction experiments have shown that ND dispersed NPC gratings are excellent candidates for use in neutron optics. Their diffraction properties for slow neutrons have reached values almost comparable to those of the best SiO$_2$ dispersed NPC gratings reported 
so far~\cite{fally2010,klepp2012a}. In particular, we have demonstrated that ND dispersed NPC gratings easily provide diffraction efficiencies higher than 20\,\% at relaxed angular selectivity, preferable for loosely collimated slow-neutron beams with rather wide wavelength distributions. ND dispersed NPC gratings presented here clearly possess a potential to outperform SiO$_2$ dispersed NPC gratings. 
We have also shown that a good prediction of the expected slow-neutron diffraction performance of NPC gratings is possible from light diffraction data only: By taking into account a refractive index profile, $a_1\Delta f$ (from light diffraction) and a literature value for SLD of dispersed nanoparticles, a solid prediction for $\Delta n_{\rm sat, \text{\tiny{N}}}$ and, thus, the slow-neutron diffraction efficiency can be made. Conversely, and probably more importantly, given some knowledge about a grating's refractive index profile from light and neutron diffraction, it is possible to yield information about the SLD of a dispersed nanomaterial such as ND, in our case. Indeed, we have developed a method to establish an estimation of SLD of the surface-modified ND, which may help to extract some information concerning the surface state.
 
\begin{acknowledgments}
\vspace*{-3mm}  
We wish to acknowledge the support of the Ministry of Education, Culture, Sports, Science and
Technology of Japan under grant 15H03576. We are grateful to M. Hino and T. Oda (Kyoto University) for providing the redirecting mirror for our slow-neutron experiments. We also thank T. Brenner for technical assistance at the Institut Laue-Langevin.
\end{acknowledgments}
\appendix
\section{REFRACTIVE INDEX MODULATION AMPLITUDE FORMED IN NPC MEDIA}

\vspace*{-3mm}  
Consider an unslanted and plane-wave phase grating recorded in an NPC medium containing the formed host polymer of the refractive index $n_p$ and dispersed spherical nanoparticles of the refractive index $n_{np}$ with their position-dependent and periodically modulated volume fractions of $f_p(x)$ and $f_{np}(x)$ at a position $x$, respectively. 
We may approximate a periodically modulated refractive index $n(x)$ with the grating vector $\bm K$ along the $x$ direction  under the Maxwell Garnett approximation~\cite{markel2016} with $0\le f_{np}(x)\ll 1$ and $n_p\approx n_{np}$ as 
\begin{eqnarray}
n(x) &=& n_{np} f_{np}(x) +n_p f_p(x) \nonumber \\
&=&n_p+f_{np}(x)(n_{np}-n_p),
\label{eq:A1}
\end{eqnarray}
where  the second equality comes from the relation that $f_p(x)+f_{np}(x)=1$ at any $x$.  
Since $n(x)$ is periodic in $x$, $f_{np}(x)$ can be written as 
\begin{equation}
f_{np}(x)=\Delta fg(x), 
\label{eq:A2}
\end{equation}
where $\Delta f$ is the modulation amplitude of $f_{np}(x)$ and $g(x)$ is a dc biased periodic function with an oscillation amplitude of $\pm1$ and 
the spatial period $\Lambda (=2\pi/|\bm K|)$. Note that $\Delta f\le \bar{f}_{np}$, with $\bar{f}_{np}$ being the spatially averaged value of $f_{np}(x)$ and $0<\bar{f}_{np}\ll1$. 

Furthermore, we note that $g(x)$ is a non-negative function and is much smaller than $1/\Delta f$, since $0\,\le\,f_{np}(x) \ll 1$. It is expressed, in terms of a Fourier series, by 
\begin{equation}
g(x)=\sum_{q=-\infty}^{\infty}g_q\exp{(iqKx)},
\label{eq:A3}
\end{equation}
where $K=|{\bm K}|$ and $g_q=g_{-q}^*$ due to $g(x)$ being a real function. 
In particular, $g_0$ is given by ${\bar f}_{np}/\Delta f$.

Likewise, $n(x)$ can be expressed in terms of a Fourier series by
\begin{equation}
n(x)= \sum_{s=-\infty}^{\infty}n_s\exp{(isKx)},
\label{eq:A4}
\end{equation}
where $n_s=n_{-s}^*$ for a pure phase grating.  
Using Eqs. (\ref{eq:A1})-(\ref{eq:A4}), we find 
\begin{eqnarray}
n_0 &=& n_p+g_0\Delta f(n_{np}-n_p) \nonumber \\
&=&n_p+{\bar f}_{np}(n_{np}-n_p) \nonumber \\
&=&n_p{\bar f}_p+n_{np}{\bar f}_{np}, 
\label{eq:A5}
\end{eqnarray}
where $\bar{f}_p=1-\bar{f}_{np}$ is used.
Also, we find from Eqs.\,(\ref{eq:A1})-(\ref{eq:A3}) that 
\begin{equation}
n_s=g_s\Delta f(n_{np}-n_p),\,\,\,\,\,\textrm{($s\ne0$)}.
\label{eq:A6}
\end{equation}
Therefore, the refractive index modulation amplitude $\Delta n_{\rm sat}$ of the 1st-order (Bragg-matched) component for $n(x)$ is given by 
\begin{eqnarray}
\Delta n_{\rm sat}&\equiv&2|n_1|\nonumber \\
&=&2|g_1|\Delta f|n_{np}-n_p|. 
\label{eq:A7}
\end{eqnarray}
When we set $2|g_1|=a_1$, we can express $\Delta n_{\rm sat}$ as
\begin{equation}
\Delta n_{\rm sat}=a_1\Delta f|n_{np}-n_p|,
\label{eq:A8}
\end{equation}
which is Eq.\,(\ref{eq:eq1}). It is straightforward to find from Eq.\,(\ref{eq:A3}) that $a_1$  is  unity  for a pure sinusoidal waveform of $n(x)$ and is $4\sin (r\pi)/\pi$ for a rectangular waveform of $n(x)$, where 

\bibliographystyle{apsrev4-2custom} 
\providecommand{\noopsort}[1]{}\providecommand{\singleletter}[1]{#1}%
%

\begin{thebibliography}{79}%
\makeatletter
\providecommand \@ifxundefined [1]{%
 \@ifx{#1\undefined}
}%
\providecommand \@ifnum [1]{%
 \ifnum #1\expandafter \@firstoftwo
 \else \expandafter \@secondoftwo
 \fi
}%
\providecommand \@ifx [1]{%
 \ifx #1\expandafter \@firstoftwo
 \else \expandafter \@secondoftwo
 \fi
}%
\providecommand \natexlab [1]{#1}%
\providecommand \enquote  [1]{``#1''}%
\providecommand \bibnamefont  [1]{#1}%
\providecommand \bibfnamefont [1]{#1}%
\providecommand \citenamefont [1]{#1}%
\providecommand \href@noop [0]{\@secondoftwo}%
\providecommand \href [0]{\begingroup \@sanitize@url \@href}%
\providecommand \@href[1]{\@@startlink{#1}\@@href}%
\providecommand \@@href[1]{\endgroup#1\@@endlink}%
\providecommand \@sanitize@url [0]{\catcode `\\12\catcode `\$12\catcode
  `\&12\catcode `\#12\catcode `\^12\catcode `\_12\catcode `\%12\relax}%
\providecommand \@@startlink[1]{}%
\providecommand \@@endlink[0]{}%
\providecommand \url  [0]{\begingroup\@sanitize@url \@url }%
\providecommand \@url [1]{\endgroup\@href {#1}{\urlprefix }}%
\providecommand \urlprefix  [0]{URL }%
\providecommand \Eprint [0]{\href }%
\providecommand \doibase [0]{https://doi.org/}%
\providecommand \selectlanguage [0]{\@gobble}%
\providecommand \bibinfo  [0]{\@secondoftwo}%
\providecommand \bibfield  [0]{\@secondoftwo}%
\providecommand \translation [1]{[#1]}%
\providecommand \BibitemOpen [0]{}%
\providecommand \bibitemStop [0]{}%
\providecommand \bibitemNoStop [0]{.\EOS\space}%
\providecommand \EOS [0]{\spacefactor3000\relax}%
\providecommand \BibitemShut  [1]{\csname bibitem#1\endcsname}%
\let\auto@bib@innerbib\@empty
\bibitem [{\citenamefont {Krueger}(2008)}]{krueger2008}%
  \BibitemOpen
  \bibfield  {author} {\bibinfo {author} {\bibfnamefont {A.}~\bibnamefont
  {Krueger}},\ }\bibfield  {title} {\bibinfo {title} {Diamond nanoparticles:
  jewels for chemistry and physics},\ }\href@noop {} {\bibfield  {journal}
  {\bibinfo  {journal} {Adv. Mater.}\ }\textbf {\bibinfo {volume} {30}},\
  \bibinfo {pages} {2445} (\bibinfo {year} {2008})}\BibitemShut {NoStop}%
\bibitem [{\citenamefont {Mochalin}\ \emph {et~al.}(2012)\citenamefont
  {Mochalin}, \citenamefont {Shenderova}, \citenamefont {Ho},\ and\
  \citenamefont {Gogotsi}}]{mochalin2012}%
  \BibitemOpen
  \bibfield  {author} {\bibinfo {author} {\bibfnamefont {V.~N.}\ \bibnamefont
  {Mochalin}}, \bibinfo {author} {\bibfnamefont {O.}~\bibnamefont
  {Shenderova}}, \bibinfo {author} {\bibfnamefont {D.}~\bibnamefont {Ho}},\
  and\ \bibinfo {author} {\bibfnamefont {Y.}~\bibnamefont {Gogotsi}},\
  }\bibfield  {title} {\bibinfo {title} {The properties and applications of
  nanodiamonds},\ }\href@noop {} {\bibfield  {journal} {\bibinfo  {journal}
  {Nat. Nanotechnol.}\ }\textbf {\bibinfo {volume} {7}},\ \bibinfo {pages} {11}
  (\bibinfo {year} {2012})}\BibitemShut {NoStop}%
\bibitem [{\citenamefont {Dolmatov}(2007)}]{dolmatov2007}%
  \BibitemOpen
  \bibfield  {author} {\bibinfo {author} {\bibfnamefont {V.~Y.}\ \bibnamefont
  {Dolmatov}},\ }\bibfield  {title} {\bibinfo {title} {Detonation-synthesis
  nanodiamonds: synthesis, structure, properties and applications},\
  }\href@noop {} {\bibfield  {journal} {\bibinfo  {journal} {Russ. Chem. Rev.}\
  }\textbf {\bibinfo {volume} {76}},\ \bibinfo {pages} {339} (\bibinfo {year}
  {2007})}\BibitemShut {NoStop}%
\bibitem [{\citenamefont {Zhang}\ \emph {et~al.}(2018)\citenamefont {Zhang},
  \citenamefont {Rhee}, \citenamefont {Hui},\ and\ \citenamefont
  {Park}}]{zhang2018}%
  \BibitemOpen
  \bibfield  {author} {\bibinfo {author} {\bibfnamefont {Y.}~\bibnamefont
  {Zhang}}, \bibinfo {author} {\bibfnamefont {K.~Y.}\ \bibnamefont {Rhee}},
  \bibinfo {author} {\bibfnamefont {D.}~\bibnamefont {Hui}},\ and\ \bibinfo
  {author} {\bibfnamefont {S.-J.}\ \bibnamefont {Park}},\ }\bibfield  {title}
  {\bibinfo {title} {A critical review of nanodiamond based nanocomposites:
  synthesis, properties and applications},\ }\href@noop {} {\bibfield
  {journal} {\bibinfo  {journal} {Compos. Part B}\ }\textbf {\bibinfo {volume}
  {143}},\ \bibinfo {pages} {19} (\bibinfo {year} {2018})}\BibitemShut
  {NoStop}%
\bibitem [{\citenamefont {Turcheniuk}\ and\ \citenamefont
  {Mochalin}(2012)}]{mochalin2017}%
  \BibitemOpen
  \bibfield  {author} {\bibinfo {author} {\bibfnamefont {K.}~\bibnamefont
  {Turcheniuk}}\ and\ \bibinfo {author} {\bibfnamefont {V.~N.}\ \bibnamefont
  {Mochalin}},\ }\bibfield  {title} {\bibinfo {title} {Biomedical applications
  of nanodiamonds},\ }\href@noop {} {\bibfield  {journal} {\bibinfo  {journal}
  {Nanotechnol.}\ }\textbf {\bibinfo {volume} {28}},\ \bibinfo {pages} {252001}
  (\bibinfo {year} {2012})}\BibitemShut {NoStop}%
\bibitem [{\citenamefont {Aharonovich}\ \emph {et~al.}(2011)\citenamefont
  {Aharonovich}, \citenamefont {Greentree},\ and\ \citenamefont
  {Prawer}}]{aharonovich2011}%
  \BibitemOpen
  \bibfield  {author} {\bibinfo {author} {\bibfnamefont {I.}~\bibnamefont
  {Aharonovich}}, \bibinfo {author} {\bibfnamefont {A.~D.}\ \bibnamefont
  {Greentree}},\ and\ \bibinfo {author} {\bibfnamefont {S.}~\bibnamefont
  {Prawer}},\ }\bibfield  {title} {\bibinfo {title} {Diamond photonics},\
  }\href@noop {} {\bibfield  {journal} {\bibinfo  {journal} {Nature Photonics}\
  }\textbf {\bibinfo {volume} {5}},\ \bibinfo {pages} {397} (\bibinfo {year}
  {2011})}\BibitemShut {NoStop}%
\bibitem [{\citenamefont {Radulaski}\ \emph {et~al.}(2019)\citenamefont
  {Radulaski}, \citenamefont {Zhang}, \citenamefont {Tzeng}, \citenamefont
  {Lagoudakis}, \citenamefont {Ishikawa}, \citenamefont {Dory}, \citenamefont
  {Fischer}, \citenamefont {Kelaita}, \citenamefont {Sun}, \citenamefont
  {Maurer}, \citenamefont {Alassaad}, \citenamefont {Ferro}, \citenamefont
  {Shen}, \citenamefont {Melosh}, \citenamefont {Chu},\ and\ \citenamefont
  {Vou\v{c}okovi\v{c}}}]{radulaski2019}%
  \BibitemOpen
  \bibfield  {author} {\bibinfo {author} {\bibfnamefont {M.}~\bibnamefont
  {Radulaski}}, \bibinfo {author} {\bibfnamefont {J.~L.}\ \bibnamefont
  {Zhang}}, \bibinfo {author} {\bibfnamefont {Y.-K.}\ \bibnamefont {Tzeng}},
  \bibinfo {author} {\bibfnamefont {K.~G.}\ \bibnamefont {Lagoudakis}},
  \bibinfo {author} {\bibfnamefont {H.}~\bibnamefont {Ishikawa}}, \bibinfo
  {author} {\bibfnamefont {C.}~\bibnamefont {Dory}}, \bibinfo {author}
  {\bibfnamefont {K.~A.}\ \bibnamefont {Fischer}}, \bibinfo {author}
  {\bibfnamefont {Y.~A.}\ \bibnamefont {Kelaita}}, \bibinfo {author}
  {\bibfnamefont {S.}~\bibnamefont {Sun}}, \bibinfo {author} {\bibfnamefont
  {P.~C.}\ \bibnamefont {Maurer}}, \bibinfo {author} {\bibfnamefont
  {K.}~\bibnamefont {Alassaad}}, \bibinfo {author} {\bibfnamefont
  {G.}~\bibnamefont {Ferro}}, \bibinfo {author} {\bibfnamefont {Z.-X.}\
  \bibnamefont {Shen}}, \bibinfo {author} {\bibfnamefont {N.~A.}\ \bibnamefont
  {Melosh}}, \bibinfo {author} {\bibfnamefont {S.}~\bibnamefont {Chu}},\ and\
  \bibinfo {author} {\bibfnamefont {J.}~\bibnamefont {Vou\v{c}okovi\v{c}}},\
  }\bibfield  {title} {\bibinfo {title} {Nanodiamond integration with photonic
  devices},\ }\href@noop {} {\bibfield  {journal} {\bibinfo  {journal} {Laser
  Photonics Rev.}\ }\textbf {\bibinfo {volume} {13}},\ \bibinfo {pages}
  {1800316} (\bibinfo {year} {2019})}\BibitemShut {NoStop}%
\bibitem [{\citenamefont {Tomita}\ \emph
  {et~al.}(2016{\natexlab{a}})\citenamefont {Tomita}, \citenamefont {Hata},
  \citenamefont {Momose}, \citenamefont {Takayama}, \citenamefont {Liu},
  \citenamefont {Chikama}, \citenamefont {Klepp}, \citenamefont {Pruner},\ and\
  \citenamefont {Fally}}]{tomita2016}%
  \BibitemOpen
  \bibfield  {author} {\bibinfo {author} {\bibfnamefont {Y.}~\bibnamefont
  {Tomita}}, \bibinfo {author} {\bibfnamefont {E.}~\bibnamefont {Hata}},
  \bibinfo {author} {\bibfnamefont {K.}~\bibnamefont {Momose}}, \bibinfo
  {author} {\bibfnamefont {S.}~\bibnamefont {Takayama}}, \bibinfo {author}
  {\bibfnamefont {X.}~\bibnamefont {Liu}}, \bibinfo {author} {\bibfnamefont
  {K.}~\bibnamefont {Chikama}}, \bibinfo {author} {\bibfnamefont
  {J.}~\bibnamefont {Klepp}}, \bibinfo {author} {\bibfnamefont
  {C.}~\bibnamefont {Pruner}},\ and\ \bibinfo {author} {\bibfnamefont
  {M.}~\bibnamefont {Fally}},\ }\bibfield  {title} {\bibinfo {title}
  {Photopolymerizable nanocomposite photonic materials and their holographic
  applications in light and neutron optics},\ }\href@noop {} {\bibfield
  {journal} {\bibinfo  {journal} {J. Mod. Opt.}\ }\textbf {\bibinfo {volume}
  {63}},\ \bibinfo {pages} {S1} (\bibinfo {year}
  {2016}{\natexlab{a}})}\BibitemShut {NoStop}%
\bibitem [{\citenamefont {Suzuki}\ \emph {et~al.}(2002)\citenamefont {Suzuki},
  \citenamefont {Tomita},\ and\ \citenamefont {Kojima}}]{suzuki2002}%
  \BibitemOpen
  \bibfield  {author} {\bibinfo {author} {\bibfnamefont {N.}~\bibnamefont
  {Suzuki}}, \bibinfo {author} {\bibfnamefont {Y.}~\bibnamefont {Tomita}},\
  and\ \bibinfo {author} {\bibfnamefont {T.}~\bibnamefont {Kojima}},\
  }\bibfield  {title} {\bibinfo {title} {Holographic recording in {TiO$_2$}
  nanoparticle-dispersed methacrylate photopolymer films},\ }\href@noop {}
  {\bibfield  {journal} {\bibinfo  {journal} {Appl.~Phys.~Lett.}\ }\textbf
  {\bibinfo {volume} {81}},\ \bibinfo {pages} {4121} (\bibinfo {year}
  {2002})}\BibitemShut {NoStop}%
\bibitem [{\citenamefont {Suzuki}\ and\ \citenamefont
  {Tomita}(2004)}]{suzuki2004}%
  \BibitemOpen
  \bibfield  {author} {\bibinfo {author} {\bibfnamefont {N.}~\bibnamefont
  {Suzuki}}\ and\ \bibinfo {author} {\bibfnamefont {Y.}~\bibnamefont
  {Tomita}},\ }\bibfield  {title} {\bibinfo {title}
  {Silica-nanoparticle-dispersed methacrylate photopolymers with net
  diffraction efficiency near 100\%},\ }\href@noop {} {\bibfield  {journal}
  {\bibinfo  {journal} {Appl.~Opt.}\ }\textbf {\bibinfo {volume} {43}},\
  \bibinfo {pages} {2125} (\bibinfo {year} {2004})}\BibitemShut {NoStop}%
\bibitem [{\citenamefont {Hata}\ \emph {et~al.}(2011)\citenamefont {Hata},
  \citenamefont {Mitsube}, \citenamefont {Momose},\ and\ \citenamefont
  {Tomita}}]{hata2011}%
  \BibitemOpen
  \bibfield  {author} {\bibinfo {author} {\bibfnamefont {E.}~\bibnamefont
  {Hata}}, \bibinfo {author} {\bibfnamefont {K.}~\bibnamefont {Mitsube}},
  \bibinfo {author} {\bibfnamefont {K.}~\bibnamefont {Momose}},\ and\ \bibinfo
  {author} {\bibfnamefont {Y.}~\bibnamefont {Tomita}},\ }\bibfield  {title}
  {\bibinfo {title} {Holographic nanoparticle-polymer composites based on
  step-growth thiol-ene photopolymerization},\ }\href@noop {} {\bibfield
  {journal} {\bibinfo  {journal} {Opt. Mater. Express}\ }\textbf {\bibinfo
  {volume} {1}},\ \bibinfo {pages} {207} (\bibinfo {year} {2011})}\BibitemShut
  {NoStop}%
\bibitem [{\citenamefont {Suzuki}\ \emph {et~al.}(2006)\citenamefont {Suzuki},
  \citenamefont {Tomita}, \citenamefont {Ohmori}, \citenamefont {Hidaka},\ and\
  \citenamefont {Chikama}}]{suzuki2006}%
  \BibitemOpen
  \bibfield  {author} {\bibinfo {author} {\bibfnamefont {N.}~\bibnamefont
  {Suzuki}}, \bibinfo {author} {\bibfnamefont {Y.}~\bibnamefont {Tomita}},
  \bibinfo {author} {\bibfnamefont {K.}~\bibnamefont {Ohmori}}, \bibinfo
  {author} {\bibfnamefont {M.}~\bibnamefont {Hidaka}},\ and\ \bibinfo {author}
  {\bibfnamefont {K.}~\bibnamefont {Chikama}},\ }\bibfield  {title} {\bibinfo
  {title} {Highly transparent {ZrO$_2$} nanoparticle dispersed acrylate
  photopolymers for volume holographic recording},\ }\href@noop {} {\bibfield
  {journal} {\bibinfo  {journal} {Opt. Express}\ }\textbf {\bibinfo {volume}
  {14}},\ \bibinfo {pages} {12712} (\bibinfo {year} {2006})}\BibitemShut
  {NoStop}%
\bibitem [{\citenamefont {Sakhno}\ \emph {et~al.}(2007)\citenamefont {Sakhno},
  \citenamefont {Goldenberg}, \citenamefont {Stumpe},\ and\ \citenamefont
  {Smirnova}}]{sakhno2007}%
  \BibitemOpen
  \bibfield  {author} {\bibinfo {author} {\bibfnamefont {O.~V.}\ \bibnamefont
  {Sakhno}}, \bibinfo {author} {\bibfnamefont {L.~M.}\ \bibnamefont
  {Goldenberg}}, \bibinfo {author} {\bibfnamefont {J.}~\bibnamefont {Stumpe}},\
  and\ \bibinfo {author} {\bibfnamefont {T.~N.}\ \bibnamefont {Smirnova}},\
  }\bibfield  {title} {\bibinfo {title} {Surface modified {ZrO$_2$} and
  {TiO$_2$} nanoparticles embedded in organic photopolymers for highly
  effective and {UV-stable} volume holograms},\ }\href@noop {} {\bibfield
  {journal} {\bibinfo  {journal} {Nanotechnol.}\ }\textbf {\bibinfo {volume}
  {18}},\ \bibinfo {pages} {105704} (\bibinfo {year} {2007})}\BibitemShut
  {NoStop}%
\bibitem [{\citenamefont {Garnweitner}\ \emph {et~al.}(2007)\citenamefont
  {Garnweitner}, \citenamefont {Goldenberg}, \citenamefont {Sakhno},
  \citenamefont {Antonietti}, \citenamefont {Niederberger},\ and\ \citenamefont
  {Stumpe}}]{garnweitner2007}%
  \BibitemOpen
  \bibfield  {author} {\bibinfo {author} {\bibfnamefont {G.}~\bibnamefont
  {Garnweitner}}, \bibinfo {author} {\bibfnamefont {L.~M.}\ \bibnamefont
  {Goldenberg}}, \bibinfo {author} {\bibfnamefont {O.~V.}\ \bibnamefont
  {Sakhno}}, \bibinfo {author} {\bibfnamefont {M.}~\bibnamefont {Antonietti}},
  \bibinfo {author} {\bibfnamefont {M.}~\bibnamefont {Niederberger}},\ and\
  \bibinfo {author} {\bibfnamefont {J.}~\bibnamefont {Stumpe}},\ }\bibfield
  {title} {\bibinfo {title} {Large‐scale synthesis of organophilic zirconia
  nanoparticles and their application in organic--inorganic nanocomposites for
  efficient volume holography},\ }\href@noop {} {\bibfield  {journal} {\bibinfo
   {journal} {Small}\ }\textbf {\bibinfo {volume} {3}},\ \bibinfo {pages}
  {1626} (\bibinfo {year} {2007})}\BibitemShut {NoStop}%
\bibitem [{\citenamefont {Fujii}\ \emph {et~al.}(2014)\citenamefont {Fujii},
  \citenamefont {Guo}, \citenamefont {Klepp}, \citenamefont {Pruner},
  \citenamefont {Fally},\ and\ \citenamefont {Tomita}}]{fujii2014}%
  \BibitemOpen
  \bibfield  {author} {\bibinfo {author} {\bibfnamefont {R.}~\bibnamefont
  {Fujii}}, \bibinfo {author} {\bibfnamefont {J.}~\bibnamefont {Guo}}, \bibinfo
  {author} {\bibfnamefont {J.}~\bibnamefont {Klepp}}, \bibinfo {author}
  {\bibfnamefont {C.}~\bibnamefont {Pruner}}, \bibinfo {author} {\bibfnamefont
  {M.}~\bibnamefont {Fally}},\ and\ \bibinfo {author} {\bibfnamefont
  {Y.}~\bibnamefont {Tomita}},\ }\bibfield  {title} {\bibinfo {title}
  {Nanoparticle polymer composite volume gratings incorporating chain transfer
  agents for holography and slow-neutron optics},\ }\href@noop {} {\bibfield
  {journal} {\bibinfo  {journal} {Opt. Lett.}\ }\textbf {\bibinfo {volume}
  {39}},\ \bibinfo {pages} {3453} (\bibinfo {year} {2014})}\BibitemShut
  {NoStop}%
\bibitem [{\citenamefont {Ni}\ \emph {et~al.}(2015)\citenamefont {Ni},
  \citenamefont {Peng}, \citenamefont {Liao}, \citenamefont {Yang},
  \citenamefont {Xue},\ and\ \citenamefont {Xie}}]{peng2015}%
  \BibitemOpen
  \bibfield  {author} {\bibinfo {author} {\bibfnamefont {M.}~\bibnamefont
  {Ni}}, \bibinfo {author} {\bibfnamefont {H.}~\bibnamefont {Peng}}, \bibinfo
  {author} {\bibfnamefont {Y.}~\bibnamefont {Liao}}, \bibinfo {author}
  {\bibfnamefont {Z.}~\bibnamefont {Yang}}, \bibinfo {author} {\bibfnamefont
  {Z.}~\bibnamefont {Xue}},\ and\ \bibinfo {author} {\bibfnamefont
  {X.}~\bibnamefont {Xie}},\ }\bibfield  {title} {\bibinfo {title} {3d image
  storage in {photopolymer/ZnS} nanocomposites tailored by
  “photoinitibitor”},\ }\href@noop {} {\bibfield  {journal} {\bibinfo
  {journal} {Macromolecules}\ }\textbf {\bibinfo {volume} {48}},\ \bibinfo
  {pages} {2958} (\bibinfo {year} {2015})}\BibitemShut {NoStop}%
\bibitem [{\citenamefont {Sakhno}\ \emph {et~al.}(2008)\citenamefont {Sakhno},
  \citenamefont {Smirnova}, \citenamefont {Goldenberg},\ and\ \citenamefont
  {Stumpe}}]{sakhno2008}%
  \BibitemOpen
  \bibfield  {author} {\bibinfo {author} {\bibfnamefont {O.~V.}\ \bibnamefont
  {Sakhno}}, \bibinfo {author} {\bibfnamefont {T.~N.}\ \bibnamefont
  {Smirnova}}, \bibinfo {author} {\bibfnamefont {L.~M.}\ \bibnamefont
  {Goldenberg}},\ and\ \bibinfo {author} {\bibfnamefont {J.}~\bibnamefont
  {Stumpe}},\ }\bibfield  {title} {\bibinfo {title} {Holographic patterning of
  luminescent photopolymer nanocomposites},\ }\href@noop {} {\bibfield
  {journal} {\bibinfo  {journal} {Mater. Sci. Eng. C}\ }\textbf {\bibinfo
  {volume} {28}},\ \bibinfo {pages} {28} (\bibinfo {year} {2008})}\BibitemShut
  {NoStop}%
\bibitem [{\citenamefont {Zhang}\ \emph {et~al.}(2019)\citenamefont {Zhang},
  \citenamefont {Yao}, \citenamefont {Zhou}, \citenamefont {Wu}, \citenamefont
  {Liu}, \citenamefont {Peng}, \citenamefont {Zhu}, \citenamefont {Smalyukh},\
  and\ \citenamefont {Xie}}]{peng2019}%
  \BibitemOpen
  \bibfield  {author} {\bibinfo {author} {\bibfnamefont {X.}~\bibnamefont
  {Zhang}}, \bibinfo {author} {\bibfnamefont {W.}~\bibnamefont {Yao}}, \bibinfo
  {author} {\bibfnamefont {X.}~\bibnamefont {Zhou}}, \bibinfo {author}
  {\bibfnamefont {W.}~\bibnamefont {Wu}}, \bibinfo {author} {\bibfnamefont
  {Q.}~\bibnamefont {Liu}}, \bibinfo {author} {\bibfnamefont {H.}~\bibnamefont
  {Peng}}, \bibinfo {author} {\bibfnamefont {J.}~\bibnamefont {Zhu}}, \bibinfo
  {author} {\bibfnamefont {I.}~\bibnamefont {Smalyukh}},\ and\ \bibinfo
  {author} {\bibfnamefont {X.}~\bibnamefont {Xie}},\ }\bibfield  {title}
  {\bibinfo {title} {Holographic polymer nanocomposites with simultaneously
  boosted diffraction efficiency and upconversion photoluminescence},\
  }\href@noop {} {\bibfield  {journal} {\bibinfo  {journal} {Compos. Sci.
  Technol.}\ }\textbf {\bibinfo {volume} {181}},\ \bibinfo {pages} {107705}
  (\bibinfo {year} {2019})}\BibitemShut {NoStop}%
\bibitem [{\citenamefont {Guo}\ \emph {et~al.}(2020)\citenamefont {Guo},
  \citenamefont {Cao}, \citenamefont {Jian}, \citenamefont {Ma}, \citenamefont
  {Wang},\ and\ \citenamefont {Zhang}}]{guo2020}%
  \BibitemOpen
  \bibfield  {author} {\bibinfo {author} {\bibfnamefont {J.}~\bibnamefont
  {Guo}}, \bibinfo {author} {\bibfnamefont {L.}~\bibnamefont {Cao}}, \bibinfo
  {author} {\bibfnamefont {J.}~\bibnamefont {Jian}}, \bibinfo {author}
  {\bibfnamefont {H.}~\bibnamefont {Ma}}, \bibinfo {author} {\bibfnamefont
  {D.}~\bibnamefont {Wang}},\ and\ \bibinfo {author} {\bibfnamefont
  {X.}~\bibnamefont {Zhang}},\ }\bibfield  {title} {\bibinfo {title}
  {Single-wall carbon nanotube promoted allylic homopolymerization for
  holographic patterning},\ }\href@noop {} {\bibfield  {journal} {\bibinfo
  {journal} {Carbon}\ }\textbf {\bibinfo {volume} {157}},\ \bibinfo {pages}
  {64} (\bibinfo {year} {2020})}\BibitemShut {NoStop}%
\bibitem [{\citenamefont {Leite}\ \emph {et~al.}(2009)\citenamefont {Leite},
  \citenamefont {Naydenova}, \citenamefont {Pandey}, \citenamefont {Babeva},
  \citenamefont {Majono}, \citenamefont {Minotova},\ and\ \citenamefont
  {Toal}}]{leite2009}%
  \BibitemOpen
  \bibfield  {author} {\bibinfo {author} {\bibfnamefont {E.}~\bibnamefont
  {Leite}}, \bibinfo {author} {\bibfnamefont {I.}~\bibnamefont {Naydenova}},
  \bibinfo {author} {\bibfnamefont {N.}~\bibnamefont {Pandey}}, \bibinfo
  {author} {\bibfnamefont {T.}~\bibnamefont {Babeva}}, \bibinfo {author}
  {\bibfnamefont {G.}~\bibnamefont {Majono}}, \bibinfo {author} {\bibfnamefont
  {S.}~\bibnamefont {Minotova}},\ and\ \bibinfo {author} {\bibfnamefont
  {V.}~\bibnamefont {Toal}},\ }\bibfield  {title} {\bibinfo {title}
  {Investigation of the light induced redistribution of zeolite {Beta}
  nanoparticles in an acrylamide-based photopolymer},\ }\href@noop {}
  {\bibfield  {journal} {\bibinfo  {journal} {J. Opt. A: Pure Appl. Opt.}\
  }\textbf {\bibinfo {volume} {11}},\ \bibinfo {pages} {024016} (\bibinfo
  {year} {2009})}\BibitemShut {NoStop}%
\bibitem [{\citenamefont {Liu}\ \emph {et~al.}(2009)\citenamefont {Liu},
  \citenamefont {Tomita}, \citenamefont {Oshima}, \citenamefont {Chikama},
  \citenamefont {Matsubara}, \citenamefont {Nakashima},\ and\ \citenamefont
  {Kawai}}]{liu2009}%
  \BibitemOpen
  \bibfield  {author} {\bibinfo {author} {\bibfnamefont {X.}~\bibnamefont
  {Liu}}, \bibinfo {author} {\bibfnamefont {Y.}~\bibnamefont {Tomita}},
  \bibinfo {author} {\bibfnamefont {J.}~\bibnamefont {Oshima}}, \bibinfo
  {author} {\bibfnamefont {K.}~\bibnamefont {Chikama}}, \bibinfo {author}
  {\bibfnamefont {K.}~\bibnamefont {Matsubara}}, \bibinfo {author}
  {\bibfnamefont {T.}~\bibnamefont {Nakashima}},\ and\ \bibinfo {author}
  {\bibfnamefont {T.}~\bibnamefont {Kawai}},\ }\bibfield  {title} {\bibinfo
  {title} {Holographic assembly of semiconductor {CdSe} quantum dots in polymer
  for volume {Bragg} grating structures with diffraction efficiency near
  100\%},\ }\href@noop {} {\bibfield  {journal} {\bibinfo  {journal} {Appl.
  Phys. Lett.}\ }\textbf {\bibinfo {volume} {95}},\ \bibinfo {pages} {261109}
  (\bibinfo {year} {2009})}\BibitemShut {NoStop}%
\bibitem [{\citenamefont {Sakhno}\ \emph {et~al.}(2009)\citenamefont {Sakhno},
  \citenamefont {Goldenberg}, \citenamefont {Stumpe},\ and\ \citenamefont
  {Smirnova}}]{sakhno2009}%
  \BibitemOpen
  \bibfield  {author} {\bibinfo {author} {\bibfnamefont {O.~V.}\ \bibnamefont
  {Sakhno}}, \bibinfo {author} {\bibfnamefont {L.~M.}\ \bibnamefont
  {Goldenberg}}, \bibinfo {author} {\bibfnamefont {J.}~\bibnamefont {Stumpe}},\
  and\ \bibinfo {author} {\bibfnamefont {T.~N.}\ \bibnamefont {Smirnova}},\
  }\bibfield  {title} {\bibinfo {title} {Effective volume holographic
  structures based on organic--inorganic photopolymer nanocomposites},\
  }\href@noop {} {\bibfield  {journal} {\bibinfo  {journal} {J. Opt. A: Pure
  Appl. Opt.}\ }\textbf {\bibinfo {volume} {11}},\ \bibinfo {pages} {024013}
  (\bibinfo {year} {2009})}\BibitemShut {NoStop}%
\bibitem [{\citenamefont {Tomita}\ \emph
  {et~al.}(2006{\natexlab{a}})\citenamefont {Tomita}, \citenamefont
  {Furushima}, \citenamefont {Ochi}, \citenamefont {Ishizu}, \citenamefont
  {Tanaka}, \citenamefont {Ozawa}, \citenamefont {Hidaka},\ and\ \citenamefont
  {Chikama}}]{tomita_hbp2006}%
  \BibitemOpen
  \bibfield  {author} {\bibinfo {author} {\bibfnamefont {Y.}~\bibnamefont
  {Tomita}}, \bibinfo {author} {\bibfnamefont {K.}~\bibnamefont {Furushima}},
  \bibinfo {author} {\bibfnamefont {K.}~\bibnamefont {Ochi}}, \bibinfo {author}
  {\bibfnamefont {K.}~\bibnamefont {Ishizu}}, \bibinfo {author} {\bibfnamefont
  {A.}~\bibnamefont {Tanaka}}, \bibinfo {author} {\bibfnamefont
  {M.}~\bibnamefont {Ozawa}}, \bibinfo {author} {\bibfnamefont
  {M.}~\bibnamefont {Hidaka}},\ and\ \bibinfo {author} {\bibfnamefont
  {K.}~\bibnamefont {Chikama}},\ }\bibfield  {title} {\bibinfo {title} {Organic
  nanoparticle (hyperbranched polymer)-dispersed photopolymers for volume
  holographic storage},\ }\href@noop {} {\bibfield  {journal} {\bibinfo
  {journal} {Appl.~Phys.~Lett.}\ }\textbf {\bibinfo {volume} {88}},\ \bibinfo
  {pages} {071103} (\bibinfo {year} {2006}{\natexlab{a}})}\BibitemShut
  {NoStop}%
\bibitem [{\citenamefont {Tomita}\ \emph
  {et~al.}(2019{\natexlab{a}})\citenamefont {Tomita}, \citenamefont {Aoi},
  \citenamefont {Oshima},\ and\ \citenamefont {Odoi}}]{tomita2019}%
  \BibitemOpen
  \bibfield  {author} {\bibinfo {author} {\bibfnamefont {Y.}~\bibnamefont
  {Tomita}}, \bibinfo {author} {\bibfnamefont {T.}~\bibnamefont {Aoi}},
  \bibinfo {author} {\bibfnamefont {J.}~\bibnamefont {Oshima}},\ and\ \bibinfo
  {author} {\bibfnamefont {K.}~\bibnamefont {Odoi}},\ }\bibfield  {title}
  {\bibinfo {title} {Hyperbranched-polymer nanocomposite gratings with
  ultrahigh refractive index modulation amplitudes for volume holographic
  optical elements},\ }\href@noop {} {\bibfield  {journal} {\bibinfo  {journal}
  {Proc. SPIE}\ }\textbf {\bibinfo {volume} {11030}},\ \bibinfo {pages}
  {1103007} (\bibinfo {year} {2019}{\natexlab{a}})}\BibitemShut {NoStop}%
\bibitem [{\citenamefont {Tomita}\ \emph {et~al.}(2008)\citenamefont {Tomita},
  \citenamefont {Nakamura},\ and\ \citenamefont {Tago}}]{tomita2008}%
  \BibitemOpen
  \bibfield  {author} {\bibinfo {author} {\bibfnamefont {Y.}~\bibnamefont
  {Tomita}}, \bibinfo {author} {\bibfnamefont {T.}~\bibnamefont {Nakamura}},\
  and\ \bibinfo {author} {\bibfnamefont {A.}~\bibnamefont {Tago}},\ }\bibfield
  {title} {\bibinfo {title} {Improved thermal stability of volume holograms
  recorded in nanoparticle-polymer composite films},\ }\href@noop {} {\bibfield
   {journal} {\bibinfo  {journal} {Opt. Lett.}\ }\textbf {\bibinfo {volume}
  {33}},\ \bibinfo {pages} {1750} (\bibinfo {year} {2008})}\BibitemShut
  {NoStop}%
\bibitem [{\citenamefont {Hata}\ and\ \citenamefont {Tomita}(2010)}]{hata2010}%
  \BibitemOpen
  \bibfield  {author} {\bibinfo {author} {\bibfnamefont {E.}~\bibnamefont
  {Hata}}\ and\ \bibinfo {author} {\bibfnamefont {Y.}~\bibnamefont {Tomita}},\
  }\bibfield  {title} {\bibinfo {title} {Order-of-magnitude
  polymerization-shrinkage suppression of volume gratings recorded in
  nanoparticle-polymer composites},\ }\href@noop {} {\bibfield  {journal}
  {\bibinfo  {journal} {Opt. Lett.}\ }\textbf {\bibinfo {volume} {35}},\
  \bibinfo {pages} {396} (\bibinfo {year} {2010})}\BibitemShut {NoStop}%
\bibitem [{\citenamefont {Smirnova}\ \emph {et~al.}(2005)\citenamefont
  {Smirnova}, \citenamefont {Sakhno}, \citenamefont {Fitio}, \citenamefont
  {Gritsai},\ and\ \citenamefont {Stumpe}}]{smirnova2005}%
  \BibitemOpen
  \bibfield  {author} {\bibinfo {author} {\bibfnamefont {T.~N.}\ \bibnamefont
  {Smirnova}}, \bibinfo {author} {\bibfnamefont {O.~V.}\ \bibnamefont
  {Sakhno}}, \bibinfo {author} {\bibfnamefont {V.~M.}\ \bibnamefont {Fitio}},
  \bibinfo {author} {\bibfnamefont {Y.}~\bibnamefont {Gritsai}},\ and\ \bibinfo
  {author} {\bibfnamefont {J.}~\bibnamefont {Stumpe}},\ }\bibfield  {title}
  {\bibinfo {title} {Nonlinear diffraction in gratings based on
  polymer--dispersed {TiO$_2$} nanoparticles},\ }\href@noop {} {\bibfield
  {journal} {\bibinfo  {journal} {Appl. Phys. B}\ }\textbf {\bibinfo {volume}
  {80}},\ \bibinfo {pages} {947} (\bibinfo {year} {2005})}\BibitemShut
  {NoStop}%
\bibitem [{\citenamefont {Tomita}(2005)}]{tomita2005a}%
  \BibitemOpen
  \bibfield  {author} {\bibinfo {author} {\bibfnamefont {Y.}~\bibnamefont
  {Tomita}},\ }\bibfield  {title} {\bibinfo {title} {Holographic manipulation
  of nanoparticle distribution morphology in photopolymers and its applications
  to volume holographic recording and nonlinear photonic crystals},\
  }\href@noop {} {\bibfield  {journal} {\bibinfo  {journal} {OSA Trends Opt.
  Photonics}\ }\textbf {\bibinfo {volume} {99}},\ \bibinfo {pages} {274}
  (\bibinfo {year} {2005})}\BibitemShut {NoStop}%
\bibitem [{\citenamefont {Liu}\ \emph {et~al.}(2010)\citenamefont {Liu},
  \citenamefont {Matsumura}, \citenamefont {Tomita}, \citenamefont {Yasui},
  \citenamefont {Kojima},\ and\ \citenamefont {Chikama}}]{liu2010}%
  \BibitemOpen
  \bibfield  {author} {\bibinfo {author} {\bibfnamefont {X.}~\bibnamefont
  {Liu}}, \bibinfo {author} {\bibfnamefont {K.}~\bibnamefont {Matsumura}},
  \bibinfo {author} {\bibfnamefont {Y.}~\bibnamefont {Tomita}}, \bibinfo
  {author} {\bibfnamefont {K.}~\bibnamefont {Yasui}}, \bibinfo {author}
  {\bibfnamefont {K.}~\bibnamefont {Kojima}},\ and\ \bibinfo {author}
  {\bibfnamefont {K.}~\bibnamefont {Chikama}},\ }\bibfield  {title} {\bibinfo
  {title} {Nonlinear optical responses of nanoparticle-polymer composites
  incorporating organic (hyperbranched polymer)-metallic nanoparticle
  complex},\ }\href@noop {} {\bibfield  {journal} {\bibinfo  {journal} {J.
  Appl. Phys.}\ }\textbf {\bibinfo {volume} {108}},\ \bibinfo {pages} {073102}
  (\bibinfo {year} {2010})}\BibitemShut {NoStop}%
\bibitem [{\citenamefont {Liu}\ \emph {et~al.}(2012)\citenamefont {Liu},
  \citenamefont {Adachi}, \citenamefont {Tomita}, \citenamefont {Oshima},
  \citenamefont {Nakashima},\ and\ \citenamefont {Kawai}}]{liu2012}%
  \BibitemOpen
  \bibfield  {author} {\bibinfo {author} {\bibfnamefont {X.}~\bibnamefont
  {Liu}}, \bibinfo {author} {\bibfnamefont {Y.}~\bibnamefont {Adachi}},
  \bibinfo {author} {\bibfnamefont {Y.}~\bibnamefont {Tomita}}, \bibinfo
  {author} {\bibfnamefont {J.}~\bibnamefont {Oshima}}, \bibinfo {author}
  {\bibfnamefont {T.}~\bibnamefont {Nakashima}},\ and\ \bibinfo {author}
  {\bibfnamefont {T.}~\bibnamefont {Kawai}},\ }\bibfield  {title} {\bibinfo
  {title} {High-order nonlinear optical response of a polymer nanocomposite
  film incorporating semiconducotor cdse quantum dots},\ }\href@noop {}
  {\bibfield  {journal} {\bibinfo  {journal} {Opt. Express}\ }\textbf {\bibinfo
  {volume} {20}},\ \bibinfo {pages} {13457} (\bibinfo {year}
  {2012})}\BibitemShut {NoStop}%
\bibitem [{\citenamefont {Dolgaleva}\ \emph {et~al.}(2009)\citenamefont
  {Dolgaleva}, \citenamefont {Shin},\ and\ \citenamefont
  {Boyd}}]{dolgaleva2009}%
  \BibitemOpen
  \bibfield  {author} {\bibinfo {author} {\bibfnamefont {K.}~\bibnamefont
  {Dolgaleva}}, \bibinfo {author} {\bibfnamefont {H.}~\bibnamefont {Shin}},\
  and\ \bibinfo {author} {\bibfnamefont {R.~W.}\ \bibnamefont {Boyd}},\
  }\bibfield  {title} {\bibinfo {title} {Observation of a microscopic cascaded
  contribution to the fifth-order nonlinear susceptibility},\ }\href@noop {}
  {\bibfield  {journal} {\bibinfo  {journal} {Phys. Rev. Lett.}\ }\textbf
  {\bibinfo {volume} {103}},\ \bibinfo {pages} {113902} (\bibinfo {year}
  {2009})}\BibitemShut {NoStop}%
\bibitem [{\citenamefont {Momose}\ \emph {et~al.}(2012)\citenamefont {Momose},
  \citenamefont {Takamaya}, \citenamefont {Hata},\ and\ \citenamefont
  {Tomita}}]{momose2012}%
  \BibitemOpen
  \bibfield  {author} {\bibinfo {author} {\bibfnamefont {K.}~\bibnamefont
  {Momose}}, \bibinfo {author} {\bibfnamefont {S.}~\bibnamefont {Takamaya}},
  \bibinfo {author} {\bibfnamefont {E.}~\bibnamefont {Hata}},\ and\ \bibinfo
  {author} {\bibfnamefont {Y.}~\bibnamefont {Tomita}},\ }\bibfield  {title}
  {\bibinfo {title} {Shift-multiplexed holographic digital data page storage in
  a nanoparticle-(thiol-ene) polymer composite film},\ }\href@noop {}
  {\bibfield  {journal} {\bibinfo  {journal} {Opt. Lett.}\ }\textbf {\bibinfo
  {volume} {37}},\ \bibinfo {pages} {2250} (\bibinfo {year}
  {2012})}\BibitemShut {NoStop}%
\bibitem [{\citenamefont {Takayama}\ \emph {et~al.}(2014)\citenamefont
  {Takayama}, \citenamefont {Nagaya}, \citenamefont {Momose},\ and\
  \citenamefont {Tomita}}]{takayama2014}%
  \BibitemOpen
  \bibfield  {author} {\bibinfo {author} {\bibfnamefont {S.}~\bibnamefont
  {Takayama}}, \bibinfo {author} {\bibfnamefont {K.}~\bibnamefont {Nagaya}},
  \bibinfo {author} {\bibfnamefont {K.}~\bibnamefont {Momose}},\ and\ \bibinfo
  {author} {\bibfnamefont {Y.}~\bibnamefont {Tomita}},\ }\bibfield  {title}
  {\bibinfo {title} {Effects of symbol modulation coding on readout fidelity of
  shift-multiplexed holographic digital data page storage in a
  photopolymerizable nanoparticle-(thiol-ene)polymer composite film},\
  }\href@noop {} {\bibfield  {journal} {\bibinfo  {journal} {Appl. Opt.}\
  }\textbf {\bibinfo {volume} {53}},\ \bibinfo {pages} {B53} (\bibinfo {year}
  {2014})}\BibitemShut {NoStop}%
\bibitem [{\citenamefont {Tomita}\ \emph
  {et~al.}(2016{\natexlab{b}})\citenamefont {Tomita}, \citenamefont {Urano},
  \citenamefont {Fukamizu}, \citenamefont {Kametani}, \citenamefont
  {Nishimura},\ and\ \citenamefont {Odoi}}]{tomita_hbp2016}%
  \BibitemOpen
  \bibfield  {author} {\bibinfo {author} {\bibfnamefont {Y.}~\bibnamefont
  {Tomita}}, \bibinfo {author} {\bibfnamefont {H.}~\bibnamefont {Urano}},
  \bibinfo {author} {\bibfnamefont {T.}~\bibnamefont {Fukamizu}}, \bibinfo
  {author} {\bibfnamefont {Y.}~\bibnamefont {Kametani}}, \bibinfo {author}
  {\bibfnamefont {N.}~\bibnamefont {Nishimura}},\ and\ \bibinfo {author}
  {\bibfnamefont {K.}~\bibnamefont {Odoi}},\ }\bibfield  {title} {\bibinfo
  {title} {Nanoparticle-polymer composite volume holographic gratings dispersed
  with ultrahigh-refractive-index hyperbranched polymer as organic
  nanoparticles},\ }\href@noop {} {\bibfield  {journal} {\bibinfo  {journal}
  {Opt. Lett.}\ }\textbf {\bibinfo {volume} {41}},\ \bibinfo {pages} {1281}
  (\bibinfo {year} {2016}{\natexlab{b}})}\BibitemShut {NoStop}%
\bibitem [{\citenamefont {Fernandez}\ \emph {et~al.}(2019)\citenamefont
  {Fernandez}, \citenamefont {Bleda}, \citenamefont {Gallego}, \citenamefont
  {Neipp}, \citenamefont {Marquez}, \citenamefont {Tomita}, \citenamefont
  {Pascual},\ and\ \citenamefont {Bel\'{e}ndez}}]{fernandez2019}%
  \BibitemOpen
  \bibfield  {author} {\bibinfo {author} {\bibfnamefont {R.}~\bibnamefont
  {Fernandez}}, \bibinfo {author} {\bibfnamefont {S.}~\bibnamefont {Bleda}},
  \bibinfo {author} {\bibfnamefont {S.}~\bibnamefont {Gallego}}, \bibinfo
  {author} {\bibfnamefont {C.}~\bibnamefont {Neipp}}, \bibinfo {author}
  {\bibfnamefont {A.}~\bibnamefont {Marquez}}, \bibinfo {author} {\bibfnamefont
  {Y.}~\bibnamefont {Tomita}}, \bibinfo {author} {\bibfnamefont
  {I.}~\bibnamefont {Pascual}},\ and\ \bibinfo {author} {\bibfnamefont
  {A.}~\bibnamefont {Bel\'{e}ndez}},\ }\bibfield  {title} {\bibinfo {title}
  {Holographic waveguides in photopolymers},\ }\href@noop {} {\bibfield
  {journal} {\bibinfo  {journal} {Opt. Express}\ }\textbf {\bibinfo {volume}
  {27}},\ \bibinfo {pages} {827} (\bibinfo {year} {2019})}\BibitemShut
  {NoStop}%
\bibitem [{\citenamefont {Smirnova}\ \emph {et~al.}(2014)\citenamefont
  {Smirnova}, \citenamefont {Sakhno}, \citenamefont {Fitio}, \citenamefont
  {Gritsai},\ and\ \citenamefont {Stumpe}}]{smirnova2014}%
  \BibitemOpen
  \bibfield  {author} {\bibinfo {author} {\bibfnamefont {T.~N.}\ \bibnamefont
  {Smirnova}}, \bibinfo {author} {\bibfnamefont {O.~V.}\ \bibnamefont
  {Sakhno}}, \bibinfo {author} {\bibfnamefont {V.~M.}\ \bibnamefont {Fitio}},
  \bibinfo {author} {\bibfnamefont {Y.}~\bibnamefont {Gritsai}},\ and\ \bibinfo
  {author} {\bibfnamefont {J.}~\bibnamefont {Stumpe}},\ }\bibfield  {title}
  {\bibinfo {title} {Simple and high performance {DFB} laser based on dye-doped
  nanocomposite volume gratings},\ }\href@noop {} {\bibfield  {journal}
  {\bibinfo  {journal} {Laser Phys. Lett.}\ }\textbf {\bibinfo {volume} {11}},\
  \bibinfo {pages} {215804} (\bibinfo {year} {2014})}\BibitemShut {NoStop}%
\bibitem [{\citenamefont {Rauch}\ and\ \citenamefont
  {Werner}(2015)}]{rauch2015}%
  \BibitemOpen
  \bibfield  {author} {\bibinfo {author} {\bibfnamefont {H.}~\bibnamefont
  {Rauch}}\ and\ \bibinfo {author} {\bibfnamefont {S.~A.}\ \bibnamefont
  {Werner}},\ }\href@noop {} {\emph {\bibinfo {title} {Neutron Interferometry,
  2nd ed.}}}\ (\bibinfo  {publisher} {Oxford University Press},\ \bibinfo
  {year} {2015})\BibitemShut {NoStop}%
\bibitem [{\citenamefont {Fally}\ \emph {et~al.}(2010)\citenamefont {Fally},
  \citenamefont {Klepp}, \citenamefont {Tomita}, \citenamefont {Nakamura},
  \citenamefont {Pruner}, \citenamefont {Ellabban}, \citenamefont {Rupp},
  \citenamefont {Bichler}, \citenamefont {Dreven\v{s}ek-Olenik}, \citenamefont
  {Kohlbrecher}, \citenamefont {Eckerlebe}, \citenamefont {Lemmel},\ and\
  \citenamefont {Rauch}}]{fally2010}%
  \BibitemOpen
  \bibfield  {author} {\bibinfo {author} {\bibfnamefont {M.}~\bibnamefont
  {Fally}}, \bibinfo {author} {\bibfnamefont {J.}~\bibnamefont {Klepp}},
  \bibinfo {author} {\bibfnamefont {Y.}~\bibnamefont {Tomita}}, \bibinfo
  {author} {\bibfnamefont {T.}~\bibnamefont {Nakamura}}, \bibinfo {author}
  {\bibfnamefont {C.}~\bibnamefont {Pruner}}, \bibinfo {author} {\bibfnamefont
  {M.~A.}\ \bibnamefont {Ellabban}}, \bibinfo {author} {\bibfnamefont {R.~A.}\
  \bibnamefont {Rupp}}, \bibinfo {author} {\bibfnamefont {M.}~\bibnamefont
  {Bichler}}, \bibinfo {author} {\bibfnamefont {I.}~\bibnamefont
  {Dreven\v{s}ek-Olenik}}, \bibinfo {author} {\bibfnamefont {J.}~\bibnamefont
  {Kohlbrecher}}, \bibinfo {author} {\bibfnamefont {H.}~\bibnamefont
  {Eckerlebe}}, \bibinfo {author} {\bibfnamefont {H.}~\bibnamefont {Lemmel}},\
  and\ \bibinfo {author} {\bibfnamefont {H.}~\bibnamefont {Rauch}},\ }\bibfield
   {title} {\bibinfo {title} {Neutron optical beam splitter from
  holographically structured nanoparticle-polymer composites},\ }\href@noop {}
  {\bibfield  {journal} {\bibinfo  {journal} {Phys. Rev. Lett.}\ }\textbf
  {\bibinfo {volume} {105}},\ \bibinfo {pages} {123904} (\bibinfo {year}
  {2010})}\BibitemShut {NoStop}%
\bibitem [{\citenamefont {Klepp}\ \emph
  {et~al.}(2012{\natexlab{a}})\citenamefont {Klepp}, \citenamefont {Pruner},
  \citenamefont {Tomita}, \citenamefont {Mitsube}, \citenamefont {Geltenbort},\
  and\ \citenamefont {Fally}}]{klepp2012a}%
  \BibitemOpen
  \bibfield  {author} {\bibinfo {author} {\bibfnamefont {J.}~\bibnamefont
  {Klepp}}, \bibinfo {author} {\bibfnamefont {C.}~\bibnamefont {Pruner}},
  \bibinfo {author} {\bibfnamefont {Y.}~\bibnamefont {Tomita}}, \bibinfo
  {author} {\bibfnamefont {K.}~\bibnamefont {Mitsube}}, \bibinfo {author}
  {\bibfnamefont {P.}~\bibnamefont {Geltenbort}},\ and\ \bibinfo {author}
  {\bibfnamefont {M.}~\bibnamefont {Fally}},\ }\bibfield  {title} {\bibinfo
  {title} {Mirrors for slow neutrons from holographic nanoparticle-polymer
  free-standing film-gratings},\ }\href@noop {} {\bibfield  {journal} {\bibinfo
   {journal} {Appl. Phys. Lett.}\ }\textbf {\bibinfo {volume} {100}},\ \bibinfo
  {pages} {214104} (\bibinfo {year} {2012}{\natexlab{a}})}\BibitemShut
  {NoStop}%
\bibitem [{\citenamefont {Klepp}\ \emph
  {et~al.}(2012{\natexlab{b}})\citenamefont {Klepp}, \citenamefont {Pruner},
  \citenamefont {Tomita}, \citenamefont {Kohlbrecher},\ and\ \citenamefont
  {Fally}}]{klepp2012b}%
  \BibitemOpen
  \bibfield  {author} {\bibinfo {author} {\bibfnamefont {J.}~\bibnamefont
  {Klepp}}, \bibinfo {author} {\bibfnamefont {C.}~\bibnamefont {Pruner}},
  \bibinfo {author} {\bibfnamefont {Y.}~\bibnamefont {Tomita}}, \bibinfo
  {author} {\bibfnamefont {J.}~\bibnamefont {Kohlbrecher}},\ and\ \bibinfo
  {author} {\bibfnamefont {M.}~\bibnamefont {Fally}},\ }\bibfield  {title}
  {\bibinfo {title} {Three-port beam splitter for cold neutrons using
  holographic nanoparticle-polymer composite diffraction gratings},\
  }\href@noop {} {\bibfield  {journal} {\bibinfo  {journal} {Appl. Phys.
  Lett.}\ }\textbf {\bibinfo {volume} {101}},\ \bibinfo {pages} {154104}
  (\bibinfo {year} {2012}{\natexlab{b}})}\BibitemShut {NoStop}%
\bibitem [{\citenamefont {Blaickner}\ \emph {et~al.}(2019)\citenamefont
  {Blaickner}, \citenamefont {Demirel}, \citenamefont {Dreven\v{s}ek-Olenik},
  \citenamefont {Fally}, \citenamefont {Flauger}, \citenamefont {Geltenbort},
  \citenamefont {Hasegawa}, \citenamefont {Kurinjimala}, \citenamefont
  {Li\v{c}en}, \citenamefont {Pruner}, \citenamefont {Sponar}, \citenamefont
  {Tomita},\ and\ \citenamefont {Klepp}}]{blaickner2019}%
  \BibitemOpen
  \bibfield  {author} {\bibinfo {author} {\bibfnamefont {M.}~\bibnamefont
  {Blaickner}}, \bibinfo {author} {\bibfnamefont {B.}~\bibnamefont {Demirel}},
  \bibinfo {author} {\bibfnamefont {I.}~\bibnamefont {Dreven\v{s}ek-Olenik}},
  \bibinfo {author} {\bibfnamefont {M.}~\bibnamefont {Fally}}, \bibinfo
  {author} {\bibfnamefont {P.}~\bibnamefont {Flauger}}, \bibinfo {author}
  {\bibfnamefont {P.}~\bibnamefont {Geltenbort}}, \bibinfo {author}
  {\bibfnamefont {Y.}~\bibnamefont {Hasegawa}}, \bibinfo {author}
  {\bibfnamefont {R.}~\bibnamefont {Kurinjimala}}, \bibinfo {author}
  {\bibfnamefont {M.}~\bibnamefont {Li\v{c}en}}, \bibinfo {author}
  {\bibfnamefont {C.}~\bibnamefont {Pruner}}, \bibinfo {author} {\bibfnamefont
  {S.}~\bibnamefont {Sponar}}, \bibinfo {author} {\bibfnamefont
  {Y.}~\bibnamefont {Tomita}},\ and\ \bibinfo {author} {\bibfnamefont
  {J.}~\bibnamefont {Klepp}},\ }\bibfield  {title} {\bibinfo {title}
  {Monte-{C}arlo simulation of neutron transmission through nanocomposite
  materials for neutron-optics applications},\ }\href@noop {} {\bibfield
  {journal} {\bibinfo  {journal} {Nucl. Instrum. Meth. Phys. Res. A}\ }\textbf
  {\bibinfo {volume} {916}},\ \bibinfo {pages} {154} (\bibinfo {year}
  {2019})}\BibitemShut {NoStop}%
\bibitem [{\citenamefont {Klepp}\ \emph
  {et~al.}(2012{\natexlab{c}})\citenamefont {Klepp}, \citenamefont {Pruner},
  \citenamefont {Tomita}, \citenamefont {Geltenbort}, \citenamefont
  {Dreven\v{s}ek-Olenik}, \citenamefont {Gyergyek}, \citenamefont
  {Kohlbrecher},\ and\ \citenamefont {Fally}}]{klepp2012c}%
  \BibitemOpen
  \bibfield  {author} {\bibinfo {author} {\bibfnamefont {J.}~\bibnamefont
  {Klepp}}, \bibinfo {author} {\bibfnamefont {C.}~\bibnamefont {Pruner}},
  \bibinfo {author} {\bibfnamefont {Y.}~\bibnamefont {Tomita}}, \bibinfo
  {author} {\bibfnamefont {P.}~\bibnamefont {Geltenbort}}, \bibinfo {author}
  {\bibfnamefont {I.}~\bibnamefont {Dreven\v{s}ek-Olenik}}, \bibinfo {author}
  {\bibfnamefont {S.}~\bibnamefont {Gyergyek}}, \bibinfo {author}
  {\bibfnamefont {J.}~\bibnamefont {Kohlbrecher}},\ and\ \bibinfo {author}
  {\bibfnamefont {M.}~\bibnamefont {Fally}},\ }\bibfield  {title} {\bibinfo
  {title} {Holographic gratings for slow-neutron optics},\ }\href@noop {}
  {\bibfield  {journal} {\bibinfo  {journal} {Materials}\ }\textbf {\bibinfo
  {volume} {2012}},\ \bibinfo {pages} {2788} (\bibinfo {year}
  {2012}{\natexlab{c}})}\BibitemShut {NoStop}%
\bibitem [{\citenamefont {Nesvizhevsky}\ \emph {et~al.}(2010)\citenamefont
  {Nesvizhevsky}, \citenamefont {Cubitt}, \citenamefont {Lychagin},
  \citenamefont {Muzychka}, \citenamefont {Nekhaev}, \citenamefont {Pignol},
  \citenamefont {Protasov},\ and\ \citenamefont
  {Strelkov}}]{nesvizhevskyMat2010}%
  \BibitemOpen
  \bibfield  {author} {\bibinfo {author} {\bibfnamefont {V.~V.}\ \bibnamefont
  {Nesvizhevsky}}, \bibinfo {author} {\bibfnamefont {R.}~\bibnamefont
  {Cubitt}}, \bibinfo {author} {\bibfnamefont {E.}~\bibnamefont {Lychagin}},
  \bibinfo {author} {\bibfnamefont {A.}~\bibnamefont {Muzychka}}, \bibinfo
  {author} {\bibfnamefont {G.}~\bibnamefont {Nekhaev}}, \bibinfo {author}
  {\bibfnamefont {G.}~\bibnamefont {Pignol}}, \bibinfo {author} {\bibfnamefont
  {K.}~\bibnamefont {Protasov}},\ and\ \bibinfo {author} {\bibfnamefont
  {A.}~\bibnamefont {Strelkov}},\ }\bibfield  {title} {\bibinfo {title}
  {Application of diamond nanoparticles in low-energy neutron physics},\ }\href
  {https://doi.org/10.3390/ma3031768} {\bibfield  {journal} {\bibinfo
  {journal} {Materials}\ }\textbf {\bibinfo {volume} {3}},\ \bibinfo {pages}
  {1768} (\bibinfo {year} {2010})}\BibitemShut {NoStop}%
\bibitem [{\citenamefont {Nesvizhevsky}\ \emph {et~al.}(2018)\citenamefont
  {Nesvizhevsky}, \citenamefont {Dubois}, \citenamefont {Gutfreund},
  \citenamefont {Lychagin}, \citenamefont {Nezvanov},\ and\ \citenamefont
  {Zhernenkov}}]{nesvizhevsky2018}%
  \BibitemOpen
  \bibfield  {author} {\bibinfo {author} {\bibfnamefont {V.~V.}\ \bibnamefont
  {Nesvizhevsky}}, \bibinfo {author} {\bibfnamefont {M.}~\bibnamefont
  {Dubois}}, \bibinfo {author} {\bibfnamefont {P.}~\bibnamefont {Gutfreund}},
  \bibinfo {author} {\bibfnamefont {E.~V.}\ \bibnamefont {Lychagin}}, \bibinfo
  {author} {\bibfnamefont {A.~Y.}\ \bibnamefont {Nezvanov}},\ and\ \bibinfo
  {author} {\bibfnamefont {K.~N.}\ \bibnamefont {Zhernenkov}},\ }\bibfield
  {title} {\bibinfo {title} {Effect of nanodiamond fluorination on the
  efficiency of quasispecular reflection of cold neutrons},\ }\href
  {https://doi.org/10.1103/PhysRevA.97.023629} {\bibfield  {journal} {\bibinfo
  {journal} {Phys. Rev. A}\ }\textbf {\bibinfo {volume} {97}},\ \bibinfo
  {pages} {023629} (\bibinfo {year} {2018})}\BibitemShut {NoStop}%
\bibitem [{\citenamefont {Tomita}\ \emph
  {et~al.}(2019{\natexlab{b}})\citenamefont {Tomita}, \citenamefont {Kageyama},
  \citenamefont {Iso}, , \citenamefont {Umemoto}, \citenamefont {Klepp},
  \citenamefont {Pruner},\ and\ \citenamefont {Fally}}]{tomita2019a}%
  \BibitemOpen
  \bibfield  {author} {\bibinfo {author} {\bibfnamefont {Y.}~\bibnamefont
  {Tomita}}, \bibinfo {author} {\bibfnamefont {A.}~\bibnamefont {Kageyama}},
  \bibinfo {author} {\bibfnamefont {Y.}~\bibnamefont {Iso}}, , \bibinfo
  {author} {\bibfnamefont {K.}~\bibnamefont {Umemoto}}, \bibinfo {author}
  {\bibfnamefont {J.}~\bibnamefont {Klepp}}, \bibinfo {author} {\bibfnamefont
  {C.}~\bibnamefont {Pruner}},\ and\ \bibinfo {author} {\bibfnamefont
  {M.}~\bibnamefont {Fally}},\ }\bibfield  {title} {\bibinfo {title}
  {Holographic nanodiamond composite gratings for light and slow-neutron beam
  control},\ }\href@noop {} {\bibfield  {journal} {\bibinfo  {journal}
  {Technical Digest of CLEO/Europe-EQEC 2019, Munich, Germany}\ ,\ \bibinfo
  {pages} {{CE}8.5}} (\bibinfo {year} {2019}{\natexlab{b}})}\BibitemShut
  {NoStop}%
\bibitem [{\citenamefont {Tomlinson}\ \emph {et~al.}(1976)\citenamefont
  {Tomlinson}, \citenamefont {Chandross}, \citenamefont {Weber},\ and\
  \citenamefont {Aumiller}}]{tomlinson1976}%
  \BibitemOpen
  \bibfield  {author} {\bibinfo {author} {\bibfnamefont {W.~J.}\ \bibnamefont
  {Tomlinson}}, \bibinfo {author} {\bibfnamefont {E.~A.}\ \bibnamefont
  {Chandross}}, \bibinfo {author} {\bibfnamefont {H.~P.}\ \bibnamefont
  {Weber}},\ and\ \bibinfo {author} {\bibfnamefont {G.~G.}\ \bibnamefont
  {Aumiller}},\ }\bibfield  {title} {\bibinfo {title} {Multicomponent
  photopolymer systems for volume phase holograms and grating devices},\
  }\href@noop {} {\bibfield  {journal} {\bibinfo  {journal} {Appl.~Opt.}\
  }\textbf {\bibinfo {volume} {15}},\ \bibinfo {pages} {534} (\bibinfo {year}
  {1976})}\BibitemShut {NoStop}%
\bibitem [{\citenamefont {Tomita}\ \emph
  {et~al.}(2005{\natexlab{a}})\citenamefont {Tomita}, \citenamefont {Suzuki},\
  and\ \citenamefont {Chikama}}]{tomita2005}%
  \BibitemOpen
  \bibfield  {author} {\bibinfo {author} {\bibfnamefont {Y.}~\bibnamefont
  {Tomita}}, \bibinfo {author} {\bibfnamefont {N.}~\bibnamefont {Suzuki}},\
  and\ \bibinfo {author} {\bibfnamefont {K.}~\bibnamefont {Chikama}},\
  }\bibfield  {title} {\bibinfo {title} {Holographic manipulation of
  nanoparticle distribution morphology in nanoparticle-dispersed
  photopolymers},\ }\href@noop {} {\bibfield  {journal} {\bibinfo  {journal}
  {Opt.~Lett.}\ }\textbf {\bibinfo {volume} {30}},\ \bibinfo {pages} {839}
  (\bibinfo {year} {2005}{\natexlab{a}})}\BibitemShut {NoStop}%
\bibitem [{\citenamefont {Tomita}\ \emph
  {et~al.}(2006{\natexlab{b}})\citenamefont {Tomita}, \citenamefont {Chikama},
  \citenamefont {Nohara}, \citenamefont {Suzuki}, \citenamefont {Furushima},\
  and\ \citenamefont {Endoh}}]{tomita2006}%
  \BibitemOpen
  \bibfield  {author} {\bibinfo {author} {\bibfnamefont {Y.}~\bibnamefont
  {Tomita}}, \bibinfo {author} {\bibfnamefont {K.}~\bibnamefont {Chikama}},
  \bibinfo {author} {\bibfnamefont {Y.}~\bibnamefont {Nohara}}, \bibinfo
  {author} {\bibfnamefont {N.}~\bibnamefont {Suzuki}}, \bibinfo {author}
  {\bibfnamefont {K.}~\bibnamefont {Furushima}},\ and\ \bibinfo {author}
  {\bibfnamefont {Y.}~\bibnamefont {Endoh}},\ }\bibfield  {title} {\bibinfo
  {title} {Two-dimensional imaging of atomic distribution morphology created by
  holographically induced mass transfer of monomer molecules and nanoparticles
  in a silica-nanoparticle-dispersed photopolymer film},\ }\href@noop {}
  {\bibfield  {journal} {\bibinfo  {journal} {Opt.~Lett.}\ }\textbf {\bibinfo
  {volume} {31}},\ \bibinfo {pages} {1402} (\bibinfo {year}
  {2006}{\natexlab{b}})}\BibitemShut {NoStop}%
\bibitem [{\citenamefont {Suzuki}\ and\ \citenamefont
  {Tomita}(2006)}]{suzukitomita2006}%
  \BibitemOpen
  \bibfield  {author} {\bibinfo {author} {\bibfnamefont {N.}~\bibnamefont
  {Suzuki}}\ and\ \bibinfo {author} {\bibfnamefont {Y.}~\bibnamefont
  {Tomita}},\ }\bibfield  {title} {\bibinfo {title} {Real-time phase-shift
  measurement during formation of a volume holographic grating in
  nanoparticle-dispersed photopolymers},\ }\href@noop {} {\bibfield  {journal}
  {\bibinfo  {journal} {Appl.~Phys.~Lett.}\ }\textbf {\bibinfo {volume} {88}},\
  \bibinfo {pages} {011105} (\bibinfo {year} {2006})}\BibitemShut {NoStop}%
\bibitem [{\citenamefont {Piao}\ and\ \citenamefont {Kim}(2014)}]{piao2014}%
  \BibitemOpen
  \bibfield  {author} {\bibinfo {author} {\bibfnamefont {M.-L.}\ \bibnamefont
  {Piao}}\ and\ \bibinfo {author} {\bibfnamefont {N.}~\bibnamefont {Kim}},\
  }\bibfield  {title} {\bibinfo {title} {Achieving high level of color
  uniformity and optical efficiency for a wedge-shaped waveguide head-mounted
  display using a photopolymer},\ }\href@noop {} {\bibfield  {journal}
  {\bibinfo  {journal} {Appl. Opt.}\ }\textbf {\bibinfo {volume} {53}},\
  \bibinfo {pages} {2180} (\bibinfo {year} {2014})}\BibitemShut {NoStop}%
\bibitem [{\citenamefont {Yeom}\ \emph {et~al.}(2015)\citenamefont {Yeom},
  \citenamefont {Kim}, \citenamefont {Kim}, \citenamefont {Zhang},
  \citenamefont {Li}, \citenamefont {Ji}, \citenamefont {Kim},\ and\
  \citenamefont {Park}}]{yeom2015}%
  \BibitemOpen
  \bibfield  {author} {\bibinfo {author} {\bibfnamefont {H.-J.}\ \bibnamefont
  {Yeom}}, \bibinfo {author} {\bibfnamefont {H.-J.}\ \bibnamefont {Kim}},
  \bibinfo {author} {\bibfnamefont {S.-B.}\ \bibnamefont {Kim}}, \bibinfo
  {author} {\bibfnamefont {H.}~\bibnamefont {Zhang}}, \bibinfo {author}
  {\bibfnamefont {B.}~\bibnamefont {Li}}, \bibinfo {author} {\bibfnamefont
  {Y.-M.}\ \bibnamefont {Ji}}, \bibinfo {author} {\bibfnamefont {S.-H.}\
  \bibnamefont {Kim}},\ and\ \bibinfo {author} {\bibfnamefont {H.-H.}\
  \bibnamefont {Park}},\ }\bibfield  {title} {\bibinfo {title} {{3D}
  holographic head mounted display using holographic optical elements with
  astigmatism aberration compensation},\ }\href@noop {} {\bibfield  {journal}
  {\bibinfo  {journal} {Opt. Express}\ }\textbf {\bibinfo {volume} {23}},\
  \bibinfo {pages} {32025} (\bibinfo {year} {2015})}\BibitemShut {NoStop}%
\bibitem [{\citenamefont {Tomita}\ \emph
  {et~al.}(2005{\natexlab{b}})\citenamefont {Tomita}, \citenamefont {Suzuki},
  \citenamefont {Furushima},\ and\ \citenamefont {Endoh}}]{tomita2005spie}%
  \BibitemOpen
  \bibfield  {author} {\bibinfo {author} {\bibfnamefont {Y.}~\bibnamefont
  {Tomita}}, \bibinfo {author} {\bibfnamefont {N.}~\bibnamefont {Suzuki}},
  \bibinfo {author} {\bibfnamefont {K.}~\bibnamefont {Furushima}},\ and\
  \bibinfo {author} {\bibfnamefont {Y.}~\bibnamefont {Endoh}},\ }\bibfield
  {title} {\bibinfo {title} {Volume holographic recording based on mass
  transport of nanoparticles doped in methacrylate photopolymers},\ }\href@noop
  {} {\bibfield  {journal} {\bibinfo  {journal} {Proc. SPIE}\ }\textbf
  {\bibinfo {volume} {5939}},\ \bibinfo {pages} {593909} (\bibinfo {year}
  {2005}{\natexlab{b}})}\BibitemShut {NoStop}%
\bibitem [{\citenamefont {Sears}(1989)}]{sears1989}%
  \BibitemOpen
  \bibfield  {author} {\bibinfo {author} {\bibfnamefont {V.~F.}\ \bibnamefont
  {Sears}},\ }\href@noop {} {\emph {\bibinfo {title} {Neutron Optics}}}\
  (\bibinfo  {publisher} {Oxford University Press},\ \bibinfo {year}
  {1989})\BibitemShut {NoStop}%
\bibitem [{\citenamefont {Avdeev}\ \emph {et~al.}(2013)\citenamefont {Avdeev},
  \citenamefont {Aksenov}, \citenamefont {Tomchuk}, \citenamefont {Bulavin},
  \citenamefont {Garamus},\ and\ \citenamefont {Osawa}}]{avdeev2013}%
  \BibitemOpen
  \bibfield  {author} {\bibinfo {author} {\bibfnamefont {M.~V.}\ \bibnamefont
  {Avdeev}}, \bibinfo {author} {\bibfnamefont {V.~L.}\ \bibnamefont {Aksenov}},
  \bibinfo {author} {\bibfnamefont {O.~V.}\ \bibnamefont {Tomchuk}}, \bibinfo
  {author} {\bibfnamefont {L.}~\bibnamefont {Bulavin}}, \bibinfo {author}
  {\bibfnamefont {V.~M.}\ \bibnamefont {Garamus}},\ and\ \bibinfo {author}
  {\bibfnamefont {E.}~\bibnamefont {Osawa}},\ }\bibfield  {title} {\bibinfo
  {title} {The spatial diamond--graphite transition in detonation nanodiamond
  as revealed by small-angle neutron scattering},\ }\href@noop {} {\bibfield
  {journal} {\bibinfo  {journal} {J. Phys.: Condens. Matter}\ }\textbf
  {\bibinfo {volume} {25}},\ \bibinfo {pages} {445001} (\bibinfo {year}
  {2013})}\BibitemShut {NoStop}%
\bibitem [{\citenamefont {Dianoux}\ and\ \citenamefont
  {Lander}(2003)}]{ILLNeutronDataBooklet2003}%
  \BibitemOpen
  \bibinfo {editor} {\bibfnamefont {A.-J.}\ \bibnamefont {Dianoux}}\ and\
  \bibinfo {editor} {\bibfnamefont {G.}~\bibnamefont {Lander}},\ eds.,\
  \href@noop {} {\emph {\bibinfo {title} {Neutron Data Booklet}}}\ (\bibinfo
  {publisher} {Old City Publishing, Copyright 2003 Institut Laue-Langevin},\
  \bibinfo {address} {Philadelphia, USA},\ \bibinfo {year} {2003})\ \bibinfo
  {note} {\url{https://www.ill.eu/quick-links/publications/}}\BibitemShut
  {NoStop}%
\bibitem [{\citenamefont {Gr\"{u}nzweig}\ \emph {et~al.}(2008)\citenamefont
  {Gr\"{u}nzweig}, \citenamefont {Pfeiffer}, \citenamefont {Bunk},
  \citenamefont {Donath}, \citenamefont {K\"{u}hne}, \citenamefont {Frei},
  \citenamefont {Dierolf},\ and\ \citenamefont
  {David}}]{gruenzweigRScIntr2008}%
  \BibitemOpen
  \bibfield  {author} {\bibinfo {author} {\bibfnamefont {C.}~\bibnamefont
  {Gr\"{u}nzweig}}, \bibinfo {author} {\bibfnamefont {F.}~\bibnamefont
  {Pfeiffer}}, \bibinfo {author} {\bibfnamefont {O.}~\bibnamefont {Bunk}},
  \bibinfo {author} {\bibfnamefont {T.}~\bibnamefont {Donath}}, \bibinfo
  {author} {\bibfnamefont {G.}~\bibnamefont {K\"{u}hne}}, \bibinfo {author}
  {\bibfnamefont {G.}~\bibnamefont {Frei}}, \bibinfo {author} {\bibfnamefont
  {M.}~\bibnamefont {Dierolf}},\ and\ \bibinfo {author} {\bibfnamefont
  {C.}~\bibnamefont {David}},\ }\bibfield  {title} {\bibinfo {title} {Design,
  fabrication, and characterization of diffraction gratings for neutron phase
  contrast imaging},\ }\href {https://doi.org/10.1063/1.2930866} {\bibfield
  {journal} {\bibinfo  {journal} {Rev. Sci. Instrum.}\ }\textbf {\bibinfo
  {volume} {79}},\ \bibinfo {pages} {053703} (\bibinfo {year}
  {2008})}\BibitemShut {NoStop}%
\bibitem [{\citenamefont {Miao}\ \emph {et~al.}(2014)\citenamefont {Miao},
  \citenamefont {Gomella}, \citenamefont {Chedid}, \citenamefont {Chen},\ and\
  \citenamefont {Wen}}]{miaoNL2014}%
  \BibitemOpen
  \bibfield  {author} {\bibinfo {author} {\bibfnamefont {H.}~\bibnamefont
  {Miao}}, \bibinfo {author} {\bibfnamefont {A.~A.}\ \bibnamefont {Gomella}},
  \bibinfo {author} {\bibfnamefont {N.}~\bibnamefont {Chedid}}, \bibinfo
  {author} {\bibfnamefont {L.}~\bibnamefont {Chen}},\ and\ \bibinfo {author}
  {\bibfnamefont {H.}~\bibnamefont {Wen}},\ }\bibfield  {title} {\bibinfo
  {title} {Fabrication of 200 nm period hard x-ray phase gratings},\ }\href
  {https://doi.org/10.1021/nl5009713} {\bibfield  {journal} {\bibinfo
  {journal} {Nano Letters}\ }\textbf {\bibinfo {volume} {14}},\ \bibinfo
  {pages} {3453} (\bibinfo {year} {2014})}\BibitemShut {NoStop}%
\bibitem [{\citenamefont {Mochalin}\ \emph {et~al.}(2008)\citenamefont
  {Mochalin}, \citenamefont {Osswald},\ and\ \citenamefont
  {Gogotsi}}]{mochalin2008}%
  \BibitemOpen
  \bibfield  {author} {\bibinfo {author} {\bibfnamefont {V.}~\bibnamefont
  {Mochalin}}, \bibinfo {author} {\bibfnamefont {S.}~\bibnamefont {Osswald}},\
  and\ \bibinfo {author} {\bibfnamefont {Y.}~\bibnamefont {Gogotsi}},\
  }\bibfield  {title} {\bibinfo {title} {Contribution of functional groups to
  the {R}aman spectrum of nanodiamond powders},\ }\href@noop {} {\bibfield
  {journal} {\bibinfo  {journal} {Chem. Mater.}\ }\textbf {\bibinfo {volume}
  {21}},\ \bibinfo {pages} {273} (\bibinfo {year} {2008})}\BibitemShut
  {NoStop}%
\bibitem [{\citenamefont {Ahn}\ \emph {et~al.}(2004)\citenamefont {Ahn},
  \citenamefont {Kim}, \citenamefont {Kim}, \citenamefont {Shim},\ and\
  \citenamefont {Cho}}]{ahn2004}%
  \BibitemOpen
  \bibfield  {author} {\bibinfo {author} {\bibfnamefont {S.~H.}\ \bibnamefont
  {Ahn}}, \bibinfo {author} {\bibfnamefont {S.~H.}\ \bibnamefont {Kim}},
  \bibinfo {author} {\bibfnamefont {B.~C.}\ \bibnamefont {Kim}}, \bibinfo
  {author} {\bibfnamefont {K.~B.}\ \bibnamefont {Shim}},\ and\ \bibinfo
  {author} {\bibfnamefont {B.~G.}\ \bibnamefont {Cho}},\ }\bibfield  {title}
  {\bibinfo {title} {Mechanical properties of silica nanoparticle reinforced
  poly(ethylene 2, 6-naphthalate)},\ }\href@noop {} {\bibfield  {journal}
  {\bibinfo  {journal} {Macromol. Res.}\ }\textbf {\bibinfo {volume} {12}},\
  \bibinfo {pages} {293} (\bibinfo {year} {2004})}\BibitemShut {NoStop}%
\bibitem [{\citenamefont {Terada}\ \emph {et~al.}(2019)\citenamefont {Terada},
  \citenamefont {Segawa}, \citenamefont {Shames}, \citenamefont {Onoda},
  \citenamefont {Ohshima}, \citenamefont {Osawa}, \citenamefont {Igarashi},\
  and\ \citenamefont {Shirakawa}}]{terada2019}%
  \BibitemOpen
  \bibfield  {author} {\bibinfo {author} {\bibfnamefont {D.}~\bibnamefont
  {Terada}}, \bibinfo {author} {\bibfnamefont {T.~F.}\ \bibnamefont {Segawa}},
  \bibinfo {author} {\bibfnamefont {A.~I.}\ \bibnamefont {Shames}}, \bibinfo
  {author} {\bibfnamefont {S.}~\bibnamefont {Onoda}}, \bibinfo {author}
  {\bibfnamefont {T.}~\bibnamefont {Ohshima}}, \bibinfo {author} {\bibfnamefont
  {E.}~\bibnamefont {Osawa}}, \bibinfo {author} {\bibfnamefont
  {R.}~\bibnamefont {Igarashi}},\ and\ \bibinfo {author} {\bibfnamefont
  {M.}~\bibnamefont {Shirakawa}},\ }\bibfield  {title} {\bibinfo {title}
  {Monodisperse five-nanometer-sized detonation nanodiamonds enriched in
  nitrogen-vacancy centers},\ }\href@noop {} {\bibfield  {journal} {\bibinfo
  {journal} {ACS Nano}\ }\textbf {\bibinfo {volume} {13}},\ \bibinfo {pages}
  {6461} (\bibinfo {year} {2019})}\BibitemShut {NoStop}%
\bibitem [{\citenamefont {Duan}\ \emph {et~al.}(2019)\citenamefont {Duan},
  \citenamefont {Tian}, \citenamefont {Zhang}, \citenamefont {Sun},
  \citenamefont {Ao}, \citenamefont {Shao},\ and\ \citenamefont
  {Wang}}]{duan2019}%
  \BibitemOpen
  \bibfield  {author} {\bibinfo {author} {\bibfnamefont {X.}~\bibnamefont
  {Duan}}, \bibinfo {author} {\bibfnamefont {W.}~\bibnamefont {Tian}}, \bibinfo
  {author} {\bibfnamefont {J.}~\bibnamefont {Zhang}}, \bibinfo {author}
  {\bibfnamefont {H.}~\bibnamefont {Sun}}, \bibinfo {author} {\bibfnamefont
  {Z.}~\bibnamefont {Ao}}, \bibinfo {author} {\bibfnamefont {Z.}~\bibnamefont
  {Shao}},\ and\ \bibinfo {author} {\bibfnamefont {S.}~\bibnamefont {Wang}},\
  }\bibfield  {title} {\bibinfo {title} {{sp$^2$/sp$^3$} framework from diamond
  nanocrystals: {A} key bridge of carbonaceous structure to cabocatalysis},\
  }\href@noop {} {\bibfield  {journal} {\bibinfo  {journal} {ACS Catalysis}\
  }\textbf {\bibinfo {volume} {9}},\ \bibinfo {pages} {7494} (\bibinfo {year}
  {2019})}\BibitemShut {NoStop}%
\bibitem [{\citenamefont {Kogelnik}(1969)}]{kogelnik1969}%
  \BibitemOpen
  \bibfield  {author} {\bibinfo {author} {\bibfnamefont {H.}~\bibnamefont
  {Kogelnik}},\ }\bibfield  {title} {\bibinfo {title} {Coupled wave theory for
  thick hologram gratings},\ }\href@noop {} {\bibfield  {journal} {\bibinfo
  {journal} {Bell Syst. Tech. J.}\ }\textbf {\bibinfo {volume} {48}},\ \bibinfo
  {pages} {2909} (\bibinfo {year} {1969})}\BibitemShut {NoStop}%
\bibitem [{\citenamefont {Uchida}(1973)}]{uchida1973}%
  \BibitemOpen
  \bibfield  {author} {\bibinfo {author} {\bibfnamefont {N.}~\bibnamefont
  {Uchida}},\ }\bibfield  {title} {\bibinfo {title} {Calculation of diffraction
  efficiency in hologram gratings attenuated along the direction perpendicular
  to the grating vector},\ }\href@noop {} {\bibfield  {journal} {\bibinfo
  {journal} {J. Opt. Soc. Am.}\ }\textbf {\bibinfo {volume} {63}},\ \bibinfo
  {pages} {280} (\bibinfo {year} {1973})}\BibitemShut {NoStop}%
\bibitem [{\citenamefont {Yeh}(1993)}]{yeh1993}%
  \BibitemOpen
  \bibfield  {author} {\bibinfo {author} {\bibfnamefont {P.}~\bibnamefont
  {Yeh}},\ }\href@noop {} {\emph {\bibinfo {title} {Introduction to
  Photorefractive Nonlinear Optics}}}\ (\bibinfo  {publisher} {Wiley},\
  \bibinfo {year} {1993})\BibitemShut {NoStop}%
\bibitem [{\citenamefont {Sharda}\ \emph {et~al.}(2003)\citenamefont {Sharda},
  \citenamefont {Soga},\ and\ \citenamefont {Jimbo}}]{sharda2003}%
  \BibitemOpen
  \bibfield  {author} {\bibinfo {author} {\bibfnamefont {T.}~\bibnamefont
  {Sharda}}, \bibinfo {author} {\bibfnamefont {T.}~\bibnamefont {Soga}},\ and\
  \bibinfo {author} {\bibfnamefont {T.}~\bibnamefont {Jimbo}},\ }\bibfield
  {title} {\bibinfo {title} {Optical properties of nanocrystalline diamond
  films by prism coupling technique},\ }\href@noop {} {\bibfield  {journal}
  {\bibinfo  {journal} {J. Appl. Phys.}\ }\textbf {\bibinfo {volume} {93}},\
  \bibinfo {pages} {101} (\bibinfo {year} {2003})}\BibitemShut {NoStop}%
\bibitem [{\citenamefont {Phillip}\ and\ \citenamefont
  {Taft}(1964)}]{phillip1964}%
  \BibitemOpen
  \bibfield  {author} {\bibinfo {author} {\bibfnamefont {H.~R.}\ \bibnamefont
  {Phillip}}\ and\ \bibinfo {author} {\bibfnamefont {E.~A.}\ \bibnamefont
  {Taft}},\ }\bibfield  {title} {\bibinfo {title} {{K}ramers-{K}ronig analysis
  of reflectance data for diamond},\ }\href@noop {} {\bibfield  {journal}
  {\bibinfo  {journal} {Phys. Rev.}\ }\textbf {\bibinfo {volume} {136}},\
  \bibinfo {pages} {A1445} (\bibinfo {year} {1964})}\BibitemShut {NoStop}%
\bibitem [{\citenamefont {Djuri\v{s}i\'{c}}\ and\ \citenamefont
  {Li}(1999)}]{djurisic1999}%
  \BibitemOpen
  \bibfield  {author} {\bibinfo {author} {\bibfnamefont {A.~B.}\ \bibnamefont
  {Djuri\v{s}i\'{c}}}\ and\ \bibinfo {author} {\bibfnamefont {E.~H.}\
  \bibnamefont {Li}},\ }\bibfield  {title} {\bibinfo {title} {Optical
  properties of graphite},\ }\href@noop {} {\bibfield  {journal} {\bibinfo
  {journal} {J. Appl. Phys.}\ }\textbf {\bibinfo {volume} {85}},\ \bibinfo
  {pages} {7404} (\bibinfo {year} {1999})}\BibitemShut {NoStop}%
\bibitem [{\citenamefont {Bodurov}\ \emph {et~al.}(2016)\citenamefont
  {Bodurov}, \citenamefont {Vlaeva}, \citenamefont {Viraneva}, \citenamefont
  {Yovcheva},\ and\ \citenamefont {Sainov}}]{bodurov2016}%
  \BibitemOpen
  \bibfield  {author} {\bibinfo {author} {\bibfnamefont {I.}~\bibnamefont
  {Bodurov}}, \bibinfo {author} {\bibfnamefont {I.}~\bibnamefont {Vlaeva}},
  \bibinfo {author} {\bibfnamefont {A.}~\bibnamefont {Viraneva}}, \bibinfo
  {author} {\bibfnamefont {T.}~\bibnamefont {Yovcheva}},\ and\ \bibinfo
  {author} {\bibfnamefont {S.}~\bibnamefont {Sainov}},\ }\bibfield  {title}
  {\bibinfo {title} {Modified design of a laser refractometer},\ }\href@noop {}
  {\bibfield  {journal} {\bibinfo  {journal} {Nanoscience \& Nanotechnology}\
  }\textbf {\bibinfo {volume} {16}},\ \bibinfo {pages} {31} (\bibinfo {year}
  {2016})}\BibitemShut {NoStop}%
\bibitem [{\citenamefont {Kasarova}\ \emph {et~al.}(2007)\citenamefont
  {Kasarova}, \citenamefont {Sultanova}, \citenamefont {Ivanov},\ and\
  \citenamefont {Nikolov}}]{kasarova2007}%
  \BibitemOpen
  \bibfield  {author} {\bibinfo {author} {\bibfnamefont {S.~N.}\ \bibnamefont
  {Kasarova}}, \bibinfo {author} {\bibfnamefont {N.~G.}\ \bibnamefont
  {Sultanova}}, \bibinfo {author} {\bibfnamefont {C.~D.}\ \bibnamefont
  {Ivanov}},\ and\ \bibinfo {author} {\bibfnamefont {I.~D.}\ \bibnamefont
  {Nikolov}},\ }\bibfield  {title} {\bibinfo {title} {Analysis of the
  dispersion of optical plastic materials},\ }\href@noop {} {\bibfield
  {journal} {\bibinfo  {journal} {Opt. Mater.}\ }\textbf {\bibinfo {volume}
  {29}},\ \bibinfo {pages} {1481} (\bibinfo {year} {2007})}\BibitemShut
  {NoStop}%
\bibitem [{\citenamefont {Poularikas}(1985)}]{poularikas1985}%
  \BibitemOpen
  \bibfield  {author} {\bibinfo {author} {\bibfnamefont {A.~D.}\ \bibnamefont
  {Poularikas}},\ }\bibfield  {title} {\bibinfo {title} {Effective index of
  refraction of isotropic media containing layered spheres},\ }\href@noop {}
  {\bibfield  {journal} {\bibinfo  {journal} {J. Appl Phys.}\ }\textbf
  {\bibinfo {volume} {58}},\ \bibinfo {pages} {1044} (\bibinfo {year}
  {1985})}\BibitemShut {NoStop}%
\bibitem [{\citenamefont {Klein}\ and\ \citenamefont {Cook}(1967)}]{klein1967}%
  \BibitemOpen
  \bibfield  {author} {\bibinfo {author} {\bibfnamefont {W.~R.}\ \bibnamefont
  {Klein}}\ and\ \bibinfo {author} {\bibfnamefont {B.~D.}\ \bibnamefont
  {Cook}},\ }\bibfield  {title} {\bibinfo {title} {Unified approach to
  ultrasonic light diffraction},\ }\href@noop {} {\bibfield  {journal}
  {\bibinfo  {journal} {IEEE Trans. Sonics Ultrason.}\ }\textbf {\bibinfo
  {volume} {SU-14}},\ \bibinfo {pages} {123} (\bibinfo {year}
  {1967})}\BibitemShut {NoStop}%
\bibitem [{\citenamefont {Moharam}\ and\ \citenamefont
  {Young}(1978)}]{moharam1978}%
  \BibitemOpen
  \bibfield  {author} {\bibinfo {author} {\bibfnamefont {M.~G.}\ \bibnamefont
  {Moharam}}\ and\ \bibinfo {author} {\bibfnamefont {L.}~\bibnamefont
  {Young}},\ }\bibfield  {title} {\bibinfo {title} {Criteria for {Bragg} and
  {Raman-Nath} regimes},\ }\href@noop {} {\bibfield  {journal} {\bibinfo
  {journal} {Appl. Opt.}\ }\textbf {\bibinfo {volume} {17}},\ \bibinfo {pages}
  {1757} (\bibinfo {year} {1978})}\BibitemShut {NoStop}%
\bibitem [{\citenamefont {Moharam}\ \emph
  {et~al.}(1980{\natexlab{a}})\citenamefont {Moharam}, \citenamefont
  {Gaylord},\ and\ \citenamefont {Magnusson}}]{moharam1980a}%
  \BibitemOpen
  \bibfield  {author} {\bibinfo {author} {\bibfnamefont {M.~G.}\ \bibnamefont
  {Moharam}}, \bibinfo {author} {\bibfnamefont {T.~K.}\ \bibnamefont
  {Gaylord}},\ and\ \bibinfo {author} {\bibfnamefont {R.}~\bibnamefont
  {Magnusson}},\ }\bibfield  {title} {\bibinfo {title} {Criteria for {Bragg}
  regime diffraction by phase gratings},\ }\href@noop {} {\bibfield  {journal}
  {\bibinfo  {journal} {Opt. Commun.}\ }\textbf {\bibinfo {volume} {32}},\
  \bibinfo {pages} {14} (\bibinfo {year} {1980}{\natexlab{a}})}\BibitemShut
  {NoStop}%
\bibitem [{\citenamefont {Moharam}\ \emph
  {et~al.}(1980{\natexlab{b}})\citenamefont {Moharam}, \citenamefont
  {Gaylord},\ and\ \citenamefont {Magnusson}}]{moharam1980b}%
  \BibitemOpen
  \bibfield  {author} {\bibinfo {author} {\bibfnamefont {M.~G.}\ \bibnamefont
  {Moharam}}, \bibinfo {author} {\bibfnamefont {T.~K.}\ \bibnamefont
  {Gaylord}},\ and\ \bibinfo {author} {\bibfnamefont {R.}~\bibnamefont
  {Magnusson}},\ }\bibfield  {title} {\bibinfo {title} {Criteria for
  {Raman-Nath} regime diffraction by phase gratings},\ }\href@noop {}
  {\bibfield  {journal} {\bibinfo  {journal} {Opt. Commun.}\ }\textbf {\bibinfo
  {volume} {32}},\ \bibinfo {pages} {19} (\bibinfo {year}
  {1980}{\natexlab{b}})}\BibitemShut {NoStop}%
\bibitem [{\citenamefont {Moharam}\ and\ \citenamefont
  {Gaylord}(1981)}]{moharam1981}%
  \BibitemOpen
  \bibfield  {author} {\bibinfo {author} {\bibfnamefont {M.~G.}\ \bibnamefont
  {Moharam}}\ and\ \bibinfo {author} {\bibfnamefont {T.~K.}\ \bibnamefont
  {Gaylord}},\ }\bibfield  {title} {\bibinfo {title} {Rigorous coupled-wave
  analysis of planar-grating diffraction},\ }\href@noop {} {\bibfield
  {journal} {\bibinfo  {journal} {J. Opt. Soc. Am.}\ }\textbf {\bibinfo
  {volume} {71}},\ \bibinfo {pages} {811} (\bibinfo {year} {1981})}\BibitemShut
  {NoStop}%
\bibitem [{\citenamefont {Oda}\ \emph {et~al.}(2017)\citenamefont {Oda},
  \citenamefont {Hino}, \citenamefont {Kitaguchi}, \citenamefont {Filter},
  \citenamefont {Geltenbort},\ and\ \citenamefont {Kawabata}}]{odaNIMA2017}%
  \BibitemOpen
  \bibfield  {author} {\bibinfo {author} {\bibfnamefont {T.}~\bibnamefont
  {Oda}}, \bibinfo {author} {\bibfnamefont {M.}~\bibnamefont {Hino}}, \bibinfo
  {author} {\bibfnamefont {M.}~\bibnamefont {Kitaguchi}}, \bibinfo {author}
  {\bibfnamefont {H.}~\bibnamefont {Filter}}, \bibinfo {author} {\bibfnamefont
  {P.}~\bibnamefont {Geltenbort}},\ and\ \bibinfo {author} {\bibfnamefont
  {Y.}~\bibnamefont {Kawabata}},\ }\bibfield  {title} {\bibinfo {title}
  {Towards a high-resolution {TOF-MIEZE} spectrometer with very cold
  neutrons},\ }\href
  {https://doi.org/https://doi.org/10.1016/j.nima.2017.03.014} {\bibfield
  {journal} {\bibinfo  {journal} {Nucl. Instrum. Meth. Phys. Res. A}\ }\textbf
  {\bibinfo {volume} {860}},\ \bibinfo {pages} {35} (\bibinfo {year}
  {2017})}\BibitemShut {NoStop}%
\bibitem [{\citenamefont {Fally}\ \emph {et~al.}(2018)\citenamefont {Fally},
  \citenamefont {Geltenbort}, \citenamefont {Jenke}, \citenamefont {Klepp},
  \citenamefont {Nesvizhevsky},\ and\ \citenamefont {Pruner}}]{dataGrating1}%
  \BibitemOpen
  \bibfield  {author} {\bibinfo {author} {\bibfnamefont {M.}~\bibnamefont
  {Fally}}, \bibinfo {author} {\bibfnamefont {P.}~\bibnamefont {Geltenbort}},
  \bibinfo {author} {\bibfnamefont {T.}~\bibnamefont {Jenke}}, \bibinfo
  {author} {\bibfnamefont {J.}~\bibnamefont {Klepp}}, \bibinfo {author}
  {\bibfnamefont {V.}~\bibnamefont {Nesvizhevsky}},\ and\ \bibinfo {author}
  {\bibfnamefont {C.}~\bibnamefont {Pruner}},\ }\bibfield  {title} {\bibinfo
  {title} {Ionic-liquids composites {(ILCs)} for holographic-grating
  neutron-optical elements},\ }\bibfield  {journal} {\bibinfo  {journal}
  {Institut Laue-Langevin (ILL): Grenoble,}\ }\href
  {https://doi.org/http://dx.doi.org/10.5291/ILL-DATA.3-14-376}
  {http://dx.doi.org/10.5291/ILL-DATA.3-14-376} (\bibinfo {year}
  {2018})\BibitemShut {NoStop}%
\bibitem [{\citenamefont {Fally}\ \emph {et~al.}(2020)\citenamefont {Fally},
  \citenamefont {Jenke}, \citenamefont {Klepp},\ and\ \citenamefont
  {Pruner}}]{dataGrating2}%
  \BibitemOpen
  \bibfield  {author} {\bibinfo {author} {\bibfnamefont {M.}~\bibnamefont
  {Fally}}, \bibinfo {author} {\bibfnamefont {T.}~\bibnamefont {Jenke}},
  \bibinfo {author} {\bibfnamefont {J.}~\bibnamefont {Klepp}},\ and\ \bibinfo
  {author} {\bibfnamefont {C.}~\bibnamefont {Pruner}},\ }\bibfield  {title}
  {\bibinfo {title} {Diamond nanoparticles in nanoparticle-polymer composite
  neutron diffraction gratings},\ }\bibfield  {journal} {\bibinfo  {journal}
  {Institut Laue-Langevin (ILL): Grenoble,}\ }\href
  {https://doi.org/http://dx.doi.org/10.5291/ILL-DATA.3-14-392}
  {http://dx.doi.org/10.5291/ILL-DATA.3-14-392} (\bibinfo {year}
  {2020})\BibitemShut {NoStop}%
\bibitem [{\citenamefont {Markel}(2016)}]{markel2016}%
  \BibitemOpen
  \bibfield  {author} {\bibinfo {author} {\bibfnamefont {V.~A.}\ \bibnamefont
  {Markel}},\ }\bibfield  {title} {\bibinfo {title} {Introduction to the
  {Maxwell Garnett} approximation: tutorial},\ }\href@noop {} {\bibfield
  {journal} {\bibinfo  {journal} {J. Opt. Soc. Am. A}\ }\textbf {\bibinfo
  {volume} {33}},\ \bibinfo {pages} {1244} (\bibinfo {year}
  {2016})}\BibitemShut {NoStop}%
\end{thebibliography}
\end{document}